\documentclass[aps,rmp,twocolumn,groupedaddress]{revtex4-2}
\usepackage[T1]{fontenc}
\usepackage[utf8]{inputenc}
\usepackage[colorlinks=true, linkcolor=blue, citecolor=blue, urlcolor=blue]{hyperref}
\usepackage{graphicx}
\usepackage{amsmath}
\usepackage{longtable}
\usepackage{braket}
\makeatletter
\renewcommand{\fnum@figure}{FIG.~\thefigure.}
\renewcommand{\fnum@table}{TABLE~\thetable.}
\makeatother

\begin{document}


\title{2D van der Waals magnets: from fundamental physics to applications}


\author{Je-Geun Park}
\email[]{jgpark10@snu.ac.kr}
\affiliation{Department of Physics and Astronomy, Seoul National University, Seoul 08826, Republic of Korea}

\author{Kaixuan Zhang}
\affiliation{Department of Physics and Astronomy, Seoul National University, Seoul 08826, Republic of Korea}

\author{Hyeonsik Cheong}
\affiliation{Department of Physics and Center for Nano Materials, Sogang University, Seoul 04107, Republic of Korea}

\author{Jae Hoon Kim }
\affiliation{Department of Physics, Yonsei University, Seoul, 03722, Republic of Korea}

\author{Carina A. Belvin}
\affiliation{Department of Physics,California Institute of Technology, Pasadena, California 91125, USA}

\author{David Hsieh}
\affiliation{Department of Physics,California Institute of Technology, Pasadena, California 91125, USA}

\author{Honglie Ning}
\affiliation{Department of Physics, Massachusetts Institute of Technology, Cambridge, Massachusetts 02139, USA}

\author{Nuh Gedik}
\affiliation{Department of Physics, Massachusetts Institute of Technology, Cambridge, Massachusetts 02139, USA}


\begin{abstract}
Magnetism has played a central role in the long and rich history of modern condensed matter physics, with many foundational insights originating from theoretical studies of two-dimensional (2D) spin systems. The discovery of 2D van der Waals (vdW) magnets has revolutionized this area by providing real, atomically thin magnetic systems for experimental investigation. Since the first experimental reports of antiferromagnetic vdW insulators in 2016—followed by studies on ferromagnetic vdW systems in 2017—the field has witnessed rapid and expansive growth, with more than two dozen vdW magnetic materials now identified, including both ferro- and antiferromagnets. In this review, we present a comprehensive overview of the major scientific and technological developments in this rapidly evolving field. These include experimental realizations of various 2D spin Hamiltonians as well as unexpected phenomena such as magnetic excitons, Floquet-engineered states, and light-induced metastable magnetic phases. In parallel, 2D vdW magnets have shown significant promise in spintronics and related applications, offering a new platform for engineering quantum functionalities. We organize this review by tracing the historical development of the field, synthesizing key milestones, and highlighting its broader impact across condensed matter physics and materials science. We conclude with an Outlook section that outlines several promising directions for future research, aiming to chart a path forward in this vibrant and still rapidly growing area.
\end{abstract}


\maketitle

\tableofcontents

\section{\label{Intro}Introduction}
Language is a unique characteristic that separates humans from all other species. Similarly, in science, magnetism has served as a universal ``language'', advancing our understanding of nature and driving technological innovations \cite{RN1}. Magnetism has played a pivotal role in physics since William Gilbert's De Magnete, Magneticisque Corporibus, et de Magno Magnete Tellure (On the Magnet and Magnetic Bodies, and on That Great Magnet the Earth), published in 1600. Over centuries, it has inspired the development of modern condensed matter physics, providing models and test beds for validating and challenging theoretical ideas. Although traditional magnets, such as horseshoe magnets, are typically bulky and three-dimensional, low-dimensional magnets have emerged as invaluable systems for exploring novel theoretical concepts.

Among these, two-dimensional magnets stand out for their role in pushing the boundaries of knowledge. Until recently, studies of two-dimensional magnets were limited to a handful of quasi-two-dimensional, natural, or artificial samples \cite{RN2, RN3}. This changed with the discovery of vdW magnets, which naturally occur as three-dimensional materials that can be exfoliated into truly two-dimensional layers. First reported in 2016, these materials have already revealed surprising and unprecedented properties, including the realization of true two-dimensional magnetism \cite{RN4, RN5, RN6, RN7, RN8}.

Uniquely, vdW magnets are intrinsically two-dimensional, offering unparalleled opportunities to observe and study diverse types of magnetism with unique properties in real materials. This review explores the scientific potential and technological applications of vdW magnets, a class of materials that has redefined our understanding of two-dimensional magnetism.

The beauty and elegance of magnetism as a research topic lie in the simplicity of the theoretical models required to understand the experimental data. This profound connection between magnetism and theoretical physics has deep roots in modern physics. More often than not, the interplay between experiments and theories has been mutually transformative and far-reaching. A prime example is the observation of resistance minima in several otherwise normal metals, such as Au, Cu, and Pb, in the early 1930s \cite{RN9}. This experimental finding eventually led to the development of spin glass physics \cite{RN10}, with the Kondo model introduced in 1964 \cite{RN11} providing its theoretical foundation. The Kondo model, in turn, became the cornerstone of heavy fermion physics \cite{RN12}, illustrating how experimental discoveries in magnetism can spawn entirely new fields. 

Spin-glass physics is notable for exploring systems with competing interactions and disorder, where frustration leads to a highly complex energy landscape. This concept, originally developed to understand magnetic materials such as disordered alloys, has profound implications beyond condensed matter physics. Building on this complexity, spin glass theory has found applications in the field of neural networks \cite{RN13}. For example, the Hopfield network leverages the energy minimization behavior observed in spin glasses to store and retrieve complex memory patterns. Thus, spin-glass physics contributes not only to understanding disordered magnetic systems but also serves as a critical framework for addressing challenges in computational and biological systems. This interdisciplinary relevance highlights the profound and far-reaching impact of magnetism and its theoretical models.

The reciprocal relationship between experimental and theoretical magnetism is striking. Over recent decades, a series of material systems have emerged, including spin glasses, heavy fermions, high-temperature superconductors, colossal magnetoresistance (CMR) manganites, and multiferroics. Theoretically, the Kondo, Hubbard, Anderson, and $t-J$ models have provided critical frameworks. A survey of historical developments reveals that each major discovery of a material has driven advances in theoretical understanding, which, in turn, has deepened the understanding of material physics. This intimate connection arises because magnetism often provides well-defined physical systems, such as Ising-like spin glasses or quantum spin liquids, that allow both qualitative and quantitative analyses of phenomena. The wealth of information gleaned from magnetic systems has been a driving force behind decades of research, bridging experiment and theory in a uniquely productive manner.

When a magnet undergoes a phase transition, its behavior can often be entirely captured by a simple theoretical model, which falls into one of a few universality classes. This remarkable feature, that diverse materials and phenomena in nature can be classified into a few universal models, is one of the most captivating aspects of phase transitions \cite{RN14}. At the heart of this understanding are the powerful concepts of scaling, universality, and renormalization, which provide the framework for describing critical phenomena together.

Although naturally occurring materials are inherently three-dimensional, two-dimensional systems hold a unique place in theoretical physics. These systems often yield analytical solutions with profound implications, which makes them a focal point for exploration. Among the key models for two-dimensional magnetism are the Ising, XY, and Heisenberg models, all described by a common spin Hamiltonian.

\begin{equation}
\label{eq:1}
H = \sum_{(i,j)} J_{i,j} \left[ (S_i^x \cdot S_j^x + S_i^y \cdot S_j^y ) + \alpha S_i^z \cdot S_j^z \right]
\end{equation}

where $(i,j)$ denotes a pair of neighboring spins, $J_{i,j}$ is the exchange integral, and $S_i^{x,y,z}$ are the $x$, $y$, and $z$ components of the spin at the $i$ site. Depending on the value of $\alpha$, this Hamiltonian describes the Ising model with $\alpha=\infty$, the XY model with $\alpha=0$, and the Heisenberg model with $\alpha=1$.

The Ising model was the first to be solved analytically. In 1944, Onsager used the transfer matrix method to derive an exact solution for magnetization as a function of temperature and the exchange integral \cite{RN15}. Today, the Ising model remains a paradigm for studying phase transitions, with applications that extend to lattice gases, percolation, and even modeling social networks.
The Heisenberg model ($\alpha=1$) presented a different challenge. In 1966, Mermin and Wagner, followed by Hohenberg in 1967, rigorously proved that continuous spin symmetries cannot be spontaneously broken at finite temperatures in two dimensions \cite{RN16, RN17}. Known as the Mermin-Wagner theorem, this result underscored the role of long-wavelength fluctuations in suppressing order. 
The XY model ($\alpha=0$) revealed its mysteries in the early 1970s through the work by Berezinskii \cite{RN18}, Kosterlitz, and Thouless \cite{RN19}. Unlike ordinary phase transitions, the two-dimensional XY model undergoes a topological transition at finite temperature, marked by a shift from bound vortex-antivortex pairs at low temperatures to unbound vortices at higher temperatures.

The theoretical successes of the three models were soon followed by experimental efforts to test their predictions. These efforts gained even greater significance after Wilson’s seminal work on renormalization group theory \cite{RN20}. Although all three models have been studied using real magnetic materials, the Ising model received the most attention with studies involving materials such as Dy(C$_2$H$_5$SO$_4$)$_3\cdot$9H$_2$O \cite{RN21}, Dy$_3$Al$_5$O$_{12}$ \cite{RN22}, DyPO$_4$ \cite{RN23}, LiTbF$_4$ \cite{RN24}, and Rb$_2$CoF$_4$ \cite{RN25}. Several systems, such as K$_2$CuF$_4$, were used to study the XY model \cite{RN26}. Despite their success, these studies faced two fundamental limitations in strictly testing the two-dimensional nature of these models.

The first limitation arises from the intrinsic dimensionality of the materials. Although systems like Rb$_2$CoF$_4$ and K$_2$CuF$_4$ exhibit weak interlayer coupling, they remain quasi-two-dimensional, with residual three-dimensional interactions influencing their critical behavior. This deviation from idealized models limits their ability to replicate two-dimensional theory predictions fully. The second challenge stems from the technical constraints of earlier experimental methods. Conventional tools like bulk magnetometry, neutron scattering, and specific heat measurements lack the spatial resolution and sensitivities necessary to probe monolayer magnetic systems. These limitations have historically left many properties of truly two-dimensional systems inaccessible.

Recent advances have begun to address these challenges. In the oxide materials community, techniques like pulsed laser deposition (PLD) now enable the fabrication of atomically thin films with exceptional precision. Notable achievements include strain-induced ferroelectricity in otherwise paraelectric SrTiO$_3$ \cite{RN27} and superconductivity at the LaAlO$_3$/SrTiO$_3$ \cite{RN28} interface. On the technical side, tools such as magneto-optical Kerr effect (MOKE) microscopy have revolutionized the field by enabling direct, atomic-scale measurements of magnetic properties.

\begin{figure}
    \includegraphics[width=\linewidth]{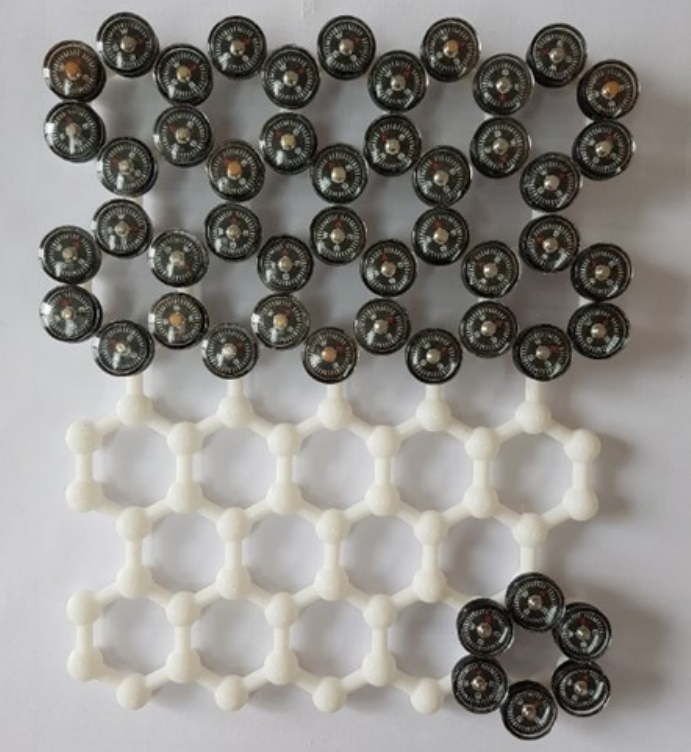}
    \caption{\label{fig:1}Conceptual model of ``magnetic graphene." The structure, created using 3D printing, features a graphene-like framework with tiny compasses representing magnetic atoms, illustrating the substitution of carbon atoms with magnetic counterparts to achieve intrinsic magnetism. Copyright: Je-Geun Park.}
\end{figure}

Although these advances represent significant progress, the challenge of isolating and characterizing truly two-dimensional magnetic systems persists. Unlike quasi-2D materials, a monolayer must entirely eliminate interlayer interactions, which requires breakthroughs in both material synthesis and experimental techniques. Interestingly, a well-known experiment of the Ising model by Kim and Chan in 1984 tested a two-dimensional liquid-vapor transition of a monolayer CH$_4$ on graphite \cite{RN29}, illustrating the feasibility of studying truly 2D systems in other contexts. The observation that there are no comprehensive studies of the XY model, both structure and dynamics, simultaneously underscores the enduring need for experimental validation of theoretical predictions in truly two-dimensional systems. The advent of vdW magnets offers a unique opportunity to address this gap, providing intrinsically two-dimensional materials with the potential to unlock unprecedented insights into magnetic phenomena.

The year 2005 marked a turning point in materials science and engineering with the first isolation of monolayer graphene, which opened up an entirely new field of two-dimensional (2D) materials \cite{RN30, RN31}. Although theoretical calculations for the electronic structure of a single graphite sheet were first performed in 1958 \cite{RN32}, it took nearly five decades for scientists to realize that single layers could be mechanically exfoliated from bulk graphite using simple adhesive tape. This discovery not only revolutionized materials science but also inspired the search for other two-dimensional materials, including magnetic counterparts. vdW magnets emerged from this exploration, offering a unique opportunity to study magnetism in truly two-dimensional systems. The field has recently expanded into new territories, including the groundbreaking discovery that twisting two layers of graphene at specific ‘magic angles’ produces a flat band, giving rise to various novel quantum phases \cite{RN33, RN34}. In particular, this phenomenon was theoretically predicted \cite{RN35} before being experimentally realized.

Although these developments are exciting, it is worth noting that the underlying physics of graphene primarily extends the framework of non-interaction electron theory, such as the tight-binding model. In contrast, decades of research on magnetic materials, especially high-temperature superconductors, have demonstrated how the inclusion of electron correlation, described by the Hubbard $U$ term, can lead to a wealth of new and intricate quantum phases \cite{RN36}. These phases are often diverse and delicate and reveal a level of complexity beyond simple band theory.

This realization naturally leads to the idea of merging the fields of 2D materials and magnetism, giving rise to the concept of ‘magnetic graphene’ (Fig. \ref{fig:1}) \cite{RN4}. In this vision, each carbon atom in graphene is replaced by a magnetic atom, creating a material with intrinsic magnetic properties. This transformation enables the study of low-dimensional magnetism in a framework as structurally simple and versatile as graphene. This hypothetical 2D magnet should be intrinsically and naturally two-dimensional and as easily exfoliable as graphene.

Despite its elegance, the concept of magnetic graphene was initially dismissed as unrealistic because of the lack of materials with the desired features. Early materials searches, constrained by familiarity with oxides, failed to yield results. A breakthrough was achieved with the rediscovery of layered compounds studied in the 1970s and 1980s, particularly transition metal phosphorus trisulfides (TMPS$_3$ with TM being transition metal elements) \cite{RN2}: Similar systems were also found in CrI$_3$ and Cr$_2$Ge$_2$Te$_6$, too. TMPS$_3$ systems were known to realize the three fundamental models of Ising, XY, and Heisenberg magnetism simply by substituting different transition metal ions (Fe, Ni, and Mn) at the TM sites \cite{RN37}. This foundational work laid the foundation for subsequent discoveries. It may be historically interesting to note that the conceptual foundation of vdW magnetic materials was already publicly articulated in 2015 at international meetings \cite{RN532,RN533}, representing one of the earliest formal introductions of the field.

\begin{figure}
    \includegraphics[width=\linewidth]{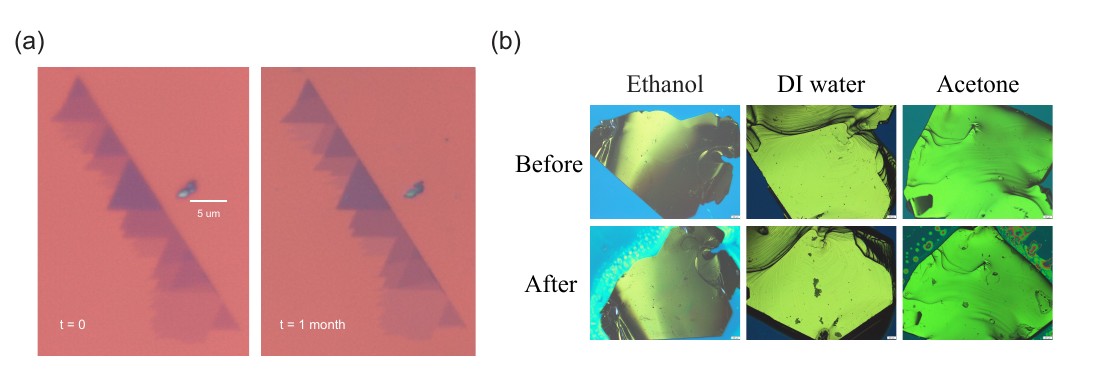}
    \caption{\label{fig:2}Demonstration of the stability of two vdW magnets. (a) CrPS$_4$ remains stable for over a month when stored in a glovebox under Ar gas. From \citet{RN45}. (b) FePS$_3$ exhibits stability in various liquid environments, highlighting its robustness under different experimental conditions. From \citet{RN58}.}
\end{figure}

\begin{figure}
    \includegraphics[width=\linewidth]{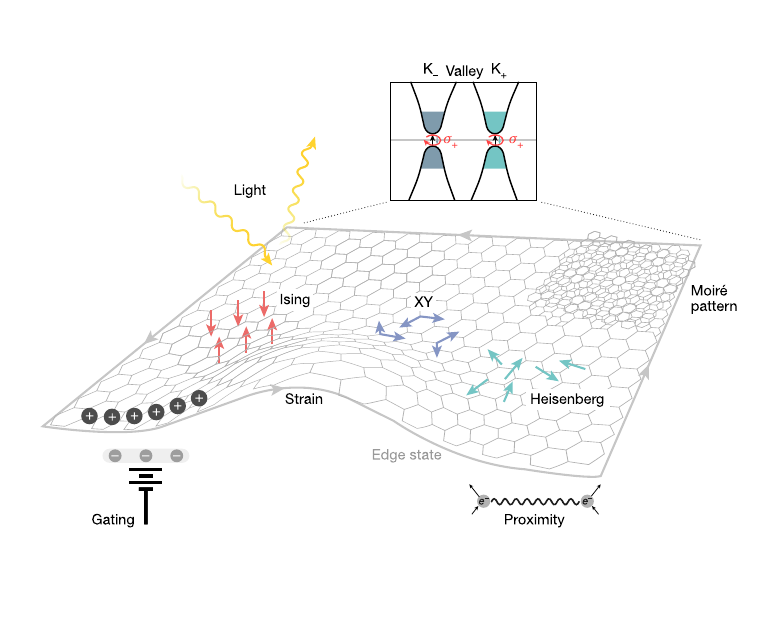}
    \caption{\label{fig:3}Physical phenomena enabled by magnetic vdW materials. These 2D magnetic systems provide an ideal platform for studying the behavior of fundamental magnetism models (Ising, XY, and Heisenberg models, represented by red, purple, and cyan arrows, respectively) in the 2D limit. Magnetic ground states in these materials can be tuned via external perturbations such as gating, strain, proximity effects, and moir\'e patterns. Furthermore, the intrinsic honeycomb lattice of these materials enables light–matter interactions through valley coupling at the K$_{-}$ and K$_{+}$ points in momentum space and edge states (gray arrows). Adapted from \citet{RN59}.}
\end{figure}

In 2016, the first report of monolayer TMPS$_3$---specifically FePS$_3$, NiPS$_3$, and MnPS$_3$---was reported~\cite{RN4,RN5,RN6,RN7}, with an independent report ~\cite{RN8}. Of particular importance was the demonstration of stable magnetism in monolayer FePS$_3$~\cite{RN7}, consistent with Onsager’s solution of the Ising model~\cite{RN15}. Similarly, the monolayers of NiPS$_3$ and MnPS$_3$ opened avenues for studying the XY and Heisenberg models, although new experimental tools were required to investigate their properties. These discoveries of antiferromagnetic (AF) monolayers were followed by reports of ferromagnetic (FM) monolayers, including Cr$_2$Ge$_2$Te$_6$ and CrI$_3$~\cite{RN38,RN39}. Magneto-optical Kerr effect (MOKE) measurements confirmed stable ferromagnetic signals down to the bilayer for Cr$_2$Ge$_2$Te$_6$ and the monolayer for CrI$_3$. Together, these works heralded the birth of the new field of vdW magnets.

Adding further depth to the field, metallic vdW magnets have emerged as critical systems. Their conductive properties allow for the use of traditional transport techniques to investigate the ground states of monolayer magnets. Notable examples include Fe$_3$GeTe$_2$, where the transition temperature is tunable by external gating~\cite{RN40}, and metallic AF systems, such as Fe$_{1/3}$TaS$_2$, synthesized through the chemical intercalation method~\cite{RN41}. The ability to control magnetic states in Fe$_3$GeTe$_2$ exemplifies how vdW magnets could enable reconfigurable spintronic devices, a key step toward energy-efficient memory and logic technologies.
Equally significant is the realization of diverse crystal lattices in 2D magnets, including honeycomb (e.g., TMPS$_3$, Cr$_2$Ge$_2$Te$_6$ and CrI$_3$), triangular (e.g., NiI$_2$)~\cite{RN42,RN43}, Kagome (e.g., Pd$_3$P$_2$S$_8$)~\cite{RN44}, and square (e.g., CrPS$_4$)~\cite{RN45}. These materials have revealed fascinating ground states, including the first magnetic topological insulator, MnBi$_2$Te$_4$~\cite{RN46}, and nanoscale skyrmion phases, as demonstrated in Co$_{1/3}$TaS$_2$~\cite{RN47,RN48}. Other promising directions include magnetoelectric effects~\cite{RN49}, such as the multiferroicity found in thin NiI$_2$~\cite{RN42,RN43}, and spintronics applications such as spin filters~\cite{RN50,RN51}, and current-controlled magnetic state~\cite{RN52}. The emergence of magnetic moir\'e systems has further expanded the field, offering platforms to study correlated phenomena and exotic spin textures~\cite{RN53,RN54,RN55}.

Despite these advances, challenges remain, particularly concerning air stability. When exposed to air, many vdW magnets decompose rapidly because of their volatile components. However, recent discoveries of stable alternatives, such as CrPS$_4$ and CrSBr, highlight the progress of the field and provide a basis for future advancements~\cite{RN45,RN56}. These materials enable experiments under ambient conditions, removing a significant barrier to the practical deployment of vdW magnets. Figure~2 demonstrates two such cases: one is the exceptional air stability of CrPS$_4$, stable for one month when kept in a glovebox with Ar gas, and another is the stability of FePS$_3$, even when exposed to different liquid environments.

With these breakthroughs in materials discovery, fundamental physics, and applications, vdW magnets are poised to transform our understanding of low-dimensional systems and drive innovations across diverse fields, from quantum technologies to spintronics.

The remainder of this article is organized as follows. Section \ref{sec:II} introduces the foundational materials and spin Hamiltonians essential to understanding vdW magnets. An extensive survey of experimental studies using optical techniques is presented in Section \ref{sec:III}. Section \ref{sec:IV} delves into light-induced out-of-equilibrium dynamics, exploring the novel phenomena that arise under nonequilibrium conditions. The applications of vdW magnets in spintronics are discussed in Section \ref{sec:V}, highlighting their potential for technological innovation. Finally, Section \ref{sec:VI} concludes the article with a summary and outlook, reflecting on current progress and future directions in this rapidly evolving field.

\section{\label{sec:II}2D vdW Magnets and Hamiltonian}
Since the first successful realization of monolayer vdW antiferromagnets in 2016, approximately two dozen materials have been identified as members of this exciting class. These materials can be categorized into three distinct groups based on their physical properties. The first classification is based on the type of magnetic ground state, distinguishing between ferromagnetic and antiferromagnetic materials. The second classification depends on their electronic properties. Although most vdW magnets are insulators, several metallic ferromagnets and antiferromagnets have also been discovered, which hold significant potential for spintronics applications. Although we have a limited number of materials with interesting properties such as multiferroic or topologically non-trivial systems, they remain equally exciting and warrant further investigation. The third category considers the lattice structure of the magnetic ions. Most vdW magnets adopt a honeycomb lattice, with the notable exception of one material forming a square lattice and a few others forming a triangular lattice. Equally intriguing classifications arise from their spin Hamiltonians, which are of great academic interest, particularly concerning the stability of magnetism in monolayer samples. Many materials exhibit behavior consistent with the Heisenberg model. In contrast, a smaller but significant subset aligns with the Ising model, and an even more limited number corresponds to the XY model. See the summary in Tables \ref{tab:1} and \ref{tab:2}.

\begin{turnpage}
\begin{table*}
\caption{\label{tab:1}Summary of the physical properties of reported vdW antiferromagnets, including their chemical formula, space group, transition temperature, electrical properties, lattice type, spin Hamiltonian type, and additional relevant information. References are provided for further details on each material.}
\begin{ruledtabular}
\begin{tabular}{c c p{2cm} p{2 cm} p{2 cm} p{2.5 cm} p{2.5 cm} p{3 cm} p{6cm}}
\hline
\textbf{Type} & \textbf{Formula} & \textbf{Space Group} & \textbf{$T_N$ (K)} & \textbf{Electrical properties} & \textbf{Lattice} & \textbf{Spin Hamiltonian} & \textbf{Others} & \textbf{Reference} \\
\hline
AFM & FePS$_3$ & C2/m & 120 & Insulator & Honeycomb & Ising & -- & \cite{RN7, RN240, RN206} \\
& MnPS$_3$ & C2/m & 78 & Insulator & Honeycomb & Heisenberg & -- & \cite{RN209,RN240} \\
& NiPS$_3$ & C2/m & 155 & Insulator & Honeycomb & XXZ & Many-body exciton & \cite{RN76, RN77, RN80, RN240} \\
& CoPS$_3$ & C2/m & 122 & Insulator & Honeycomb & Kitaev-Heisenberg & -- & \cite{RN212, RN88, RN211} \\
& FePSe$_3$ & $R\bar{3}$ & 119 & Insulator & Honeycomb & Biquadratic & -- & \cite{RN213, RN214} \\
& MnPSe$_3$ & $R\bar{3}$ & 74 & Insulator & Honeycomb & Easy axis Heisenberg & -- & \cite{RN210, RN214} \\
& CrPS$_4$ & C2/m & 41 & Insulator & Square & Easy axis Heisenberg & -- & \cite{RN45,RN215,RN94} \\
& NiI$_2$ & $R\bar{3}$ & 76, 59.5 & Insulator & Triangular & -- & Multiferroic & \cite{RN42,RN216} \\
& CoI$_2$ & $P\bar{3}m1$ & 11, 9.4 & Insulator & Triangular & Kitaev-Heisenberg & Multiferroic & \cite{RN217,RN216} \\
& CrSBr & Pmmn & 132 & Semiconductor & Triangular & Heisenberg & Layer-dependent ferromagnetism & \cite{RN56,RN178,RN218} \\
& MnBi$_2$Te$_4$ & $R\bar{3}m$ & 25 & Insulator & Triangular & Easy axis Heisenberg & Chern insulator (odd layer), axion insulator (even) & \cite{RN46, RN219, RN230,RN241} \\
& CrCl$_3$ & $R\bar{3}$ & 14, 17 & Insulator & Honeycomb & Heisenberg & 2D XY model in monolayer & \cite{RN134,RN111,RN220,RN112} \\
& RuCl$_3$ & C2/m & 7 & Insulator & Honeycomb & Kitaev-Heisenberg & -- & \cite{RN221, RN222} \\
& Co$_{1/3}$TaS$_2$ & P6$_3$22 & 26.5, 38 & Metal & Triangular & Biquadratic & Weyl semimetal, triple-Q & \cite{RN203, RN47} \\
& Ni$_{1/3}$TaS$_2$ & P6$_3$22 & 158 & Metal & Triangular & Easy axis Heisenberg, DM & -- & \cite{RN223} \\
& Ni$_{1/3}$NbS$_2$ & P6$_3$22 & 84 & Metal & Triangular & Heisenberg, DM & -- & \cite{RN223} \\
& Co$_{1/3}$NbS$_2$ & P6$_3$22 & 27.5 & Metal & Triangular & -- & -- & \cite{RN224,RN225} \\
& Cr$_{1/3}$TaS$_2$ & P6$_3$22 & 150 & Metal & Triangular & -- & Magnetic vortex domain, chiral soliton & \cite{RN226,RN227} \\
& Fe$_{1/3}$NbS$_2$ & P6$_3$22 & 43 & Metal & Triangular & -- & Spin glass & \cite{RN228, RN229, RN225} \\
\hline
\end{tabular}
\end{ruledtabular}
\end{table*}
\end{turnpage}

\begin{turnpage}
\begin{table*}
\caption{\label{tab:2}Summary of the physical properties of reported vdW ferromagnets, including their chemical formula, space group, transition temperature, electrical properties, lattice type, spin Hamiltonian type, and additional relevant information. References are provided for further details on each material.}
\begin{ruledtabular}
\begin{tabular}{c c p{2cm} p{2 cm} p{2 cm} p{2.5 cm} p{2.5 cm} p{3 cm} p{6cm}}
\hline
\textbf{Type} & \textbf{Formula} & \textbf{Space Group} & \textbf{$T_c$ (K)} & \textbf{Electrical Properties} & \textbf{Lattice} & \textbf{Spin Hamiltonian} & \textbf{Others} & \textbf{Reference} \\
\hline
FM & Cr$_2$Si$_2$Te$_6$ & $R\bar{3}$ & 33 & Semiconductor & Honeycomb & Easy axis Heisenberg, DM & Topological magnon insulator & \cite{RN231,RN232} \\
& Cr$_2$Ge$_2$Te$_6$ & $R\bar{3}$ & 66 & Semiconductor & Honeycomb & Easy axis Heisenberg, DM & Topological magnon insulator & \cite{RN38,RN232,RN233,RN234,RN100,RN235} \\
& Fe$_3$GeTe$_2$ & P6$_3$mmc & 223 & Metal & Honeycomb & Easy axis Heisenberg & Nodal line semimetal & \cite{RN143,RN165,RN236,RN237} \\
& Fe$_3$GaTe$_2$ & P6$_3$mmc & 350$\sim$380 & Metal & Honeycomb & -- & Nodal line semimetal & \cite{RN155,RN238} \\
& Fe$_4$GeTe$_2$ & $R\bar{3}$m & 270 & Metal & Honeycomb & -- & -- & \cite{RN149} \\
& Fe$_5$GeTe$_2$ & $R\bar{3}$m & 310 & Metal & Honeycomb & -- & -- & \cite{RN150,RN239} \\
& 1T-VSe$_2$ & $P\bar{3}$m1 & Above R.T. & Metal & Triangular & -- & Ferromagnetic ordering in few monolayers & \cite{RN242,RN243} \\
& 1T-CrTe$_2$ & $P\bar{3}$m1 & Above R.T. & Metal & Triangular & -- & -- & \cite{RN244,RN245,RN246,RN247} \\
& Fe$_{1/3}$TaS$_2$ & P6$_3$22 & 38 & Metal & Triangular & -- & -- & \cite{RN248} \\
& VI$_3$ & $R\bar{3}$ & 55 & Insulator & Honeycomb & Kitaev-Heisenberg & -- & \cite{RN135,RN249,RN250} \\
& CrBr$_3$ & $R\bar{3}$ & 37 & Insulator & Honeycomb & Easy axis Heisenberg & -- & \cite{RN251,RN125,RN252} \\
& CrI$_3$ & $R\bar{3}$ & 61 & Insulator & Honeycomb & Heisenberg, DM & Layer dependent ferromagnetism & \cite{RN253,RN254,RN39} \\
\hline
\end{tabular}
\end{ruledtabular}
\end{table*}
\end{turnpage}

\subsection{AFM insulators \& metals}
The first demonstration of monolayer magnetism was achieved using antiferromagnetic vdW magnets, such as FePS$_3$. However, because of the intrinsically net-zero magnetic moment of the antiferromagnetic ground state, initial measurements relied on Raman spectroscopy, which leverages spin-lattice or magnon-phonon coupling. Suitable experimental tools for studying vdW antiferromagnets remain limited to dates, highlighting the urgent need to develop new techniques tailored to these materials.
Despite these challenges, vdW antiferromagnets have garnered significant attention for several reasons. One notable advantage is their ability to readily realize key spin Hamiltonians, such as the Ising, XY, and Heisenberg models. In addition, these materials hold great promise for spintronics applications. Antiferromagnetic (AFM) spintronics, in particular, offers distinct advantages over its ferromagnetic (FM) counterpart, not least because it avoids issues related to stray magnetic fields and can be used for higher frequency devices. As a result, spintronic devices based on vdW antiferromagnets represent a highly promising area for future exploration. Another notable point is that these samples are mostly stable in air.

\subsubsection{\texorpdfstring{TMPS$_3$}{TMPS3}}
\begin{figure}
    \includegraphics[width=\linewidth]{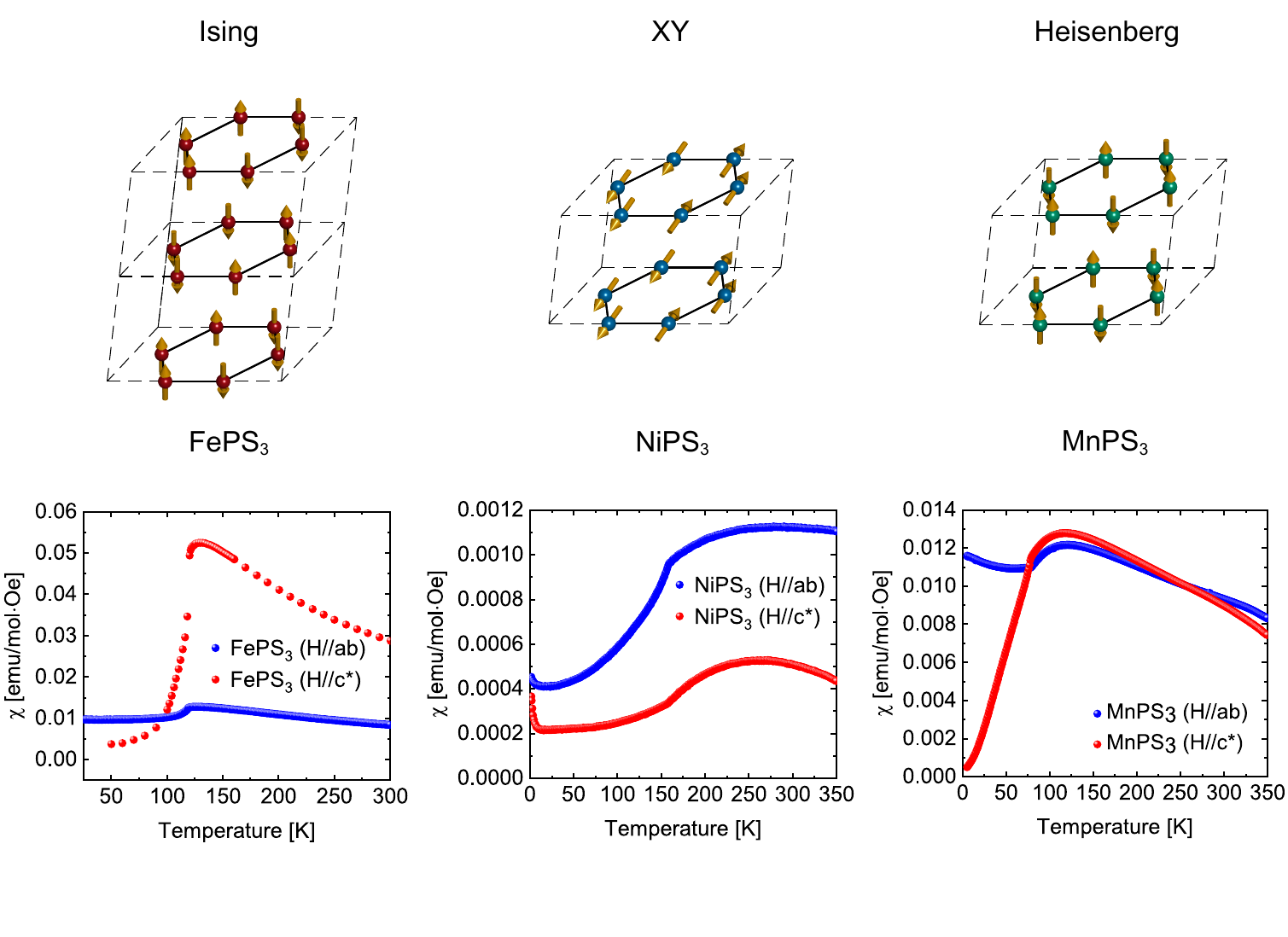}
    \caption{\label{fig:4}Magnetic structures of TMPS$_3$ (TM = Fe, Ni, Mn) alongside their respective magnetic susceptibility data. The illustrations depict the distinct magnetic configurations of each compound, highlighting the variations in spin alignments. Corresponding susceptibility measurements provide experimental evidence of their unique magnetic behaviors across different temperature ranges.}
\end{figure}

Transition metal phosphorus trisulfides are isostructural insulators characterized by a metal-ion honeycomb lattice interspersed with (P$_2$S$_6$)$^{2-}$-complex anions. Magnetic coupling in these materials occurs via a superexchange pathway involving sulfur atoms, extending up to the third-nearest interactions, denoted as $J_1$, $J_2$, and $J_3$, from the closest neighbor to the furthest neighbor~\cite{RN60}. These interactions are crucial in determining the magnetic order~\cite{RN61}. The spin direction varies between the different compounds: in MnPS$_3$ and FePS$_3$, it is perpendicular to the layer, while in NiPS$_3$, it lies along the layer with a small canting towards the $c$-axis. Furthermore, the magnetic behavior of these materials is described by a distinct spin Hamiltonian: MnPS$_3$ conforms to an isotropic Heisenberg Hamiltonian, FePS$_3$ aligns with the Ising model, and NiPS$_3$ follows an XXZ Hamiltonian with XY-like behavior at a low energy limit~\cite{RN62} (see Fig. \ref{fig:4}).

\textbf{FePS$_3$}

Iron phosphorus trisulfide (FePS$_3$) represents a Mott insulator with an antiferromagnetic Néel temperature $T_N$ of 118~K~\cite{RN63}. The weak interlayer Fe-Fe interaction and large anisotropy along the $c$ direction result in long-range magnetic order, leading to a magnetic unit cell that is twice the size of the crystallographic cell~\cite{RN63}. In 2016, intrinsic antiferromagnetic order in monolayer FePS$_3$ was reported, for the first time, through observation of the zone-folded phonon mode due to the doubling of the unit cell in the ordered phase emerging at the transition temperature. When the intensity of this Raman mode is taken as an order parameter~\cite{RN7}, its temperature dependence follows a universal curve from bulk to monolayer, consistent with the Onsager solution for the 2D order-disorder transition. The little dependence of the transition temperature on thickness further highlights the weak interlayer interaction of this 2D magnet~\cite{RN7}. Using neutron powder diffraction combined with a new ultra-high-pressure technique (up to 183~kbar) and low temperature (down to 80~K), a transition from antiferromagnetic to ferromagnetic interplanar interaction was observed, along with a shift from 2D-like to 3D-like magnetic character. At high pressure, a second structural transition was identified, along with an insulator-metal transition, the suppression of long-range order, and the emergence of short-range magnetic order, which persists above room temperature. These phenomena were modeled using Monte Carlo simulations of spin configuration~\cite{RN64}.

The easy-axis magnetic anisotropy of Fe$^{2+}$ is estimated to be 22~meV per Fe ion by two different measurements of photoelectron emission microscopy (PEEM)~\cite{RN205} and inelastic neutron scattering experiments~\cite{RN206}. This large magnetic anisotropy is mainly due to the trigonal distortion of the Fe$^{2+}$ 3$d$ bands, which produces a significant anisotropy. Furthermore, a comprehensive 3D magnetic anisotropy map of FePS$_3$ was constructed using a combination of magnetostatic models, relativistic density functional theory (DFT) calculations, and experimental torque magnetometry measurements. This map confirmed that the easy-axis lies perpendicular to the $ab$ plane and revealed anisotropic behavior within the $ab$, $ac$, and $bc$ planes~\cite{RN66}. Most intriguingly, recent high-resolution X-ray studies have uncovered spin-mediated shear oscillators~\cite{RN67} and their dynamical criticality~\cite{RN68} in FePS$_3$, opening new avenues for understanding and utilizing this material.

\textbf{MnPS$_3$}

Manganese phosphorus trisulfide (MnPS$_3$) is a two-dimensional (2D) antiferromagnetic Heisenberg-type material with a N\'eel temperature of 78~K in bulk~\cite{RN69}. However, the $T_N$ of MnPS$_3$ varies depending on the number of material layers, decreasing to 66~K in the trilayer sample due to interlayer vdW interaction~\cite{RN70}. Raman spectroscopy studies of MnPS$_3$ from bulk to bilayer reveal changes in phonon modes associated with Mn ion vibrations as the temperature decreases, crossing $T_N$. These changes, characterized by a red shift in the Raman peaks at low temperatures, are consistent with first-principle calculations~\cite{RN71}. This suggests that the magnetic order in MnPS$_3$ remains stable down to the bilayer limit. MnPS$_3$ exhibits three distinct phases, identified by Raman spectroscopy at different temperatures. Beyond the N\'eel temperature, a second transition at 120~K corresponds to a magnetic phase transition driven by 2D spin critical fluctuations~\cite{RN72}. The N\'eel-ordered state in MnPS$_3$ enables a linear magnetoelectric phase by breaking spatial inversion and time-reversal symmetries. This magnetoelectric phase remains stable to a thickness of 5.3~nm, as demonstrated by optical second-harmonic studies (SHG)~\cite{RN73}. The same technique reveals that the long-range N\'eel antiferromagnetic order persists from the bulk down to the bilayer, while the short-range order and magnetoelastic coupling emerge below 200~K~\cite{RN74}. Most notably, MnPS$_3$ exhibits a giant modulation of optical nonlinearity through Floquet engineering, marking a significant advance in its potential applications~\cite{RN75}.

\textbf{NiPS$_3$}

Nickel phosphorus trisulfide (NiPS$_3$) is an XXZ-type antiferromagnetic material with a N\'eel temperature of 155~K~\cite{RN76,RN77} and spin $S = 1$ on a honeycomb lattice~\cite{RN78}. The magnetic behavior of NiPS$_3$ is highly dependent on its thickness, becoming particularly unstable in monolayer form~\cite{RN74}. Thin layer flakes of NiPS$_3$, including the monolayer, have been successfully mechanically exfoliated and analyzed using Raman spectroscopy, which revealed distinct differences between thin layers and bulk material below the N\'eel temperature~\cite{RN5}. Further investigation utilizing spectroscopic ellipsometry, X-ray absorption, photoemission spectroscopy, and density functional theory has identified an anomalous shift at the magnetic ordering temperature, suggesting a strong correlation between electronic and magnetic structures~\cite{RN79}. The intrinsic Zhang-Rice singlet also generates a spin-orbit exciton state that originates from a coherent Zhang-Rice-triplet-to-singlet transition. This exciton state, characterized by a narrow linewidth, has been observed through X-ray scattering, photoluminescence, and optical absorption measurements~\cite{RN80}. These highly anisotropic quantum excitons coexist with multiple phonon-bound states~\cite{RN81} and spin-induced linear polarization of photoluminescence under external magnetic fields~\cite{RN82}.

The magnetic structure of NiPS$_3$ features $a$-axis magnetic moments and isotropic magnetic susceptibility~\cite{RN77}. Inelastic neutron scattering detected a magnetic excitation at the Brillouin zone center at 7~meV. Fitting these experimental data to the Heisenberg Hamiltonian yielded exchange parameters of up to three layers: $J_1 = 1.9$~meV, $J_2 = -0.1$~meV, and $J_3 = -6.9$~meV~\cite{RN77}. Pressure has shown potential in tuning the magnetic behavior of NiPS$_3$, inducing a collapse of magnetic order and leading to an insulator-metal transition~\cite{RN83}. Electron doping also modulates its properties, with antiferromagnetic–ferrimagnetic transitions occurring at a doping level of 0.2 to 0.5 electrons/cell~\cite{RN84}. Substituting Ni with Mn induces a spin-flop transition, which offers the potential for application in 2D magnets~\cite{RN85}. In addition, ultrafast terahertz experiments reveal exciton-driven antiferromagnetic metal states~\cite{RN86} and magnetically brightened dark electron-phonon bound states~\cite{RN87} in NiPS$_3$, highlighting its multifaceted magnetic and electronic properties.

\textbf{CoPS$_3$}

Cobalt phosphorus trisulfide (CoPS$_3$) exhibits a N\'eel temperature $T_{\mathrm{N}} = 120$~K and XY-like anisotropy, with no phase transitions observed between 2 and 300~K under an external magnetic field of up to 10~T~\cite{RN88}. A detailed investigation using powder-inelastic neutron scattering, combined with an XXZ-type Hamiltonian, revealed the magnetic excitation of CoPS$_3$. The exchange parameters $J_1$, $J_2$, and $J_3$ were estimated as $-2.08$, $-0.26$, and $4.21$~meV with strong easy-axis anisotropy $K = -2.06$~meV~\cite{RN89}. The ratio between the interlayer and intralayer exchange parameters $J'/J$ was also estimated to be 0.03~\cite{RN90}.

\textbf{CrPS$_4$}

Chromium thiophosphate (CrPS$_4$) has a monoclinic crystal structure, with Cr atoms occupying octahedral sites formed by surrounding S atoms, which enclose P atoms in tetrahedral coordination. It hosts an A-type antiferromagnetic order with a N\'eel temperature of 36~K~\cite{RN91,RN92}. The antiferromagnetic structure consists of out-of-plane ferromagnetic monolayers coupled antiferromagnetically between layers~\cite{RN93}. Electrical resistivity measurements confirm its semiconducting behavior with a band gap of 0.166~eV. Magnetic exchange constants $J_1$ and $J_c$ were determined to be 0.143 and $-0.955$~meV, respectively~\cite{RN94}. In the atomically thin limit, CrPS$_4$ shows crystallographic orientation along the diagonal rows of the Cr atom, exhibits strong in-plane optical anisotropy, and reveals lattice vibrations highly dependent on thickness. The photoluminescence peak at 1.31~eV arises from a $d$-$d$ type transition associated with Cr$^{3+}$ ions~\cite{RN95}. Furthermore, CrPS$_4$ has demonstrated stability against air exposure for up to one day in monolayer, bilayer, and trilayer forms while maintaining its magnetic order~\cite{RN45}.

\textbf{TM$_{1/3}$MS$_2$: TM = transition metal elements, M = Ta, Nb}

TM$_{1/3}$MS$_2$ (TM = 3$d$ transition metals, M = Ta, Nb) are promising materials for studying metallic antiferromagnetism in a triangular lattice~\cite{RN96,RN97}. These compounds are derived from the well-known metallic 2H-MS$_2$, with triangular layers of transition metals (TM) intercalated in the vdW gap of 2H-MS$_2$. This intercalation results in a chiral structure with a non-centrosymmetric $P6_3 22$ space group (No.~182) and induces inversion symmetry breaking~\cite{RN98}. Charge transfer from TM atoms to 2H-MS$_2$ leads to divalent (TM$^{2+}$) or trivalent (TM$^{3+}$) TM ions, localizing magnetic moments at TM sites. Remarkably, these materials possess diverse magnetic ground states depending on the specific TM and M elements used. Furthermore, TM$_{1/3}$MS$_2$ can be mechanically exfoliated down to a few atomic layers~\cite{RN99}, allowing the exploration of noncollinear antiferromagnetism and associated topological physics in a truly two-dimensional framework.

\subsection{FM insulators \& metals}

Unlike antiferromagnetic vdW magnets, ferromagnetic vdW magnets offer a distinct advantage: it is relatively easier to measure their ferromagnetic signals. For instance, optical probes such as the magneto-optical Kerr effect (MOKE) technique have been widely used in studying ferromagnetic vdW magnets, leading to the first demonstration of monolayer ferromagnetic CrI$_3$ and bilayer Cr$_2$Ge$_2$Te$_6$. Since then, several other ferromagnetic vdW magnets with diverse spin Hamiltonians have been discovered (see Table \ref{tab:3}). More recently, attention has been shifted to room-temperature ferromagnetic systems, such as Fe$_5$GeTe$_2$. These ferromagnetic materials, such as spin filters in the early stages, have been explored for spintronics applications. With the discovery of room-temperature ferromagnetism, various attempts have been made to develop spintronic devices that operate at room temperature.

\begin{table*}
\centering
\caption{\label{tab:3}This table summarizes the key properties of ferromagnetic vdW magnets, including their Curie temperature ($T_\mathrm{c}$), coercive field ($H_\mathrm{c}$), and saturated magnetization ($M_\mathrm{S}$). These parameters are crucial for understanding these materials' magnetic behavior and potential applications in spintronic and quantum technologies. Full references are provided for each entry to guide further reading and verification of the data.}
\begin{tabular}{c c c c c}
\hline
Formula & $T_C$ (K) & $H_C$ (Oe) & Saturation magnetization & References \\
 & & & ($\mu_B$/magnetic Ion) & \\
\hline
Cr$_2$Si$_2$Te$_6$ & 33 & 3270 at 2 K & 3.08 & \cite{RN255,RN256} \\
Cr$_2$Ge$_2$Te$_6$ & 66 & 312 at 2 K & 3.24 & \cite{RN38,RN255} \\
Fe$_3$GeTe$_2$ & 223 & 33 at 2 K* & 1.6 & \cite{RN237,RN257} \\
Fe$_3$GaTe$_2$ & 350$\sim$380 & 1014 at 3 K & 1.68 & \cite{RN155} \\
Fe$_4$GeTe$_2$ & 270 & 1000 at 2 K & 1.8 & \cite{RN149} \\
Fe$_5$GeTe$_2$ & 310 & 3000 at 2 K & 2 & \cite{RN150} \\
1T-VSe$_2$ & Above R.T. & 112 at 10 K & 15 $\mu_B$ / f.u. & \cite{RN242} \\
1T-CrTe$_2$ & Above R.T. & 50 at 2 K* & 1.7 & \cite{RN244} \\
Fe$_{1/3}$TaS$_2$ & 38 & 19000 at 2 K & 4.1 & \cite{RN248} \\
VI$_3$ & 55 & 9100 at 2 K & 1.3 & \cite{RN135} \\
CrBr$_3$ & 37 & 200 at 2 K & 3.8 & \cite{RN128,RN125} \\
CrI$_3$ & 61 & 200 at 10 K & 3.1 & \cite{RN254,RN258} \\
\hline
\multicolumn{5}{l}{\footnotesize * : Note this is bulk; few-layer shows larger coercivity}
\end{tabular}
\end{table*}

\textbf{Cr$_2$Ge$_2$Te$_6$ and Cr$_2$Si$_2$Te$_6$}

Chromium germanium telluride (Cr$_2$Ge$_2$Te$_6$ or simply CrGeTe$_3$) belongs to the M$_2$X$_2$Te$_6$ chalcogenides family and has a rhombohedral structure, as confirmed by X-ray single crystal diffraction. Neutron powder diffraction reveals its ferromagnetic order with a Curie temperature $T_\mathrm{C}$ of about 61~K, below which a weak out-of-plane anisotropy exists~\cite{RN100}. Micro-Raman measurement across a wide range of temperatures (from 10 to 325~K) on thin flakes of CrGeTe$_3$ confirms strong spin-phonon coupling in the ferromagnetic phase below Curie temperature~\cite{RN101}. Since 2017, CrGeTe$_3$ has become an important ferromagnetic material due to direct observation of long-range ferromagnetic order by MOKE in nanoflakes of a few layers~\cite{RN38}. The Curie temperature decreases as the thickness reduces, but increases under higher external magnetic fields. This temperature dependence differs from the 3D bulk counterpart and can be considered an intrinsic property of a soft 2D vdW ferromagnet~\cite{RN38}. The material's small out-of-plane magnetocrystalline anisotropy arises from a slight distortion of the Cr-Te$_6$ octahedral cage and spin-orbit coupling on Cr ions. Due to strong interlayer magnetic coupling, the effects of dimensionality are highly dependent on temperature and thickness, as shown by the Curie temperature dependence~\cite{RN102}. Furthermore, microarea Kerr measurements on five-layer flakes encapsulated between two hexagonal boron nitride (hBN) layers reveal a strong field effect. Below $T_\mathrm{C}$, tunable magnetization loops are observed at different gate voltages, which may rebalance the spin-polarized band structure~\cite{RN103}.

Cr$_2$Si$_2$Te$_6$ hosts a similar structure but a larger easy-axis magnetic anisotropy and a lower bulk Curie temperature of 33~K~\cite{RN104}. Its large band gap, similar to Cr$_2$Ge$_2$Te$_6$, significantly suppresses electronic conduction and can be reduced by inducing short-range crystal disorder through defects. For example, applying a high pressure above 10~GPa can induce an insulator-to-metal transition at low temperatures~\cite{RN105}. Both positive and negative magnetoresistance coexist in this compound, reaching a ratio of up to 60\% under an in-plane field and 1000\% under an out-of-plane field~\cite{RN106}. This nonlinear magnetoresistance effect depends on the thickness, magnetic field, and temperature of the sample. The nonlinear behavior transitions to a linear one as the sample thickness changes in a temperature range above the Curie temperature~\cite{RN107}. Furthermore, proton fluence can tune the magnetic properties of Cr$_2$Si$_2$Te$_6$, especially the saturation magnetization and magnetic anisotropy, as revealed by electron paramagnetic resonance spectroscopy~\cite{RN108}.

\textbf{CrX$_3$}

The structure of chromium trihalides (CrX$_3$, e.g. CrI$_3$, CrBr$_3$, and CrCl$_3$) is isostructural, consisting of a honeycomb octahedral network of Cr ions, each coordinated by six monovalent halide anions at the corners~\cite{RN109,RN110}. These compounds stack in a monoclinic phase at high temperatures (220~K in CrI$_3$~\cite{RN111}, 420~K in CrBr$_3$~\cite{RN112}, and 240~K in CrCl$_3$~\cite{RN111}) and in a rhombohedral phase at lower temperatures. On the magnetism side, the superexchange coupling dominates the exchange pathways in CrX$_3$ through the halide ions. The ferromagnetic order within a single layer arises from the $\sim$90$^\circ$ bond angle between two Cr$^{3+}$ and one halide ion, which aligns with the prediction of the Goodenough-Kanamori-Anderson rule~\cite{RN109,RN113}. While CrI$_3$ and CrBr$_3$ exhibit out-of-plane easy-axis anisotropy~\cite{RN110}, CrCl$_3$ exhibits easy-plane anisotropy~\cite{RN111}. Efforts were also made to control the magnetic anisotropy characteristic by adjusting the Br doping concentration in CrCl$_{3-x}$Br$_x$~\cite{RN114}.

\textbf{CrI$_3$}

CrI$_3$ is another important material with fascinating scientific examples~\cite{RN39}. In bulk CrI$_3$~\cite{RN39}, the interlayer exchange is ferromagnetic, with a Curie temperature of 61~K. However, the situation is more complex and differs significantly for a few-layer CrI$_3$. The monolayer CrI$_3$ has been identified as an Ising ferromagnet with an out-of-plane spin orientation, as confirmed by MOKE and a Curie temperature of 45~K, which is lower than the bulk value due to weak interlayer coupling~\cite{RN39} and strong anisotropy down to the single layer~\cite{RN115}. The bilayer CrI$_3$ forms an A-type Ising antiferromagnet, where the intralayer spins are ferromagnetically coupled, but the interlayer spins are antiferromagnetically aligned, both layers preferring the out-of-plane easy-axis~\cite{RN39}.

Similarly, the spins in the third layer align parallel to the intralayer spins but antiparallel to the second layer along the out-of-plane direction, and the pattern continues in further layers. Consequently, the bilayer CrI$_3$ exhibits a metamagnetic effect, while the trilayer CrI$_3$ restores the net magnetization~\cite{RN39}. Magnetic order in CrI$_3$ can be detected via magnetic circular dichroism~\cite{RN39} and electron tunneling~\cite{RN50,RN51,RN116}, which can be tuned by electrostatic gating~\cite{RN103,RN117,RN118,RN119}. The modulation of magnetic order by hydrostatic pressure has also been reported. For example, the bilayer CrI$_3$ can transform from a layered antiferromagnetic phase to a ferromagnetic phase, while the trilayer CrI$_3$ can have three coexisting phases: ferromagnetic and antiferromagnetic~\cite{RN120}. A spin-flip transition in bilayer CrI$_3$ occurs when a 0.7~T out-of-plane magnetic field is applied, and in thicker flakes, increasing the magnetic field can induce layer-by-layer switching~\cite{RN50}. 

In addition, an acoustic magnon mode at 0.3~meV and an optical magnon mode at 17~meV were observed in the bilayer and bulk CrI$_3$, which were absent in the monolayer counterpart~\cite{RN121}. Additionally, CrI$_3$ has a centrosymmetric structure, which theoretically forbids SHG. However, a large emergence of SHG has been reported, originating in the layered antiferromagnetic order~\cite{RN122}. Furthermore, spontaneous helical light emission (at 1.1~eV) varies with different interlayer magnetic order: 50\% helicity for the ferromagnetic monolayer, almost fully spin polarized for the ferromagnetic bilayer, and a vanishing value for the antiferromagnetic bilayer~\cite{RN123}. These magneto-optical effects have been used to tune inelastic scattering light by controlling the odd or even layers of the scattering surface~\cite{RN124}.

\textbf{CrBr$_3$}

In bulk chromium bromide (CrBr$_3$), the interlayer exchange between two adjacent layers is also ferromagnetic, with a Curie temperature of 37~K~\cite{RN39,RN125}. The monolayer CrBr$_3$ is ferromagnetic, but its Curie temperature is slightly lower at 30~K, as determined by various independent methods~\cite{RN126,RN127,RN128}. CrBr$_3$ maintains its ferromagnetic order regardless of thickness~\cite{RN129}. Unlike CrI$_3$, the spontaneous emission of helical light at 1.35~eV has been observed in monolayer CrBr$_3$ with $\sim$20\% circular polarization, and it also displays spin-polarized behavior in the bilayer~\cite{RN127}. To preserve these air-sensitive materials down to the monolayer limits for spintronics application, MgO was developed as a removable combined protection-encapsulation passivation layer. This layer protects samples from air exposure for up to 1 year, making it suitable for vertical devices~\cite{RN130}.

\textbf{CrCl$_3$}

Successful exfoliation of chromium chloride (CrCl$_3$)  has been reported down to the monolayer limit. Two anomalies in heat capacity at 14 and 17~K confirm the occurrence of two distinct stages during cooling: Specifically, a ferromagnetic correlation is formed before the long-range antiferromagnetic order is established. Additionally, the anomaly in magnetic susceptibility suggests spin-lattice coupling, which was confirmed by first-principle calculation~\cite{RN111}. Antiferromagnetic interlayer exchange was observed in CrCl$_3$ with a magnetic ordering temperature of 17~K in few-layer CrCl$_3$~\cite{RN131}. CrCl$_3$ exhibits persistent layered antiferromagnetism down to the bilayer, with easy-plane anisotropy normal to the c-axis, showing no preferences for the in-plane polarizing direction~\cite{RN132}. The dominant form of magnetic anisotropy is the shape anisotropy in CrCl$_3$~\cite{RN132}. The insulation properties of CrCl$_3$ were studied using ligand-field photoluminescence, and a phase diagram for a bilayer CrCl$_3$ was summarized~\cite{RN133}. Furthermore, a nearly ideal easy-plane, single monolayer of CrCl$_3$ constructed on graphene/6H-SiC(0001) exhibited the critical scaling characteristic of a 2D-XY system, indicating the realization of the Berezinskii-Kosterlitz-Thouless phase transition~\cite{RN134}.

\textbf{VI$_3$}

Similar compounds can be explored by replacing Cr with other transition metals, such as VI$_3$, another type of trihalide. Magnetic measurements and calculation of the band structure revealed a transition temperature of 79~K, where structural and heat capacity changes occur. This hard ferromagnetic Mott insulator exhibits Ising-type spins along the easy-axis ($c$-axis) and an optical band gap of 0.67~eV. It also has a high degree of anisotropy and pressure-dependent magnetic properties~\cite{RN135}. Magnetocrystalline anisotropy was thoroughly investigated by applying a magnetic field in the $ab$ plane and an orthogonal plane to the $ab$ plane. In the orthogonal plane, two-fold symmetry persists at different temperatures, with a maximum tilt of 40$^\circ$ from the normal of the basal $ab$ plane.

In contrast, in-plane measurements show a transition from a two-fold-like angular signal to a six-fold-like angular signal at $T_c$, and another six-fold-like signal appears as the temperature approaches $T_c = 26$~K~\cite{RN136}. Furthermore, two more transition temperatures were observed at 54.5 and 53~K, corresponding to the onset of ferromagnetism on the crystal surface. $T_c$ strongly depends on pressure, reaching 99~K at 7.3~GPa~\cite{RN137}. Measurements from various methods confirmed the strong moisture degradation and showed that the magnetic properties of single crystal VI$_3$ did not change significantly after the first two hours of exposure~\cite{RN138}.

\textbf{Fe$_3$GeTe$_2$}

Iron germanium telluride (Fe$_3$GeTe$_2$, or FGT), a ferromagnetic material with strong out-of-plane anisotropy~\cite{RN139}, features Fe$_3$Ge slabs staying within the vdW gap of Te layers, with a bulk Curie temperature of approximately 220~K~\cite{RN140,RN141}. Exfoliating Fe$_3$GeTe$_2$ is mechanically challenging, although this can be facilitated by techniques assisted with gold~\cite{RN142} or alumina-assisted techniques~\cite {RN40}. Moreover, $T_c$ increases as the Fe composition increases, decreasing the lattice constant $a$ and increasing the lattice constant $c$~\cite{RN129}. $T_c$ also depends on the number of layers, dropping from 207 to 130~K when the sample thickness is reduced to fewer than five layers~\cite{RN143}. Due to the difficulty in material growth, efforts have been made to prepare a larger monolayer of FGT. In particular, a method with a 6\%-monolayer yield was reported in 2022 using a liquid-phase exfoliation assisted by three-stage sonication ~\cite{RN144}.

Several efforts have been made to modulate the magnetic characteristics of Fe$_3$GeTe$_2$. These properties can be tuned by cobalt doping~\cite{RN145}, applying high pressure~\cite{RN146}, etc. Magnetic circular dichroism spectroscopy has revealed a decrease in exchange interaction and magnetocrystalline anisotropy under pressure and temperature~\cite{RN147}. The magnetic anisotropy energy of FGT can be modulated by the electric-field-driven 3$d$-orbital occupancy, leading to a tunable negative differential conductance~\cite{RN148}. Remarkably, ionic liquid gating has been used to enhance the Curie temperature of a four-layer Fe$_3$GeTe$_2$ to approximately 305~K, well above the bulk Curie temperature of around 205~K. Many attempts have been made to increase $T_c$ under normal conditions to realize room-temperature applications of 2D magnetic materials. A direct strategy to achieve this is to add more Fe atoms to the slabs, as seen in Fe$_4$GeTe$_2$ and Fe$_5$GeTe$_2$~\cite{RN149}, where $T_c$ can reach 310~K in bulk and 280~K in a thin flake, as confirmed by anomalous Hall effect measurements~\cite{RN150,RN151}. In Fe$_5$GeTe$_2$, $T_c$ can also be enhanced, and its ferromagnetic states can even be changed to an antiferromagnetic phase using cobalt substitution~\cite{RN152}. This family of materials holds promise for applications, such as using few-layer film Fe$_{5+x}$GeTe$_2$ grown by molecular beam epitaxy to make planar spiral inductors~\cite{RN153}, or understanding the nontrivial electronic topology in Fe$_{3-x}$GeTe$_2$~\cite{RN154}. In addition, replacing Ge with Ga atoms results in another room temperature vdW magnet, Fe$_3$GaTe$_2$, with $T_c$ ranging from 350 to 380~K~\cite{RN155}.

As the first vdW ferromagnetic metal to be investigated, Fe$_3$GeTe$_2$ is extremely suitable for developing spintronic concepts and applications. Large tunable tunneling magnetoresistance effects have been widely studied in various Fe$_3$GeTe$_2$-based spin valve devices~\cite{RN156}, such as Fe$_3$GeTe$_2$/hBN/Fe$_3$GeTe$_2$~\cite{RN157}, Fe$_3$GeTe$_2$/MoS$_2$/Fe$_3$GeTe$_2$~\cite{RN158}, Fe$_3$GeTe$_2$/InSe/Fe$_3$GeTe$_2$~\cite{RN159}, and Fe$_3$GaTe$_2$/GaSe/Fe$_3$GaTe$_2$~\cite{RN160}. Furthermore, a plateau-like spin resistance has been unexpectedly observed in twisted FGT/FGT homojunctions without spacer~\cite{RN161,RN162}. Spin-orbit torque and current-driven switching have also been reported in Fe$_3$GeTe$_2$/Pt systems~\cite{RN163,RN164}. Furthermore, Fe$_3$GeTe$_2$ exhibits a topological nodal line with a large Berry curvature, leading to a large anomalous Hall effect~\cite{RN165}, an anomalous Nernst effect~\cite{RN166}, and most importantly, gigantic intrinsic spin-orbit torque~\cite{RN52,RN167}. A series of works has focused on this direction. Zhang \textit{et al.} reported that an in-plane current can tune the hard spin state to a soft state in nanoscale FGT devices by substantially reducing the coercive field. This surprising finding is possible because the in-plane current produces an unusually large spin-orbit torque in a single FGT without any heavy metal layer. This spin-orbit torque is directly related to the large Berry curvature and its band topology, which can alter the spin-related free energy landscape~\cite{RN52}.

Using this principle, a new spin device model was demonstrated, in which highly efficient nonvolatile switching and multilevel states are achieved with a tiny current~\cite{RN168}. They further developed this spin model into an all-vdW classic three-terminal SOT device using FGT/hBN/FGT heterostructure, where the spin configurations are written by the spin-orbit torque and read by the tunneling spin-resistance separately in a decoupled way~\cite{RN169}. More recently, it was reported that inversion symmetry breaking occurs through SHG in this previously thought centrosymmetric material~\cite{RN170}, adding another contribution to spin-orbit torque and resolving the remaining puzzle. Independent groups have also contributed to this direction in various ways: SHG measurements to support the intrinsic spin-orbit torque in bulk Fe$_3$GeTe$_2$~\cite{RN171}; imaging of current-driven switching and determination of motion velocity~\cite{RN172}; and three studies employing Fe$_3$GaTe$_2$, with the same structure as Fe$_3$GeTe$_2$, demonstrated similar phenomena and broadened the scope of giant intrinsic spin-orbit torque and device switching~\cite{RN173,RN174,RN175}.

\textbf{CrSBr}

Recently, chromium sulfur bromide (CrSBr) has attracted attention due to the coupling of its electronic properties and the layered A-type antiferromagnetic order~\cite{RN176}, which sensitively affects its Wannier excitons~\cite{RN177}. CrSBr is structurally orthorhombic with the D$_{2h}$ space group and exhibits a N\'eel temperature of 132~K in bulk~\cite{RN178}. The monolayers of CrSBr display intralayer ferromagnetism and interlayer antiferromagnetism. As the temperature decreases between 100 and 40~K, the magnetic fluctuation of CrSBr slows down, leading to an intermediate long-range magnetically ordered state, with spin freezing occurring below 40~K. CrSBr has uniaxial anisotropy in the plane, which is inherited from the anion structure~\cite{RN179,RN180}. One promising application of CrSBr is as a 2D magnetic semiconductor. The band gap has been experimentally found to be 1.51~eV at 200~K and computationally estimated as 2.1~eV~\cite{RN181}. Electron transport in CrSBr is tunable by controlling the carrier concentration through an electrostatic gate, revealing the strong coupling between magnetic order and charge transport~\cite{RN182}. Structural control of CrSBr has been used to tune the material's physical properties. For example, an electron beam was used to drive a local phase transformation, where the new configuration exhibits in-plane ferromagnetism and weak interlayer antiferromagnetism~\cite{RN183}.

\textbf{CrTe$_2$ and 1T-VSe$_2$}

In addition to the classic materials mentioned above, other materials with significant potential remain to be further explored despite fewer reports. One such material is CrTe$_2$, which maintains ferromagnetic long-range orders up to 300~K and down to the ultrathin limit. Furthermore, the magnetoresistance of CrTe$_2$ changes sign from $-0.6\%$ at 300~K to $+5\%$ at 10~K~\cite{RN184}. Another material of interest is 1T-VSe$_2$, which has been investigated using x-ray magnetic circular dichroism. Although no long-range ferromagnetism was observed, the study revealed the presence of short-range ferromagnetism and antiferromagnetic interactions between neighboring vanadium ions~\cite{RN185}.

\subsection{Multiferroic materials}

The magnetoelectric (ME) effect is a fundamental concept in modern condensed matter physics, referring to the ability to control magnetic polarization electrically or electric polarization magnetically. Two-dimensional (2D) vdW magnets have emerged as a new class of materials, demonstrating novel ME effects with diverse manifestations. Several important vdW magnets exhibit multiferroicity in two dimensions, including spin-charge correlations, the atomic ME effect, current-induced intrinsic spin-orbit torque, and the electrical gating and magnetic control of their electronic properties. Due to their intrinsic two-dimensional nature, vdW magnets with these ME effects represent an exciting frontier in condensed matter research, offering vast potential for innovative scientific exploration and technological development.

\begin{figure}
    \includegraphics[width=\linewidth]{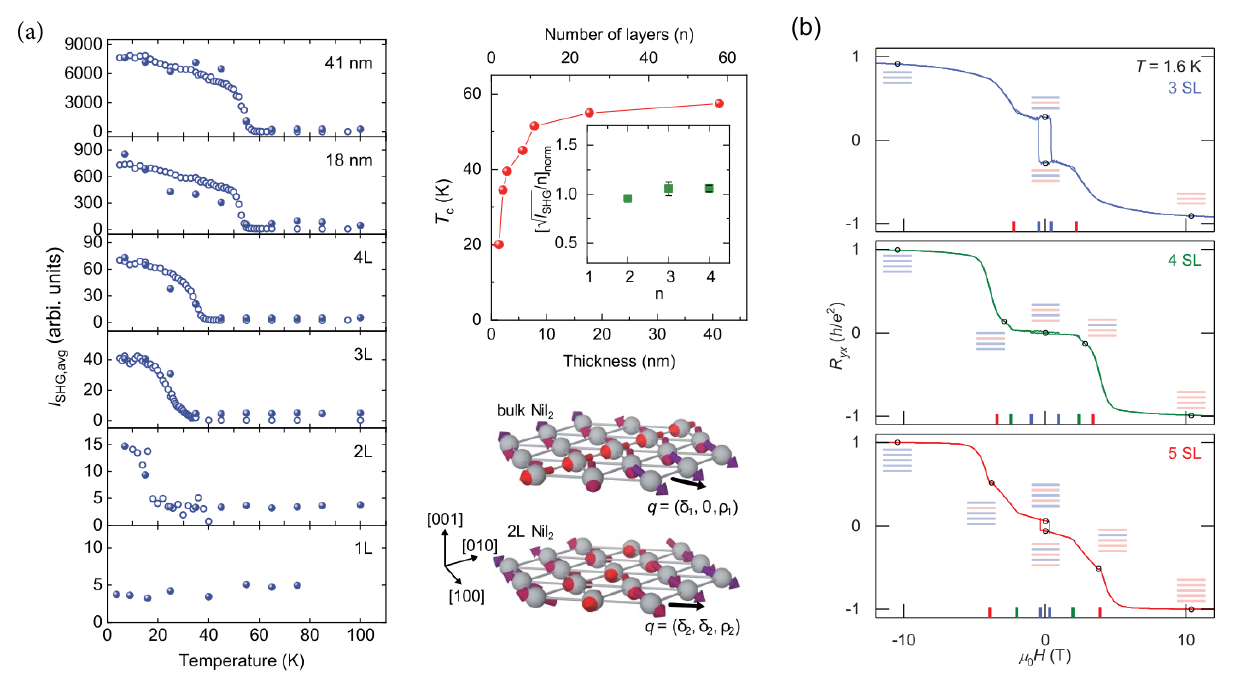}
    \caption{\label{fig:5}Two examples of vdW magnets exhibiting exotic phases: (a) a multiferroic state in NiI$_2$. From \citet{RN42}, and (b) a magnetic topological insulating state in MnBi$_2$Te$_4$. From \citet{RN46}. }
\end{figure}

\textbf{NiI$_2$: Atomically thin multiferroic NiI$_2$}

Bulk nickel iodide (NiI$_2$) undergoes a transition from monoclinic to trigonal layered structure at 59.5~K. A paramagnetic to antiferromagnetic transition occurs at a N\'eel temperature $T_{\mathrm{N}}$ of 75~K, which decreases to 35~K in trilayer flakes and 20~K in bilayer (see Fig. \ref{fig:5}(a), highlighting the role of interlayer coupling~\cite{RN42}. Monolayer and few-layer flakes of this material can be grown in h-BN and SiO$_2$/Si and applied in transistor fabrication, achieving a high on/off ratio of up to $10^6$~\cite{RN186}. In 2021, Ju \textit{et al.} initially proposed that the multiferroic order persists down to bilayer exfoliated nanoflakes of natural NiI$_2$~\cite{RN42}, while another group reported similar SHG in monolayer NiI$_2$ grown on h-BN substrates by physical vapor deposition~\cite{RN43}. However, many ongoing efforts on this compound remain controversial and have yet to be clarified. Multiferroic-enabled magnetic excitons may also appear in a 2D quantum-entangled system such as NiI$_2$~\cite{RN187}. The symmetry breaking in this material could enable magnetic excitons, which arise from Zhang-Rice-triplet and Zhang-Rice-singlet states. This mechanism produces an optical absorption peak at 1.384~eV with a 5~meV linewidth. Another interesting report was recently made using optical tools that NiI$_2$ hosts large chiral magnetoelectric correlations in two dimensions with a record dynamical magnetoelectric coupling constant $\alpha(\omega)$ in the terahertz spectral range~\cite{RN188}.

\textbf{NiPS$_3$: Atomic magnetoelectric effect in NiPS$_3$}

NiPS$_3$ is a charge transfer (CT) insulator characterized by zigzag antiferromagnetic ordering~\cite{RN76,RN77,RN79}. Previous studies have estimated that CT energy is negative~\cite{RN79} or within a small positive range, spanning 0.9 to 2.5~eV~\cite{RN80,RN189,RN190}. The low CT energy, coupled with strong $pd$-$\sigma$ hybridization, allows the doubly degenerate unoccupied Ni $e_g$ orbitals to mix with the occupied ligand $p$ orbitals of the six surrounding sulfur atoms. This leads to the spontaneous formation of self-doped ligand holes~\cite{RN191}. The interaction between an electron at the Ni site and a self-doped ligand hole results in spin-orbital entangled many-body states within the NiS$_6$ octahedron, manifesting as spin-triplet and orbital-singlet states~\cite{RN80,RN86,RN190,RN192}. These CT states are similar to the Zhang-Rice triplet (ZRT) states predicted for high-$T_c$ cuprates~\cite{RN193,RN194}. When ZRT states are formed in each NiS$_6$ octahedron, two types of orthonormal S 3$p$ holes can populate sulfur sites, forming spin-triplet states with neighboring Ni 3$d$ orbitals. Under zigzag magnetic ordering, two distinct edge-shared sulfur sites emerge, defined by the spin alignments of adjacent Ni ions: S1 sites with parallel spin alignments and S2 sites with antiparallel alignments. At S1 sites, the electron-hole orbitals share the same spin, whereas at S2 sites the spins are opposite. Charge disproportionation between these sulfur sites induces charge-stripe modulation, which may result in local electronic dipole polarization. Many-body calculations suggest that exchange interactions (Hund's coupling) between self-doped holes can create charge deviations of approximately 0.002–0.003$|e|$ between S1 and S2 sites, generating a local dipole polarization of about 0.1–0.2~$\mu$C/cm$^2$ per NiS$_6$ octahedron~\cite{RN80}. This atomic magnetoelectric effect plays a significant role in the spin-charge coupling observed in optical excitations~\cite{RN79,RN87,RN195}.

\textbf{Fe$_3$GeTe$_2$: Gigantic Intrinsic SOT in Fe$_3$GeTe$_2$}

The spin-orbit torque effect (SOT) in vdW ferromagnetic materials, particularly Fe$_3$GeTe$_2$, represents a transformative development in spintronics. SOT arises from spin polarizations generated by spin-orbit coupling mechanisms such as the spin Hall or Edelstein effect. This torque enables efficient manipulation of magnetization, which traditionally requires ferromagnet/heavy-metal bilayer structures~\cite{RN196,RN197}. Fe$_3$GeTe$_2$—a vdW ferromagnetic metal with a high Curie temperature ($\sim$200~K) and strong perpendicular magnetic anisotropy~\cite{RN139,RN143}—exhibits intrinsic SOT without the need for heavy metal layers. Recent investigations have discovered exceptionally efficient intrinsic SOT in Fe$_3$GeTe$_2$. Symmetry analysis reveals that only one specific SOT coefficient ($\Gamma_0$) remains active due to the crystalline symmetries of the material~\cite{RN167}. This coefficient significantly alters the spin-related free-energy landscape, reducing magnetic anisotropy and switching barriers. The theoretical atomic magnetoelectric (ME) coefficient is approximately 30~Oe/(mA/$\mu$m$^2$), which closely matches experimental measurements of around 50~Oe/(mA/$\mu$m$^2$)~\cite{RN52}. This large ME coefficient is attributed to the material's substantial Berry curvature, stemming from its topological nodal-line band structures.

Experimental findings have demonstrated a dramatic reduction in the coercivity driven by SOT, validated through coercivity measurements, theoretical simulations, and angular magnetoresistance studies. Fe$_3$GeTe$_2$ shows an SOT efficiency that exceeds that of conventional heavy metals~\cite{RN52}, such as Pt and Ta, by a factor of 100, highlighting the strong ME effect rooted in its unique band topology. Independent studies using second-harmonic electrical transport techniques have further confirmed these results~\cite{RN171}. The SOT effect in Fe$_3$GeTe$_2$ is closely related to its symmetry characteristics. Although initially considered centrosymmetric, the broken inversion symmetry, likely induced by Fe vacancies, has been identified in Fe-deficient samples using the SHG technique~\cite{RN170}. This defect-induced symmetry breaking influences the material’s band topology and quantum behaviors, enabling phenomena such as non-reciprocal transport and non-linear optical responses. The intrinsic SOT and ME effects in Fe$_3$GeTe$_2$ have opened new avenues in magnetic memory technology. Devices utilizing this material achieve reductions in switching current density and power dissipation by factors of 400 and 4000, respectively, compared to Pt-based systems~\cite{RN168}. Furthermore, these devices support multilevel states, increasing data capacity with up to eight states per device~\cite{RN168}. Recent demonstrations using Fe$_3$GaTe$_2$ at room temperature have replicated similar physical and memory functionalities, reinforcing its potential for industrial-scale SOT-MRAM applications~\cite{RN173,RN175,RN198}.

\subsection{Magnetic topological materials}

Magnetic topological materials combine magnetic ordering with topologically protected electronic states, leading to unique phenomena such as the quantum anomalous Hall effect and magnetic Weyl semimetals. The interplay of symmetry, magnetism, and topology makes these materials a central focus in condensed matter physics. For this reason, intensive efforts have been made to discover new vdW magnets with nontrivial topology. Although the list is relatively short, they present exciting new scientific opportunities and warrant continuing interest.

\textbf{MnBi$_2$Te$_4$}

Manganese bismuth telluride (MnBi$_2$Te$_4$) has been confirmed as an antiferromagnetic topological insulator through \textit{ab initio} calculations and experimental techniques (see Fig. \ref{fig:5}(b)) \cite{RN199}. The surface of symmetry breaking (0001), reported in the same study, indicates a large bandgap in the topological surface state. Various methods have determined that a Bi-Te flux-grown MnBi$_2$Te$_4$ sample exhibits a Néel temperature $T_\mathrm{N}$ of 24 K. Below $T_\mathrm{N}$, MnBi$_2$Te$_4$ displays A-type antiferromagnetic order along the $c$-axis. The material also exhibits strong spin-lattice coupling, as demonstrated by the effects of critical scattering on thermal conductivity near $T_\mathrm{N}$ \cite{RN200}. As the temperature increases, MnBi$_2$Te$_4$ undergoes a transition from a canted antiferromagnetic phase to a ferromagnetic phase. Spin excitations (magnons) and fluctuations also depend on the thickness of the flake \cite{RN69}. Moreover, the potential for interlayer magnetophotonics has been explored through anomalies observed in magneto-Raman spectroscopy near the phase transition \cite{RN61}. Interdisciplinary research combining experiments and first-principles calculations has highlighted the potential of Mn(Sb$_x$Bi$_{1-x}$)$_2$Te$_4$ as an intrinsic magnetic topological insulator within an optimal region of the Sb-Bi phase diagram \cite{RN201}. In the MnBi$_{2n}$Te$_{3n+1}$ family, the dominance of the topological insulator Bi$_2$Te$_3$ was investigated using cryogenic low-frequency Raman spectroscopy for $n=1,2,3,4$ \cite{RN202}.

\textbf{Co$_{1/3}$TaS$_2$}

The layered metallic triangular lattice antiferromagnet Co$_{1/3}$TaS$_2$ has garnered significant recent interest due to its distinctive non-coplanar triple-$\mathbf{Q}$ magnetic ground state. This material undergoes two antiferromagnetic phase transitions at $T_\mathrm{N1} = 38$~K and $T_\mathrm{N2} = 26.5$~K. Neutron diffraction measurements have revealed that both phases develop magnetic Bragg peaks at the M-points of the Brillouin zone ($\mathbf{Q}_\nu = \mathbf{G}_\nu/2$, $\nu = 1,2,3$), where $\mathbf{G}_\nu$ are reciprocal lattice vectors related by 120$^\circ$ rotations about the $c$-axis \cite{RN47,RN48}. A key observation below $T_\mathrm{N2}$ is the emergence of a large spontaneous Hall conductivity, $\sigma_{xy}(H=0)$, which rules out the possibility of a multi-domain single-$\mathbf{Q}$ magnetic order for $T < T_\mathrm{N2}$. Such a configuration would violate the symmetry constraints imposed by time-reversal symmetry combined with lattice translations ($\tau_1 a T$) \cite{RN47,RN48}, making it incompatible with the observed Hall conductivity. Instead, triple-$\mathbf{Q}$ ordering below $T_\mathrm{N2}$ has been conclusively established through a combined analysis of neutron diffraction and bulk electrical transport measurements, clarifying the symmetry of the magnetic state. The potential for obtaining atomically thin flakes of Co$_{1/3}$TaS$_2$ through mechanical exfoliation \cite{RN203} or chemical intercalation \cite{RN204,RN65} highlights its promise as a versatile platform for studying the two-dimensional limit of triple-$\mathbf{Q}$ magnetism. In this regime, the material could exhibit topologically nontrivial spin textures, offering exciting opportunities to explore novel magnetic and electronic phenomena.

\subsection{Spin Hamiltonian}
The spin Hamiltonian is essential for understanding the magnetic properties and interactions in vdW magnets. It provides a theoretical framework for describing complex spin dynamics, anisotropies, and exchange interactions in these low-dimensional systems. VdW magnets, with their intrinsic two-dimensionality and tunable properties, host a variety of exotic magnetic states, such as skyrmions and noncollinear spin textures, which can be captured and analyzed through the spin Hamiltonian. This understanding is crucial for designing vdW magnetic materials for spintronics applications and exploring novel quantum phenomena in reduced dimensions.

A generic spin Hamiltonian can be written in the following formula: 
\begin{align}
H &= \sum \left[ J\left( S_{i,x} S_{i+1,x} + S_{i,y} S_{i+1,y} \right) 
+ \alpha S_{i,z} S_{i+1,z} \right] \nonumber \\
&\quad + K \sum \left( S_{i,z} \right)^2,
\end{align}
where \( S_{i,x(y,z)} \) is the x(y,z) component of the magnetic moment at the i$^{th}$ site. J is the exchange interaction between the neighboring magnetic moments, and \( \alpha \) is the ratio between two exchange integrals for the x/y and z components. K represents the magnetic anisotropy, with a positive value indicating an in-plane anisotropy and a negative value indicating an out-of-plane anisotropy. The following diagram illustrates how the generic formula can realize the three fundamental spin Hamiltonian.

\begin{figure}
    \includegraphics[width=\linewidth]{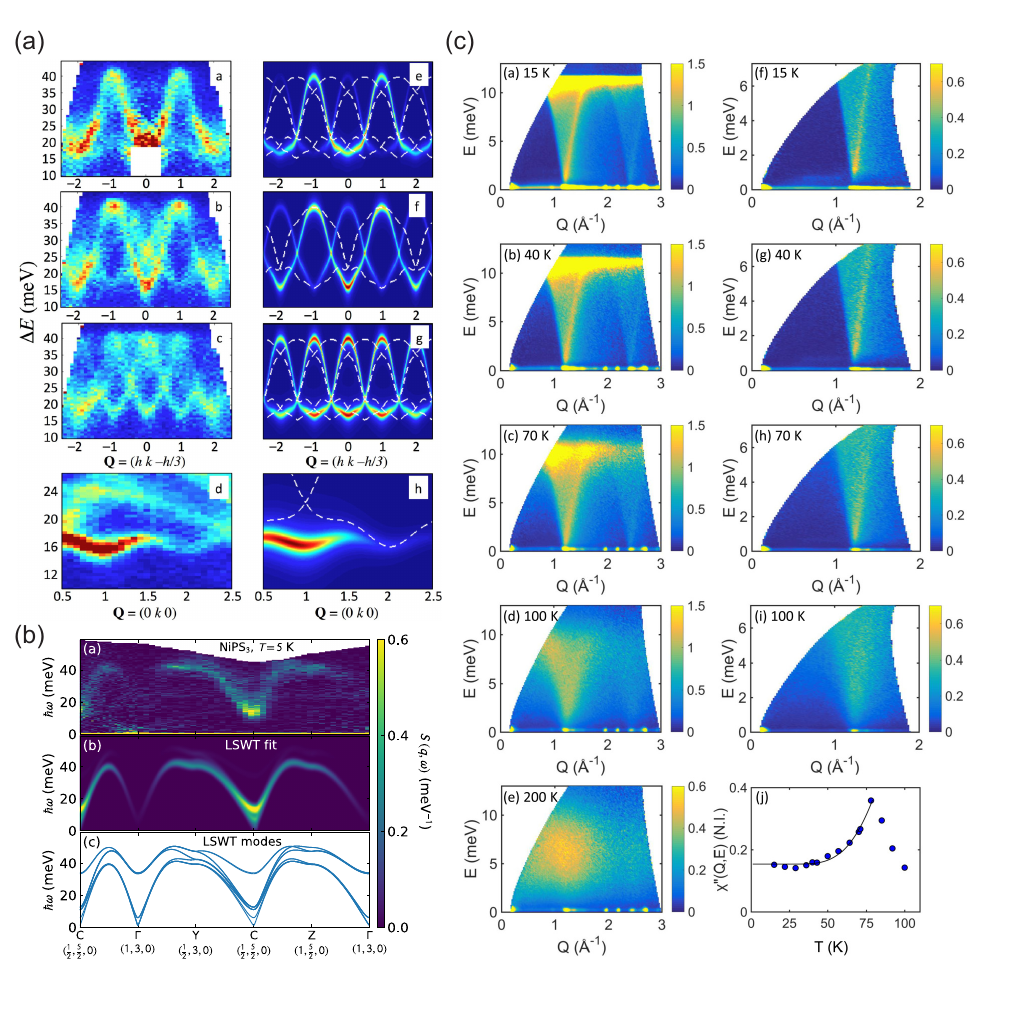}
    \caption{\label{fig:6}Representative inelastic neutron scattering data showcasing three types of spin Hamiltonians in vdW magnets: (a) Ising Hamiltonian in FePS$_3$. From \citet{RN206}; (b) XY Hamiltonian in NiPS$_3$. From \citet{RN208}; and (c) Heisenberg Hamiltonian in MnPS$_3$. From \citet{RN210}.}
\end{figure}

\textbf{Ising Model}

One of the most important intellectual motivations behind the study of vdW magnets is to test the theoretical predictions experimentally made decades ago regarding the three fundamental spin Hamiltonian. Among these, the Ising model has played a pivotal role in shaping the field due to its symbolic position in the history of magnetism. The Ising model can be realized in the earlier generic model by taking the limit of $\alpha = \infty$. 

The Ising model was the first to be solved analytically. In 1943, Onsager solved the Ising model for two dimensions: Onsager used the transfer matrix method to drive an exact solution for magnetization as a function of temperature and the exchange integral \cite{RN15}. This result not only demonstrated a phase transition at finite temperature, but also introduced concepts such as critical exponents and universality, which have become foundational in modern statistical mechanics.

Despite its seminal contribution, no real experimental tests have been conducted using monolayer magnetic materials. This changed in 2016 when FePS$_3$ demonstrated stable magnetism in monolayer form using Raman spectroscopy \cite{RN7}. Interestingly, FePS$_3$ exhibits an exceptionally large out-of-plane magnetic anisotropy of 22 meV per Fe ion, as observed in both photoelectron emission microscopy (PEEM) \cite{RN205} and inelastic neutron scattering experiments (see Fig. \ref{fig:6}(a)) \cite{RN206}. This large magnetic anisotropy is understood to arise from the trigonal distortion of the Fe$^{2+}$ 3d bands, which produces a significant anisotropy.

Afterwards, similar efforts have been made with vdW ferromagnets, with notable examples of Cr$_2$Ge$_2$Te$_6$ \cite{RN38} and CrI$_3$ \cite{RN39}. Both systems have used MOKE measurements to detect ferromagnetic signals from thin samples. The magnetism remained stable to the bilayer of Cr$_2$Ge$_2$Te$_6$, while stable magnetism is observed in the monolayer CrI$_3$. Another interesting vdW ferromagnet is Fe$_3$GeTe$_2$, which has been shown to exhibit stable magnetism in the monolayer form \cite{RN40}. Furthermore, it was shown that the transition temperature of few-layer Fe$_3$GeTe$_2$ can be controlled and even tuned to room temperature.

\textbf{XY Model}

When the $\alpha$ term in the generic model becomes zero, the spin Hamiltonian reduces to the XY model, where the spin fluctuates only within the XY plane. This XY model was theoretically investigated in the early 1970s by two groups \cite{RN18,RN19}, leading to an astonishing discovery of topological order. In many ways, this discovery heralded the beginning of what would later become the vibrant field of topological physics. This Berezinskii-Kosterlitz-Thouless (BKT) transition introduced the concept of topological order, profoundly impacting our understanding of 2D systems. Its insights have been extended to superfluidity, superconductivity, and even quantum computing technologies, its importance being recognized by the 2016 Nobel Prize.

An experimental realization of the XY model was achieved in NiPS$_3$, where the order parameter, measured by Raman spectroscopy, remained stable down to the bilayer before disappearing in the monolayer form \cite{RN76}. A notable recent development: recent inelastic neutron scattering experiments successfully pinned down the basic Hamiltonian of NiPS$_3$ (see Fig. \ref{fig:6}(b)) \cite{RN207,RN208}. According to these experiments, an XXZ model better describes NiPS$_3$, although it can be approximated as an XY system in the low-energy limit. A similar XXZ model has been observed in several compounds, including CoPS$_3$, through inelastic neutron scattering experiments \cite{RN89}. Another interesting report concerns the ferromagnetic CrCl$_3$, for which a monolayer was grown by molecular beam epitaxy on graphene \cite{RN134}. The X-ray magnetic circular dichroism (XMCD) measurements found that the monolayer CrCl$_3$ is a ferromagnet, in contrast to the bulk CrCl$_3$, which is antiferromagnetic and exhibits critical XY behavior.

\textbf{Heisenberg Model}

When magnetic moments are allowed to fluctuate along the three principal axes with $\alpha = 1$ in the generic model, it becomes the Heisenberg model. The Heisenberg model ($\alpha = 1$) presented a different challenge. In 1966, Mermin and Wagner, followed by Hohenberg in 1967, rigorously proved that continuous spin symmetries cannot be spontaneously broken at finite temperatures in two dimensions. The Heisenberg model has been instrumental in advancing our understanding of quantum spin systems, particularly in identifying spin liquid states and exploring quantum phase transitions, with direct implications for experiments in quantum magnets and ultra-cold-atom systems.

Due to the continuous spin symmetry in the three directions, achieving stable magnetism in thin samples becomes more challenging. The intensive theoretical studies, now known as the Mermin-Wagner theorem \cite{RN16}, dictate that no stable magnetism can exist at finite temperature in ideal two dimensions for the Heisenberg model due to strong spin fluctuations. Generally speaking, most vdW magnets, both ferromagnets and antiferromagnets, fall into this category of spin Hamiltonian. When real vdW magnets with the Heisenberg model were experimentally examined, it was found that such systems do not exhibit stable magnetism. For example, inelastic neutron scattering of MnPS$_3$ revealed that it follows the Heisenberg model (see Fig. \ref{fig:6}(a)) \cite{RN209}. Raman and SHG studies of MnPS$_3$ showed that the antiferromagnetic order is rapidly suppressed with decreasing thickness, eventually disappearing in thin samples \cite{RN71,RN73}.

\textbf{Magnon as quasiparticle}

Magnons are fundamental quasiparticles in magnetically ordered materials, representing collective spin-wave modes that carry spin angular momentum. Their dispersion is well defined as a function of wave vector $q$ and angular frequency $\omega$. Although magnons can be directly investigated using inelastic neutron scattering, they can also be effectively probed through Raman spectroscopy and terahertz (THz) spectroscopy. The magnon dispersion, along with their interactions—both among themselves and with other quasiparticles such as phonons and excitons—provides crucial insights into the fundamental magnetic structure and allows for the experimental determination of the exchange constants underlying the model Hamiltonian of the magnetic system under study. It should be noted that optical tools have played an important role in the micro-spectral study of very thin samples.

In ferromagnets, where the unit cell typically contains a single magnetic ion, the magnon spectrum features a single branch near the $\Gamma$ point of the Brillouin zone. The energy of this magnon mode at $q = 0$, often referred to as the magnon 'gap,' arises primarily from magnetic anisotropy and can be directly probed using terahertz light, whose wavelength is of the order of the submillimeter. This $q = 0$ mode, also known as the uniform mode, represents a collective in-phase precession of all spins in the system, effectively forming a rotating macroscopic moment. This behavior aligns with a semiclassical description of magnetization dynamics. The uniform mode is also called the ferromagnetic resonance (FMR), with its characteristic frequency typically falling within the microwave region, i.e. a few GHz or higher. As a result, microwave spectroscopy has traditionally been the tool of choice for investigating FMRs in ferromagnetic systems. In contrast, terahertz spectroscopy has been particularly useful for studying magnon modes at 50 GHz or higher for exceptional cases of ferromagnets with high-FMR frequencies.

However, antiferromagnets inherently contain at least two magnetic sublattices per unit cell, giving rise to at least two magnon branches near the $\Gamma$ point. In a two-sublattice antiferromagnet, the lower-energy magnon mode is primarily determined by magnetic anisotropy, while the higher-energy mode also reflects the strength of the exchange interaction. The corresponding antiferromagnetic resonances (AFMRs) span frequencies from approximately 100 GHz to several THz, making THz spectroscopy an ideal technique for their investigation. Most antiferromagnets are collinear or weakly ferromagnetic (WF). In collinear antiferromagnets, the AFMR modes correspond to oscillations of the antiferromagnetic order parameter, often described as the Néel vector $\mathbf{L}$, representing the difference between the two sublattice magnetizations. In this case, $\mathbf{L}$ oscillates in magnitude while maintaining a fixed orientation in space. At zero field, two AFMR modes are degenerate. In contrast, weakly ferromagnetic antiferromagnets exhibit a small net magnetization due to spin canting, where the sublattice moments do not fully cancel. This spin canting introduces a weak ferromagnetic component, splitting the two AFMRs into two distinct modes: the quasi-ferromagnetic resonance (qFMR) and the quasi-antiferromagnetic resonance (qAFMR). Although both modes are technically still AFMR modes, they can be interpreted differently. The qFMR mode resembles a conventional FMR where the weak ferromagnetic moments precess in phase, akin to the behavior of a ferromagnet. Meanwhile, the qAFMR mode retains the characteristics of an AFMR in a collinear antiferromagnet, where the oscillation of the two nonparallel sublattice magnetizations resembles the dynamics of the Néel vector $\mathbf{L}$.

\section{\label{sec:III}Optical Studies}
Although traditional characterization tools such as magnetic susceptibility measurements or inelastic neutron scattering (INS) can be used to probe the magnetic states of bulk crystals of 2D vdW magnetic materials, such tools become impractical when it comes to thin samples of these materials due mainly to the small sample volume. It has been demonstrated that various optical measurement techniques can be reliably applied to study the magnetic states of thin samples of vdW magnetic materials. These include second-harmonic generation (SHG), Raman scattering, magneto-optical Kerr effect (MOKE), terahertz spectroscopy, photoluminescence (PL), and optical absorption, and magneto-optical measurements. Optical methods have the advantages of nondestructive measurements and high sensitivity, although the correlation between the optical data and the magnetic state is sometimes indirect. In this section, the results of the optical measurements on 2D vdW magnetic materials are summarized.

\subsection{Second-Harmonic Generation}

\subsubsection{SHG technique}

Second-Harmonic Generation (SHG) is a special case of sum-frequency generation in which the two electromagnetic waves are identical \cite{RN259}. When a strong pulse of a laser with frequency $\omega$ is incident in a nonlinear medium, the second-order induced polarization can be written as
\[
P_i^{(2\omega)} = \epsilon_0 \sum_{jk} \chi_{ijk}^{(2)} E_j^{(\omega)} E_k^{(\omega)},
\]
where $i$, $j$, and $k$ refer to the Cartesian components and $\chi_{ijk}^{(2)}$ are the components of the $3 \times 6$ tensor of the second-order susceptibility \cite{RN260}. The symmetry of the system is reflected in the susceptibility tensor $\chi_{ijk}^{(2)}$ and consequently in the polarization dependence of the SHG signal. Therefore, examination of the polarization dependence of the SHG signal is an effective way to monitor the symmetry changes.

In the dominant electric dipole term, the second-order signal is nonzero only if $\chi^{(2)}$ is finite. In centrosymmetric systems with inversion symmetry, $\chi^{(2)}$ is zero due to symmetry, so the SHG signal is non-zero only in systems without inversion symmetry \cite{RN260}. In magnetic materials, second-order susceptibility can be further written as $\chi^{(2)} = \chi_i^{(2)} + \chi_c^{(2)}$, where the time-invariant tensor $\chi_i^{(2)}$ is responsible for the crystallographic contribution (i-type) and the time-noninvariant tensor $\chi_c^{(2)}$ for the spin-dependent contribution (c-type) \cite{RN261}. Ferromagnetic (FM) ordering does not inherently break the inversion symmetry of the crystal, whereas it is possible to break the inversion symmetry with antiferromagnetic (AFM) ordering. Therefore, SHG is often used to monitor magnetic ordering in AFM materials.

Since the SHG signal is proportional to the square of the incident light intensity, pulsed lasers with either picosecond or femtosecond pulse widths are used to achieve high peak intensity while keeping the time-integrated intensity on the sample low to avoid damages to the sample. In a typical SHG experiment, a picosecond YAG laser with a wavelength of 1064 nm is used as the source, and the signal at 532 nm is measured. For samples on substrates, the backscattering geometry is used. In this case, an appropriate filter should be used to block the reflected fundamental wave at 1064 nm from entering the detection system. In addition, a polarizer for the incident beam and an analyzer for the SHG signal are used to examine the polarization dependence of the SHG.

\subsubsection{\texorpdfstring{CrI$_3$}{CrI3}}

CrI$_3$ is one of the most studied magnetic vdW materials and exhibits layer-dependent Ising ferromagnetism \cite{RN39}. Each layer of CrI$_3$ is ferromagnetically ordered with magnetic moments in the out-of-plane direction, whereas adjacent layers are antiferromagnetically coupled. Regarding atomic structure, the bilayer CrI$_3$ is centrosymmetric regardless of any rigid translation between the two sheets, which prohibits i-type SHG \cite{RN122}. However, antiferromagnetic ordering breaks both time-reversal and spatial-inversion symmetries, which leads to the c-type SHG. Using this unique property, Sun \textit{et al.} observed a strong SHG signal below the critical temperature \cite{RN122}. They also showed that the polarization dependence of the SHG signal exhibits the underlying C$_{2h}$ crystallographic symmetry due to the monoclinic stacking of the layers. This indicates that polarization-resolved SHG measurements can be used to monitor the magnetic transitions in vdW antiferromagnets and also to probe the crystallographic symmetry.

\subsubsection{\texorpdfstring{MnPS$_3$ and MnPSe$_3$}{MnPS3 and MnPSe3}}

\begin{figure}
    \includegraphics[width=\linewidth]{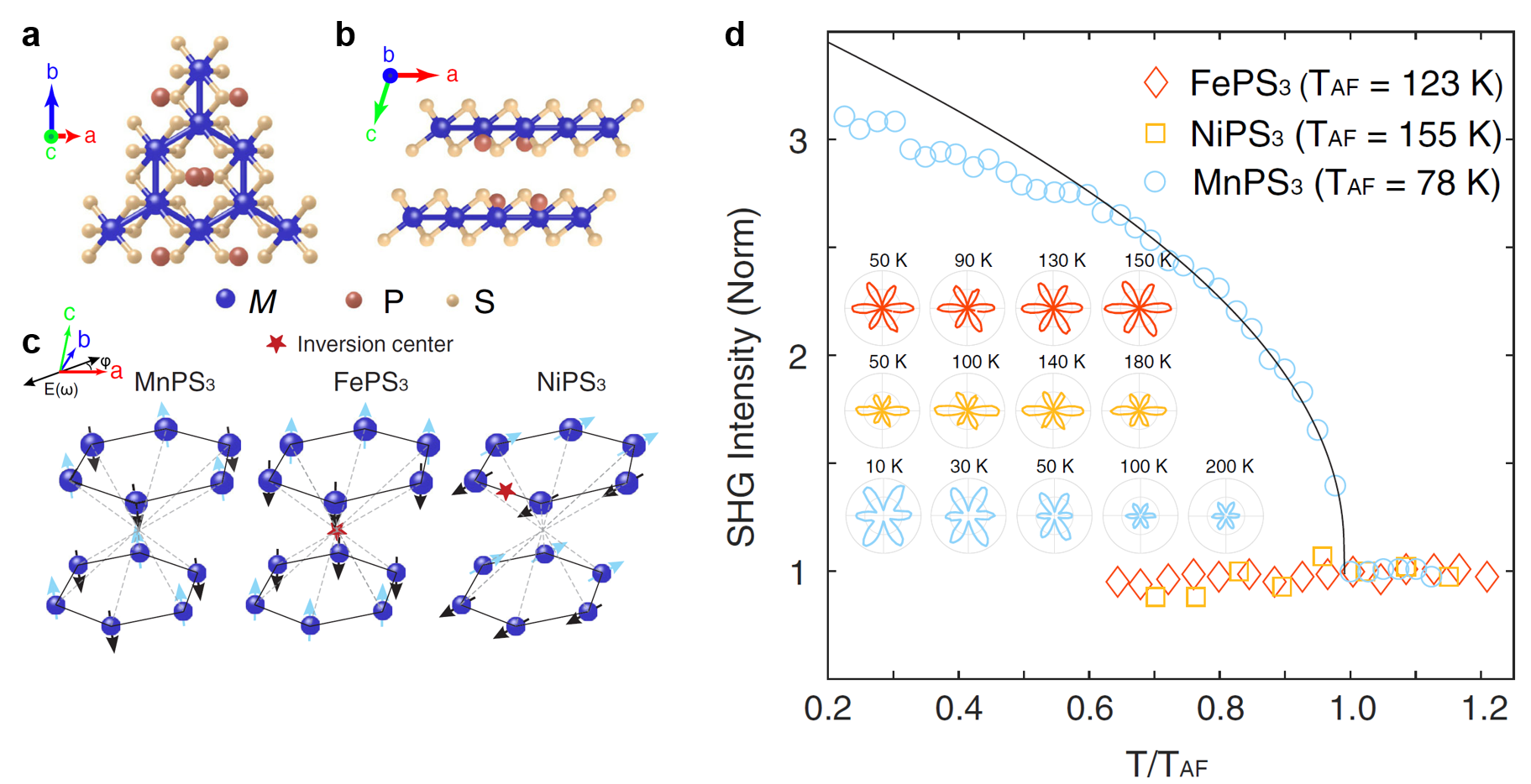}
    \caption{\label{fig:7}Crystal structure of MPS$_3$ (M = Mn, Fe, Ni) viewed along the (a) $c$- and (b) $b$- axis. (c) AFM ordered structures of MPS$_3$ with the arrows indicating the spin directions. FePS$_3$ and NiPS$_3$ have inversion symmetry, but MnPS$_3$ does not have an inversion center. (d) Temperature dependence of the SHG intensity from MPS$_3$. Adapted from \citet{RN73}.}
\end{figure}

MnPS$_3$ is a Heisenberg-type antiferromagnet that exhibits a Néel-type ordering below the Néel temperature of 78 K \cite{RN62}. Its crystal structure is centrosymmetric and identical to FePS$_3$ and NiPS$_3$ without magnetic ordering. However, unlike the other two, MnPS$_3$ loses inversion symmetry upon AFM ordering, as illustrated in Fig. \ref{fig:7} \cite{RN73}. As shown in Fig. \ref{fig:7}(d), all three materials show weak SHG signals above the Néel temperature, since all of them are centrosymmetric. This is due to the contribution of the electric quadrupole to the SHG, which can be finite even in centrosymmetric systems \cite{RN262}. When the temperature is lowered below the Néel temperature, only MnPS$_3$ exhibits an enhancement of the SHG signal due to the loss of inversion symmetry in the AFM ordered structure \cite{RN73}. This work clearly demonstrates that the critical behavior of the SHG signal below the Néel temperature is a key signature of the breaking of the inversion symmetry in the AFM state.

Ni $et$ $al.$ improved the sensitivity of the SHG measurements and successfully measured the SHG signals from samples of a few layers of MnPS$_3$ \cite{RN74}. They were able to observe the enhancement of the SHG signal at low temperatures down to 2 layers. The SHG signal was considerably suppressed for the 2-layer sample and was unobservable for the monolayer. They were also able to determine the Néel temperature from the temperature dependence of the SHG intensity: the lowering of the Néel temperature was observed for thicknesses below 10 layers. They also used the improved sensitivity to obtain SHG intensity maps and observed domain structures in bulk samples and samples with few layers \cite{RN74}. 

MnPSe$_3$ also exhibits inversion-breaking Néel-type AFM ordering below the Néel temperature of 68 K \cite{RN263}. Unlike MnPS$_3$, however, MnPSe$_3$ has in-plane spins with very large XY anisotropy \cite{RN214}. Similarly to MnPS$_3$, the electric quadrupole SHG signal was observed above the Néel temperature with a 6-fold polarization dependence pattern. Below the Néel temperature, strong enhancement of the SHG signal was observed, but with a 2-fold polarization dependence. Using SHG imaging, they observed domains with strong SHG signals separated by thin boundaries with weak SHG signals \cite{RN263}. The strong SHG domains all showed the same orientation of the 2-fold polarization dependences. Ni \textit{et al.} interpreted this 2-fold polarization dependence as due to Ising-like ordering induced by residual strain in the samples. They tested this hypothesis by applying an in-plane strain to the sample, which rotated the orientation of the 2-fold polarization pattern. They were also able to observe a transition in a monolayer sample. Magnetic ordering is not expected in the 2D limit in Heisenberg- or XY-type AFM. This observation of a transition in the monolayer could be another indication of the Ising-like order, possibly due to strain. This differs from the reports of suppression of magnetic ordering in the monolayer limit in MnPS$_3$ \cite{RN74} or NiPS$_3$ \cite{RN76}.

\subsubsection{CrSBr}

CrSBr has rectangular layers that exhibit an in-plane anisotropic FM order but AFM coupling between the layers, with a Néel temperature of 132 K \cite{RN176}. CrSBr is centrosymmetric regardless of the number of layers in the paramagnetic state. Below Néel temperature, the AFM order between layers in even numbers of layers breaks the inversion symmetry and the time-reversal symmetry similar to the case of bilayer CrI$_3$. In monolayer or odd numbers of layers, the inversion symmetry is maintained, and so a strong enhancement of the SHG signal is not expected below the Néel temperature. However, Lee \textit{et al.} reported that the SHG signal showed a clear enhancement below a critical temperature of $\sim$146 K in the ferromagnetic monolayer CrSBr \cite{RN178}. This observation can only be explained by invoking higher-order contributions to SHG. Since there is a clear correlation with magnetic ordering, Lee \textit{et al.} proposed that the magnetic dipole contribution \cite{RN264} is responsible for the SHG below the critical temperature \cite{RN178}. They also observed that the transition temperature increases with decreasing layer numbers in contrast to other 2D vdW magnets. This is one of few examples of SHG enhancement due to ferromagnetic ordering.

\subsubsection{\texorpdfstring{CrPS$_4$}{CrPS4}}

CrPS$_4$ has monoclinic symmetry with space group C2 which does not include inversion symmetry. Below the N\'eel temperature of 38 K, each layer has FM ordering, but bulk CrPS$_4$ shows AFM ordering with alternating layers having opposite spin directions, similar to the case of CrI$_3$. The lack of inversion symmetry leads to an i-type SHG signal, and its polarization dependence reflects the 2-fold symmetry of the monoclinic structure \cite{RN265}. Since AFM ordering does not introduce inversion symmetry breaking for this particular material, no significant change in the SHG signal is observed near the N\'eel temperature. However, the SHG signal increases dramatically below a critical temperature of 25 K in the monolayer. The polarization dependence pattern also changes as the temperature is changed through the critical temperature. This result was interpreted as due to the additional nonzero component $\chi^{i}$ of the susceptibility tensor due to FM ordering \cite{RN265}. This differs from the case of CrSBr in that the additional SHG signal due to FM ordering can be described within the electric dipole approximation without invoking higher-order contributions.

\subsubsection{\texorpdfstring{NiI$_2$}{NiI2}}

NiI$_2$ has a centrosymmetric crystal structure with space group \textit{R$\bar{3}$m}. Bulk NiI$_2$ exhibits two magnetic transitions: as the temperature decreases, there is a transition to a collinear AFM phase at 76 K (T$_\mathrm{N1}$), and then a transition to a helimagnetic state of ferroelectric order occurs at 58 K (T$_\mathrm{N2}$) \cite{RN266,RN216}. The ferroelectric order induces a strong SHG signal below T$_\mathrm{N2}$ \cite{RN42,RN43}. In few-layer NiI$_2$, the critical temperature for the enhancement of the SHG signal decreases as the thickness is decreased. Ju $et$ $al.$ reported that the critical temperature was $\sim$20 K for the bilayer, but no enhancement of SHG was observed for the monolayer \cite{RN42}. They interpreted the lack of ferroelectric order in the monolayer as a result of the Mermin-Wagner theorem \cite{RN16} since the magnetic Hamiltonian of NiI$_2$ is of Heisenberg type. On the other hand, Song $et$ $al.$ reported that they observed the transition for all thicknesses down to the monolayer, with the critical temperature of $\sim$20 K for the monolayer \cite{RN43}. They explained that the anisotropy in the Heisenberg-type magnetic Hamiltonian supports magnetic ordering at finite temperature in the monolayer. This disagreement has yet to be reconciled. A possible explanation is the different sample preparations: Ju $et$ $al.$ prepared the monolayer sample on SiO$_2$-covered Si substrates by exfoliation of bulk crystals, whiles Song $et$ $al.$ grew a few-layer samples on hBN substrates by physical vapor deposition. Different interactions with the substrate may result in different degrees of anisotropy in the Heisenberg-type magnetic interaction that is necessary for the magnetic ordering. Another, though less likely, possibility is the misidentification of the number of layers by one of the two groups because optical contrast or atomic force microscopy are known to be less reliable for identifying monolayers of 2D materials.

\subsection{Raman spectroscopy}

\subsubsection{Raman scattering}

Raman spectroscopy is one of the most widely used experimental tools in 2D material research \cite{RN267,RN268}: it has been successfully used to determine the number of layers, strain, doping, crystallographic orientation, etc. In Raman scattering \cite{RN269}, the incident photons interact with phonons in the target material and lose or gain energy depending on whether a phonon is generated or absorbed in the scattering process, respectively. By analyzing the energy difference between the incident and scattered photons, one obtains information about the phonons and the crystal symmetry. However, the scattering effect is not limited to phonons, but other quasiparticles, such as magnons or electronic excitations, can also produce similar inelastic light scattering effects. It should be noted that, because the momentum of light is negligible in comparison with the size of the Brillouin zone, only those quasiparticles with zero momenta (or a group of quasiparticles with net momentum of zero) can be probed by Raman scattering. 

In vdW magnetic materials research, Raman spectroscopy has successfully monitored the magnetic ordering in atomically thin layers. The signatures of magnetic ordering in the Raman spectra can be categorized into three cases: (1) changes in the symmetry, including the doubling of the unit cell due to magnetic ordering, may lead to the appearance (or disappearance) of Raman peaks; (2) magnetic ordering modifies the interaction between the atoms in the crystal resulting in the shift or splitting of some phonon modes; and (3) genuinely magnetic phenomena such as magnons or spin fluctuations show up with magnetic ordering. Raman spectroscopy is particularly useful for studying antiferromagnetic materials because ferromagnetic materials can be directly probed with MOKE. In the following sections, some representative works on Raman spectroscopy of vdW magnetic materials are summarized.

\subsubsection{\texorpdfstring{FePS$_3$ and FePSe$_3$}{FePS3, FePSe3}}

\begin{figure}
    \includegraphics[width=\linewidth]{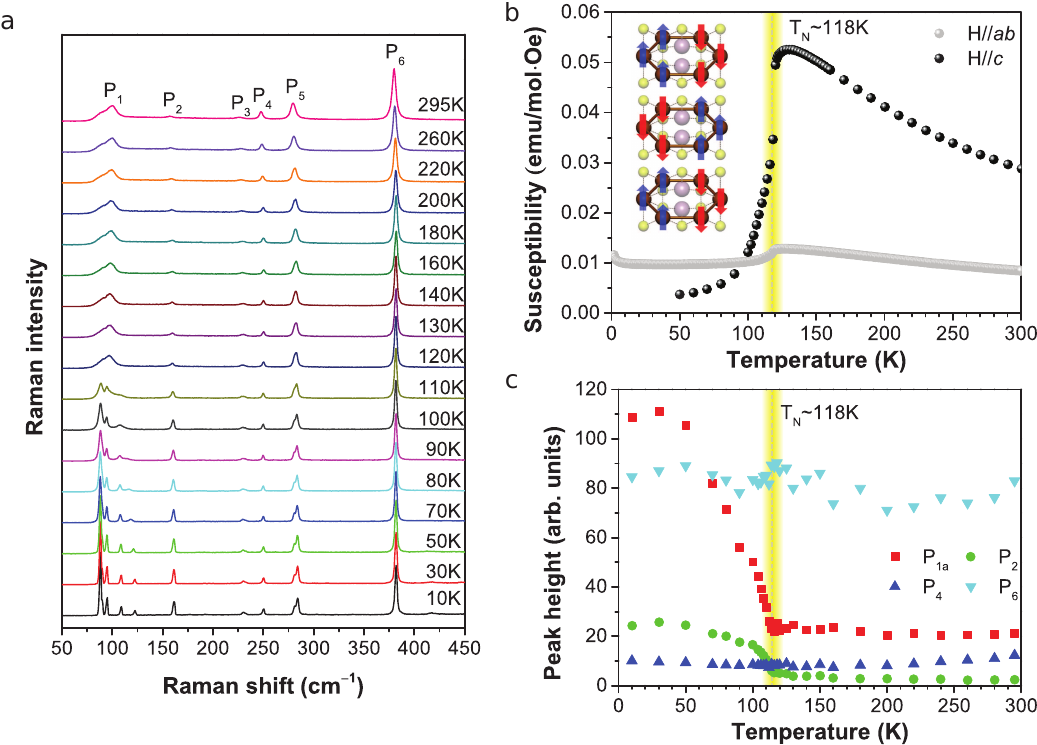}
    \caption{\label{fig:8}(a) Temperature dependence of the Raman spectrum for bulk FePS$_3$. (b) Temperature dependence of the magnetic susceptibility along $a$ or $b$ (black spheres) and $c$ (gray spheres) axes. (c) Temperature dependence of the intensities of several Raman peaks. From \citet{RN7}.
}
\end{figure}

FePS$_3$ is a zigzag-type Ising AFM with a N\'eel temperature (T$_\mathrm{N}$) of 118 K for the bulk. As shown in Fig. \ref{fig:7}, because the inversion symmetry is maintained in the AFM phase, the magnetic ordering cannot be monitored by SHG. However, the in-plane unit cell doubles in the AFM phase, leading to zone-folding of the in-plane Brillouin zone. Upon zone-folding, some zone-boundary phonons are folded back onto the zone center and become accessible by Raman scattering. In the Raman spectrum of the bulk FePS$_3$, most of the peaks change little as the temperature is reduced. However, a broad peak at $\sim$100 cm$^{-1}$ undergoes a dramatic change as the temperature is lowered through T$_\mathrm{N}$ \cite{RN7,RN270}. Four sharp peaks emerge below T$_\mathrm{N}$ as shown in Fig. \ref{fig:8}(a). The intensity of the peak at the lowest frequency, P$_{1a}$, shows a good correlation with the AFM ordering measured by magnetic susceptibility (Fig. \ref{fig:8}(b) and (c)). The highest frequency peak of the four new peaks has a temperature dependence somewhat different from those of the other 3 peaks and is determined to be the magnon mode \cite{RN271}. This magnon mode splits under a magnetic field applied in the direction normal to the layers. Similar trends were observed in atomically thin layers down to the monolayer, supporting the prediction of the Ising model that the ordering is possible in the 2D limit \cite{RN7,RN270}. This was the first demonstration that Raman spectroscopy can be reliably used to monitor AFM ordering in atomically thin vdW magnets.

FePSe$_3$ is similar to FePS$_3$, except for the chalcogen element. It also exhibits Ising-type zigzag AFM ordering below T$_\mathrm{N}$. In the Raman spectrum, two Raman modes at $\sim$75 and 115 cm$^{-1}$ are strongly enhanced below T$_\mathrm{N}$ down to the monolayer \cite{RN272}. Under a magnetic field, the peak at $\sim$75 cm$^{-1}$ does not change appreciably, but the one at $\sim$115 cm$^{-1}$ shows a prominent field dependence, indicating contributions of magnons. Luo $et$ $al.$ discovered that this peak has a fine structure of five peaks that are resolved as the field increases. From the field dependences of these peaks, they concluded that four of them originate from hybridization of magnons and phonons.

\subsubsection{\texorpdfstring{MnPS$_3$ and MnPSe$_3$}{MnPS3, MnPSe3}}

MnPS$_3$ is a Heisenberg-type AFM with spins pointing normal to the layers in the ordered state. In the Raman spectrum of MnPS$_3$, most of the blue modes shift monotonically as a result of volume contraction as the temperature is lowered. However, a peak at $\sim$155 cm$^{-1}$ exhibits a sudden redshift near T$_\mathrm{N}$, accompanied by a broadening of the line \cite{RN71,RN72}. This peculiar behavior of the mode was speculated to be the result of large fluctuations near the critical temperature and spin-phonon coupling \cite{RN71}. The correlation of this peak's frequency and linewidth with T$_\mathrm{N}$ measured by magnetic susceptibility measurements allows us to use them as indicators of magnetic ordering. Unfortunately, however, due to the weak intensity of this peak, the T$_\mathrm{N}$ of the few-layer MnPS$_3$ could only be estimated down to the bilayer \cite{RN70}. Because MnPS$_3$ is a Heisenberg-type AFM, no magnetic ordering is expected in the monolayer. 

In a related MnPSe$_3$ material, the Raman signatures of magnetic ordering are more subtle \cite{RN273,RN274}. Hybridization of a phonon at $\sim$130 cm$^{-1}$ and a two-magnon excitation in the AFM state \cite{RN274} and small redshifts of several Raman peaks below T$_\mathrm{N}$ \cite{RN273} have been reported.

\subsubsection{\texorpdfstring{NiPS$_3$}{NiPS3}}

\begin{figure}
    \includegraphics[width=\linewidth]{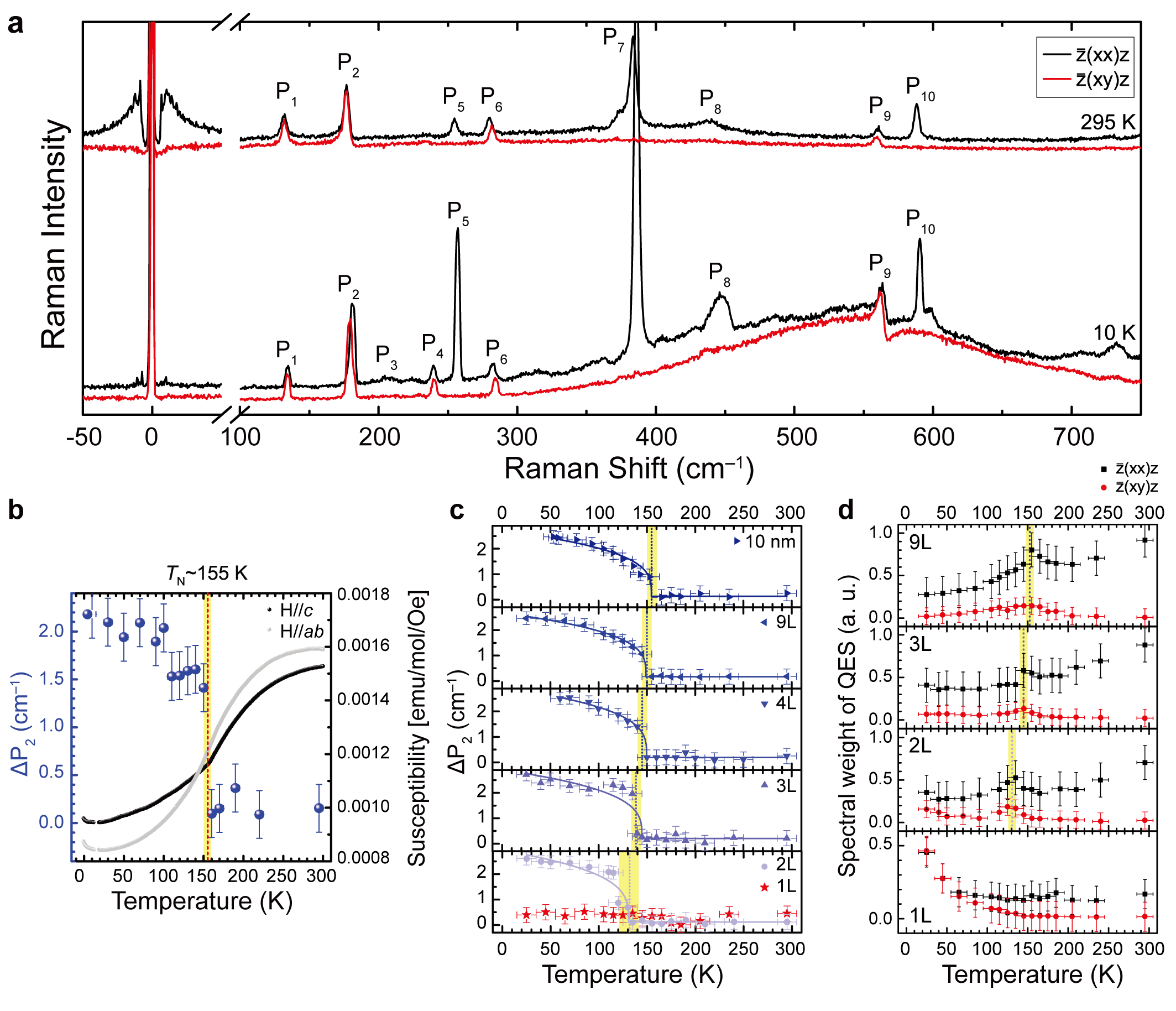}
    \caption{\label{fig:9}(a) Raman spectra measured at 10 and 295 K in parallel [$\bar{z}$(xx)z] (black) and cross [$\bar{z}$(xy)z] (red) polarization configurations. (b) Temperature dependences of $\Delta P_2$ (circles) and susceptibility of bulk NiPS$_3$. (c) Temperature dependence of magnetic-order-induced frequency difference $\Delta P_2$ for various thicknesses. (d) Spectral weight of QES between 11 and 40 cm$^{-1}$ as a function of temperature for various thicknesses for parallel (black squares) and cross (red circles) polarization scattering configurations. From \citet{RN76}.}
\end{figure}

NiPS$_3$ is an XXZ-type AFM with a T$_\mathrm{N}$ of $\sim$155 K in the bulk. Its Raman spectrum exhibits a diverse set of changes throughout the magnetic phase transition. Figure~\ref{fig:9}(a) compares the Raman spectra of NiPS$_3$ measured at 295 and 10 K \cite{RN76}. Several changes are observed in the low temperature AFM phase: (1) a broad peak centered at $\sim$550 cm$^{-1}$ due to two-magnon scattering appears; (2) the peak labelled P$_9$ exhibits a Breit-Wigner-Fano asymmetric line shape due to a Fano resonance; (3) the P$_2$ peak at $\sim$180 cm$^{-1}$ appears at slightly different Raman shifts for two different polarizations; and (4) the low-frequency signal below $\sim$50 cm$^{-1}$ is suppressed. The two-magnon signal is due to a double spin-flip process via the exchange mechanism and is an example of a genuinely magnetic phenomenon. Although it is a hallmark of magnetic ordering, some two-magnon signal persists beyond T$_\mathrm{N}$, so it is difficult to pinpoint the transition temperature from it. The Fano resonance is due to quantum interference between $E_g$-like in-plane phonon modes and the continuum of the two-magnon excitation. This line shape change is also gradual as the temperature is reduced through T$_\mathrm{N}$. Near the frequency of P$_2$, there are two phonon modes with almost the same frequencies but different symmetries in the paramagnetic phase. In the AFM phase, the two vibration modes are slightly modified because of the coupling with the aligned Ni spins, resulting in the splitting of the frequencies. As a result, the peak appears at different Raman shifts when measured in different polarization configurations. In particular, the lower-frequency mode of the two is strong when the incident and scattered photon polarizations are aligned along the $a$- or $b$-axis of the crystal, which can be used to determine the crystal axis of NiPS$_3$. This splitting ($\Delta$P$_2$) correlates well with the magnetic transition determined from magnetic susceptibility measurements [Fig.~\ref{fig:9}(b)]. Figure~\ref{fig:9}(c) shows the temperature dependence of $\Delta$P$_2$ for few-layer flakes down to the monolayer. The transition temperature is reduced slightly down to the bilayer, but is not observed in the monolayer. The low-frequency signal centered at 0 cm$^{-1}$ observed at room temperature is attributed to ‘quasi-elastic scattering (QES)’ due to spin fluctuations, which is reduced at low temperatures because magnetic ordering suppresses spin fluctuations. If the Bose-Einstein factor is considered \cite{RN275}, the QES signal is enhanced near T$_\mathrm{N}$ in both polarization configurations. The temperature dependence of the QES signal is summarized in Fig.~\ref{fig:9}(d). The maximum spectral weight of the QES signal is in agreement with the transition temperature determined in Fig.~4.3(c). For the monolayer, the QES signal increases as the temperature approaches 0 K, but never reaches a maximum. This, together with the temperature dependence of $\Delta$P$_2$, leads to the conclusion that the magnetic ordering is suppressed in the 2D limit of a monolayer, as the XY model predicts.

\begin{figure}
    \includegraphics[width=\linewidth]{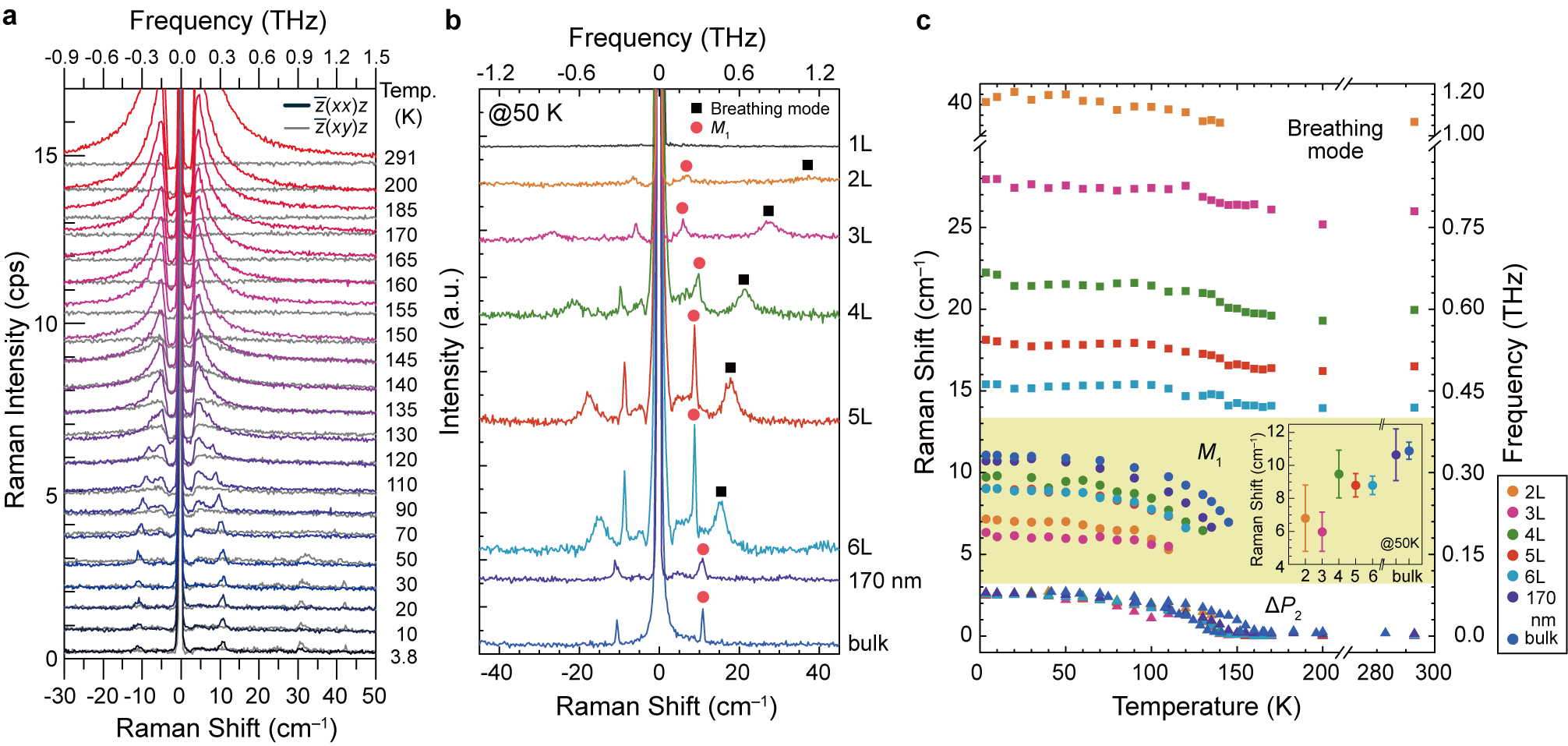}
    \caption{\label{fig:10}Temperature dependence of the polarized Raman spectra of bulk ($\sim$170 nm) NiPS$_3$ in parallel [z(xx)$\bar{z}$] and cross [z(xy)$\bar{z}$] polarization configurations. (b) Layer-number dependence of low-temperature Raman spectra. The M$_1$ and interlayer breathing modes are marked with red circles and black square symbols, respectively. (c) Temperature dependence of the Raman modes in the low-frequency range for different numbers of layers. From \citet{RN276}.}
\end{figure}

Furthermore, several ultra-low frequency features are identified in the AFM phase of NiPS$_3$. A sharp peak is observed at $\sim$11~cm$^{-1}$ (M$_1$) when the incident and scattered photon polarizations are aligned at 45$^\circ$ with respect to the $a$- or $b$-axis [$\bar{z}(xx)z$] as shown in Fig.~\ref{fig:10}(a). When the analyzer is rotated 90$^\circ$ [$\bar{z}(xy)z$], much weaker peaks are identified at 31 (M$_2$) and 42~cm$^{-1}$ (M$_3$) \cite{RN276}. Compared with the data from inelastic neutron scattering and magneto-Raman measurements \cite{RN276,RN277}, the peaks at 11 and 42~cm$^{-1}$ are identified as one-magnon excitations. The origin of the peak at 31~cm$^{-1}$ is still unclear: its temperature dependence is distinct from that of the other two. Because the polarization dependences of the magnon peaks do not follow the predictions of the traditional model for one-magnon Raman scattering \cite{RN278}, a more detailed theoretical model based on spin exchanges between neighboring Ni atoms was developed with consideration of the crystal structure to explain the observed polarization dependences \cite{RN276}. The M$_1$ mode is interpreted as due to the spin exchanges between Ni atoms along the ferromagnetically aligned zigzag direction. In contrast, the 42~cm$^{-1}$ mode is due to exchange between Ni atoms along the antiferromagnetically aligned armchair direction. For few-layer NiPS$_3$, only the M$_1$ mode can be resolved due to the signal [Fig.~\ref{fig:10}(b)]. This mode redshifts as the thickness is reduced but is not observed in the monolayer, which is consistent with the suppression of the magnetic ordering in the monolayer limit. As the temperature increases, this mode redshifts with the [1-T/T$_\mathrm{N}$]$^{0.23}$ dependence of the XY model [Fig.~\ref{fig:10}(c)]. Interestingly, the breathing mode, because of the rigid vibration of the individual layers in the out-of-plane direction, also blue shifts at low temperatures. This results from the stiffening of the out-of-plane interlayer coupling due to the interlayer magnetic ordering.

\subsubsection{\texorpdfstring{CrI$_3$}{CrI3}}

The monolayer CrI$_3$ has an optical A$_{1g}$ mode at $\sim127$~cm$^{-1}$ in the paramagnetic state. This corresponds to the vibrations of the top and bottom iodine layers in opposite directions, with the Cr atoms stationary. This mode is strong when the incident and scattered photons have the same polarizations, and weak if they are orthogonal. Upon magnetic ordering, the polarization of the scattered photons is rotated $\sim 40^\circ$ with respect to that of the incident photons, with the direction of rotation depending on the magnetization direction (up/down). This rotation of the polarization exhibits a good correlation with the magnetic transition \cite{RN124}. In the bilayer, the interlayer coupling splits the in- and out-of-phase vibrations of the two layers, which is called Davydov splitting \cite{RN279}. Davydov splitting separates the A$_{1g}$ mode into a Raman active A$_{1g}$ mode and an infrared active A$_{2u}$ mode. In the paramagnetic state, the A$_{2u}$ mode is Raman inactive due to the inversion symmetry of the bilayer CrI$_3$. In the antiferromagnetic state, however, the inversion symmetry is lifted, so this mode becomes Raman active. If a strong magnetic field is applied so that the spins in the two layers are aligned in the same direction, the inversion symmetry is restored, and this mode is suppressed in the Raman spectrum \cite{RN124}. Therefore, if the Raman spectrum is measured with appropriate polarizations as the temperature is lowered, a new Raman peak appears as the temperature drops below $T_\mathrm{N}$.

Magnon modes are also observed in the Raman spectra of CrI$_3$ \cite{RN121}. In the FM monolayer, an ultralow-frequency Raman mode is observed under a magnetic field. This mode shifts linearly with the magnetic field and extrapolates to $\sim2.4$~cm$^{-1}$ at zero field. This mode is interpreted as the in-phase acoustic magnon mode. In the AFM bilayer, a magnon mode is observed that extrapolates to $\sim3$~cm$^{-1}$ at zero field. As the magnetic field increases in the out-of-plane direction, it linearly shifts up to $\sim0.7$~T at which the bilayer undergoes a metamagnetic transition to an FM-like state, and an abrupt jump to a lower frequency is observed. In addition to these ultralow-frequency magnons, a weak shoulder-like peak at 148~cm$^{-1}$ splits under magnetic field. The authors interpreted this mode as a high-frequency, out-of-phase optical magnon mode, which is symmetry forbidden in the monolayer but allowed in the bilayer. Another vdW magnet that has been studied with Raman spectroscopy is RuCl$_3$, for example, in \cite{RN280,RN281,RN282}. There are also Raman studies on the vdW magnet MnBi$_2$Te$_4$ \cite{RN283,RN69}.

\subsection{Magneto-Optical Kerr Effect (MOKE) and Reflective Magnetic Circular Dichroism (RMCD)}

\begin{figure}
    \includegraphics[width=\linewidth]{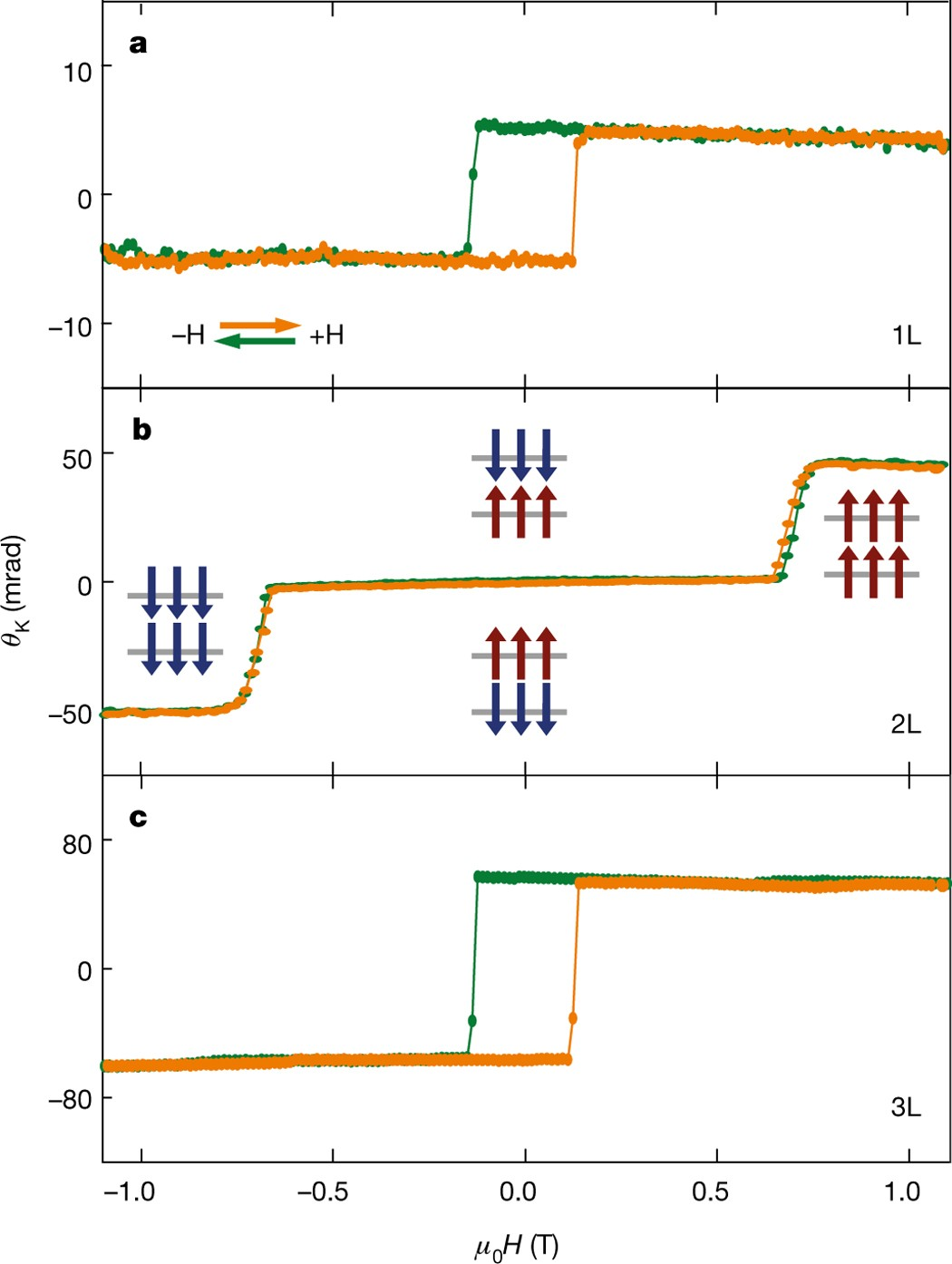}
    \caption{\label{fig:11}MOKE signal on a monolayer (a), bilayer (b), and trilayer (c) CrI$_3$ flakes. From \citet{RN39}.}
\end{figure}

When light travels in a medium under a magnetic field or net magnetization, the left- and right-circularly polarized components experience different dielectric constants \cite{RN284}, showing magnetic circular birefringence and magnetic circular dichroism \cite{RN143}. In the magneto-optical Kerr effect (MOKE) measurements, a linearly polarized incident light is reflected off the surface of the specimen, and the rotation of the polarization plane as a result of the magnetic circular birefringence is measured. By monitoring this polarization rotation, one can monitor the magnetization of the specimen. The MOKE signal is strongest when the magnetization (or the external magnetic field) is along the direction of the light beam. If the magnetization is in the plane of the specimen, the MOKE signal can still be measured if the light is incident at an oblique angle with the incident plane aligned along the magnetization direction \cite{RN285}. In reflective magnetic circular dichroism (RMCD) measurements, the reflectivity difference between the two circularly polarized light components is monitored, producing a result similar to MOKE. Both MOKE and RMCD are nondestructive and can be measured with a focused laser beam, making them appropriate for measuring small specimens of 2D vdW magnetic materials. Because they are sensitive to net magnetization, these techniques are primarily used to study FM materials.  MOKE measurements were used to detect FM ordering in thin atomic layers of Cr$_2$Ge$_2$Te$_6$ \cite{RN38} and CrI$_3$ \cite{RN39}, and RMCD for Fe$_3$GeTe$_2$ \cite{RN42}, and VI$_3$ \cite{RN534}. In vdW magnets that exhibit monolayer FM and interlayer AFM coupling, such as CrI$_3$ \cite{RN39} and CrPS$_4$ \cite{RN45}, nonzero MOKE signals are observed at zero field only for odd-numbered layers. The MOKE signal vs.\ applied magnetic field curves (Fig.~\ref{fig:11}) for the monolayer (1L) and the trilayer (3L) show ferromagnetic behavior with a finite remanent magnetization, whereas no MOKE signal is observed at zero field for the bilayer (2L). At high magnetic fields, the two layers of the 2L flake are ferromagnetically aligned, and a finite MOKE signal is observed \cite{RN39}. In a twisted bilayer of CrI$_3$, the magnetic field dependence of the RMCD was interpreted to be due to coexisting FM-AFM states \cite{RN55}. CrSBr has FM monolayers with in-plane spins and AFM coupling between the layers. With the application of an in-plane magnetic field, the spins can be made to align ferromagnetically, which can be monitored by MOKE. Yu \textit{et al.} used this technique for 2D imaging of the AFM-FM transition in bulk CrSBr \cite{RN286}.

\subsection{Terahertz dynamics of magnons}

\subsubsection{Magnon dynamics in the terahertz region}
All magnon modes in both ferromagnetic and antiferromagnetic materials exhibit coupling to the magnetic field component of the probing THz light. This interaction can be understood from two perspectives: semiclassically as the exertion of magnetic torque on individual spins, and quantum mechanically as the induction of magnetic dipole transitions between spin-wave eigenstates. In THz time-domain spectroscopy (THz-TDS), pulsed THz radiation serves as an ultrafast excitation source. The magnetic field component of the THz pulse acts as a half-cycle transient perturbation, displacing spin moments from their equilibrium orientations, which are initially aligned along their internal effective magnetic field, which governs the magnetic ordering of the material. Following this perturbation, the spins relax back toward equilibrium, precessing around the effective magnetic field axis. This precessional motion results in the emission of electromagnetic radiation, a process known as free induction decay. The emitted THz radiation is then recorded in a time-resolved fashion by a detector, encoding the dynamical response of the system. By applying a Fourier transform to the recorded time-domain signal, the corresponding frequency-domain absorption spectra of magnon excitations can be extracted. More precisely, this analysis provides access to the system’s dynamical susceptibility, which can be interpreted within the linear response theory framework to gain insight into the magnetic interactions and anisotropies present in the material. In the lower-frequency GHz regime, magnon modes are typically observed, reflecting low-energy collective spin excitations in ordered magnets. Complementary information on such spin dynamics can be obtained via ferromagnetic resonance (FMR) and antiferromagnetic resonance (AFMR) techniques. These methods are based on the observation of the resonant absorption of continuous microwave radiation at the GHz frequency while sweeping an external static magnetic field, with the objective of matching the driving frequency with the natural precession frequencies of the magnetization vector $\mathbf{M}$ or the precession / vibrational frequencies of the antiferromagnetic vector $\mathbf{L}$.

\subsubsection{\texorpdfstring{CrI$_3$}{CrI3}}

\begin{figure}
    \includegraphics[width=\linewidth]{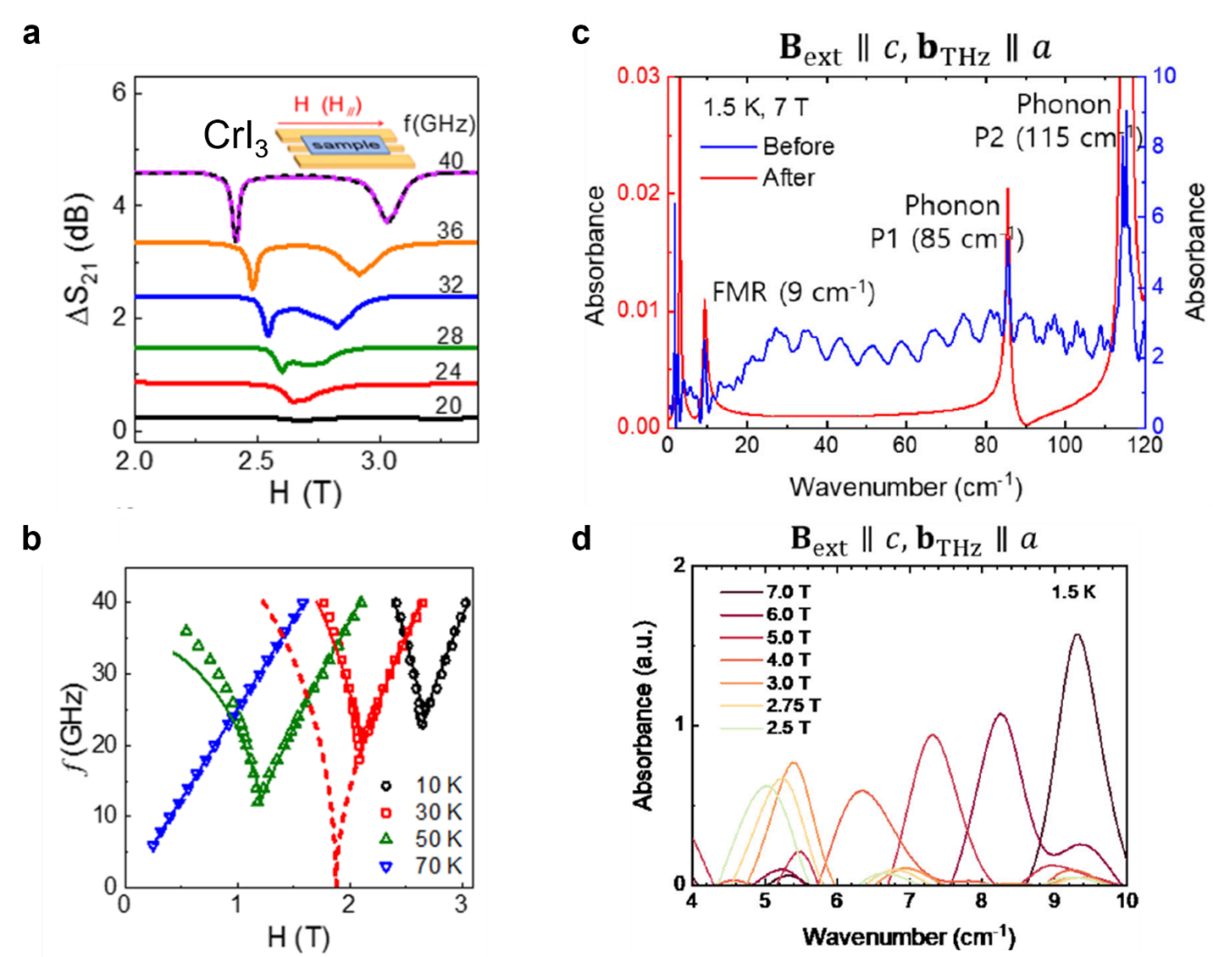}
    \caption{\label{fig:12}(a) Typical FMR spectra at 10~K for CrI$_3$. (b) FMR frequency vs. applied magnetic field ($f$--$H$) dispersions at various temperatures for CrI$_3$. From \citet{RN289}. (c) Absorbance spectra of a 217~$\mu$m-thick bulk CrI$_3$ crystal at 7~T and 1.5~K. Blue and red curves represent raw and post-processed data, respectively. (d) Absorbance spectra at 1.5~K near FMR with an external magnetic field aligned along the $c$-axis. From \citet{RN290}.}
\end{figure}

CrI$_3$ is a 2D vdW insulator in which genuine 2D ferromagnetism was demonstrated down to the monolayer limit \cite{RN39}. The bulk phase of this material was extensively studied in the 1960s, when it was recognized as a rare example of a ferromagnetic insulator. Given its robust ferromagnetism from bulk to monolayer, FMR studies have been actively pursued over the years. Interestingly, CrI$_3$ has two Cr$^{3+}$ ions per unit cell, and therefore two branches of spin waves are expected, corresponding to an acoustic mode with in-phase precession and an optical mode with out-of-phase precession \cite{RN121}. The first bulk FMR measurement, reported in 1965, revealed zero-field FMR at 2.2~cm$^{-1}$ (0.27~meV) for the lower-branch magnon \cite{RN287}. A recent bulk FMR study analyzed the magnetic field dependence and determined key magnetic parameters relevant to the magnetic Hamiltonian for bulk CrI$_3$ \cite{RN288}. Surprisingly, angle-dependent FMR measurement revealed a large Kitaev exchange interaction parameter $K = -5.2$~meV, which is approximately 25 times larger than the Heisenberg exchange interaction $J \approx -0.2$~meV. Furthermore, symmetric off-diagonal anisotropy $\Gamma \sim -67.5~\mu$eV was identified, which plays a crucial role in stabilizing the ferromagnetic order. Moreover, magnetic inhomogeneity was considered in the FMR linewidth analysis, which accounted for the multidomain magnetic structures of CrI$_3$ \cite{RN289} (Fig.~\ref{fig:12}(a) and (b)).

Although the above studies were conducted using traditional microwave resonance techniques, in which FMRs were detected by varying the external magnetic field at a discrete set of microwave frequencies, a fully spectroscopic FMR study was recently conducted using THz-TDS \cite{RN290} (Fig.~\ref{fig:12}(c) and (d)). By analyzing the magnetic field dependence of the FMR, various magnetic parameters, including the Kitaev interaction parameter $K = 0.2132$~meV and the off-diagonal interaction parameter $\Gamma = -0.0599$~meV, were extracted. The significant discrepancy in the $K$-values between THz-TDS and previous microwave-based studies highlights the experimental challenges to accurately determine the parameters of exchange interaction in CrI$_3$.

A single-crystal sample of CrI$_3$ in a few layers cannot be studied in the microwave or THz regions due to the long wavelength of probing light. Therefore, Raman spectroscopy has been used to track FMRs and surface magnon excitations in the 2D limit of CrI$_3$ \cite{RN121}. (See Raman section for further details.)

\subsubsection{\texorpdfstring{MnPS$_3$ and MnPSe$_3$}{MnPS3 and MnPSe3}}

Both MnPS$_3$ and MnPSe$_3$ possess two Mn$^{2+}$ ions per unit cell, leading to distinct magnon branches. These magnon modes are influenced by the materials' anisotropic properties, which affect their overall spin-wave dynamics. The difference in spin alignment between these two materials arises from their contrasting exchange interactions and anisotropic characteristics. In MnPS$_3$, the spins align perpendicular to the honeycomb plane, forming an easy-axis antiferromagnetic structure due to a combination of single-ion anisotropy and dipolar interactions. In contrast, MnPSe$_3$ exhibits easy-plane antiferromagnetic ordering, where weaker single-ion anisotropy and stronger dipolar interaction contribute to stabilizing the in-plane spin configuration \cite{RN210}.

GHz absorption experiments in MnPS$_3$ have identified magnon gaps at 3.8~cm$^{-1}$ (0.47~meV) and 3.4~cm$^{-1}$ (0.42~meV), also confirming its classification as a biaxial antiferromagnet \cite{RN291}. Experiments using Raman scattering, far-infrared transmission measurements, and GHz (microwave) absorption have identified two nonzero magnon gaps at 14~cm$^{-1}$ (1.7~meV) and 0.7~cm$^{-1}$ (0.09~meV) in MnPSe$_3$, confirming its classification as a biaxial antiferromagnet \cite{RN292}.

A key difference between MnPS$_3$ and MnPSe$_3$ lies in their exchange interactions. MnPS$_3$ is characterized by dominant intralayer exchange, reinforcing its quasi-2D nature, whereas MnPSe$_3$ exhibits stronger interlayer interactions, which are an order of magnitude larger than in MnPS$_3$, leading to a more 3D-like magnetic behavior \cite{RN293}. This distinction influences magnon dispersion and anisotropy effects in both materials. Magnetic parameters derived from experiments suggest competing exchange interactions, affecting their overall spin dynamics.

Although Raman and GHz absorption have successfully characterized magnon gaps, THz spectroscopy remains underexplored in these materials. Although direct THz spectroscopic studies on MnPS$_3$ and MnPSe$_3$ have yet to be extensively conducted, existing magneto-spectroscopy studies in the GHz and FIR ranges suggest that THz techniques could provide complementary insights into their magnon dynamics. In particular, the upper magnon gap at 14~cm$^{-1}$ (1.7~meV) in MnPSe$_3$ falls within the detectable range of THz spectroscopy, making it a strong candidate for future investigations \cite{RN291}.

\subsubsection{CrSBr}

This material exhibits a rare form of triaxial magnetocrystalline anisotropy \cite{RN294}. The hard-axis is the $c$-axis, the intermediate axis is the $a$-axis, and the easy axis is the $b$-axis. Moments lie along the $b$-axis within the vdW plane, and moments on neighboring vdW planes lie in the opposite direction, thereby forming intralayer FM and interlayer AFM. The interlayer exchange and the three anisotropy fields are weak, so the AFMR frequencies are low, ranging from 1--40 GHz under an external magnetic field up to 0.5 T \cite{RN295}. Such soft magnetic characteristics yield a rich variety of AFM magnon dynamics in this material. At 5 K under zero magnetic field, one observes two magnon branches, acoustic and optical. Only a weak magnetic dispersion appears under a magnetic field along the $c$-axis (the hard-axis), but under an in-plane magnetic field, one observes a hybridization between these two AFMR modes. At around 0.4 T in-plane magnetic field, there is an abrupt change in magnon spectra due to a spin-flop transition \cite{RN295}. All these interesting features of CrSBr imply that its magnetic characteristics are highly tunable with a relatively low magnetic field under 0.5 T. Combined with its exfoliatability and air stability down to a few layers \cite{RN56}, CrSBr is considered a promising candidate for practical applications in the field of 2D nanospintronics.

\subsubsection{\texorpdfstring{NiI$_2$}{NiI2}}

\begin{figure}
    \includegraphics[width=\linewidth]{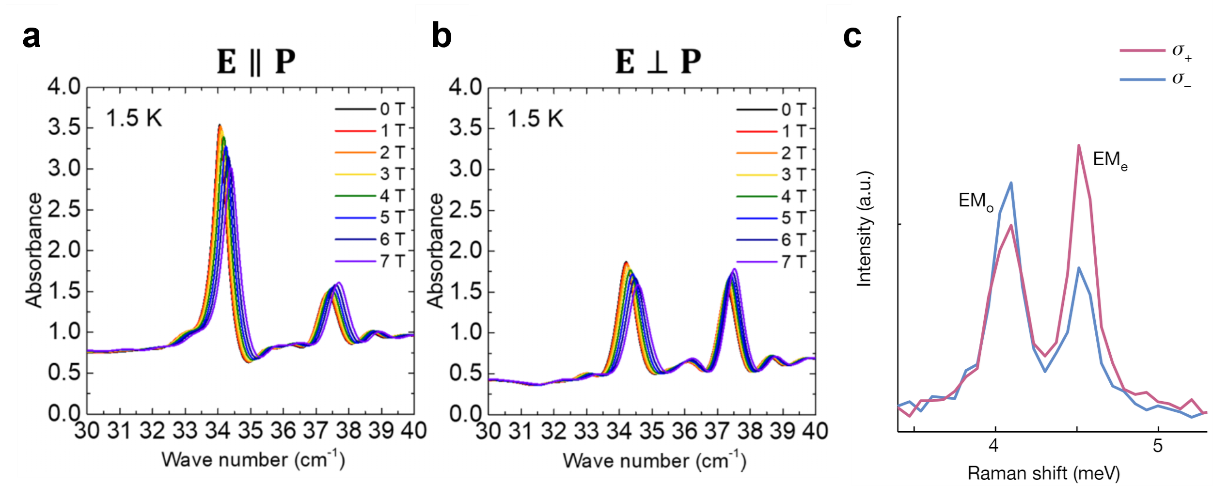}
    \caption{\label{fig:13}Out of plane magnetic field dependences of the absorption spectra of NiI$_2$ at 1.5~K with (a) \textbf{E} $\parallel$ \textbf{P} and (b) \textbf{E} $\perp$ \textbf{P} polarizations. From \citet{RN296}. (c) Low-frequency spontaneous Raman scattering measurement of exfoliated NiI$_2$ with $\sigma^+$ (red) and $\sigma^-$ (blue) circularly polarized light obtained at 2.4~K. The two peaks at 4.09~meV and 4.51~meV are the EM$_o$ and EM$_e$ electromagnons, respectively. From \citet{RN188}.}
\end{figure}

NiI$_2$ is a layered transition metal halide known for its helimagnetic ordering below its second antiferromagnetic transition at $T_\mathrm{N}$ = 59.5 K. However, the propagation vector for this screw rotation is not aligned to the crystallographic axes, and the planes of spin rotation are not perpendicular to the propagation vector. Interestingly, the material spontaneously develops ferroelectricity upon entering into this helimagnetic phase, with the induced electric polarization in the $ab$ plane. Several studies addressed the issue of magnon dynamics of this material in the THz range. A Raman study reported magnon modes at 31 and 37 cm$^{-1}$ \cite{RN43}. The authors claimed the presence of ferroelectric polarization even in the monolayer limit. Because of the simultaneous presence of noncollinear spin order and ferroelectric polarization, the nature of magnons was to be elucidated in the context of electromagnon excitations. Definitive evidence for the electromagnon nature of these two modes came from a THz study which showed, by combining transmission and reflection measurements, that the magnon response occurs in the electric dipolar (permittivity) rather than magnetic dipolar (permeability) channel \cite{RN296} (Fig.~\ref{fig:13}(a) and (b)). A recent Raman study further reported giant optical activity at THz frequencies, which appeared to arise primarily from relativistic spin-orbit coupling due to the noncollinear spin texture \cite{RN188} (Fig.~\ref{fig:13}(c)).  Another vdW magnet that has been studied with THz spectroscopy is RuCl$_3$, for example, in \cite{RN297,RN298,RN299,RN300}.

\subsubsection{\texorpdfstring{RuCl$_3$}{RuCl3}}

RuCl$_3$ is a layered transition-metal halide crystallizing in a honeycomb lattice structure. Below $T_N \sim 7$ K, it exhibits zigzag antiferromagnetic ordering. This material has been extensively studied using time-domain THz spectroscopy, which has revealed rich low-energy magnetodynamics sensitive to magnetic field, polarization, and symmetry-breaking interactions. At zero magnetic field, a prominent AFMR appears at approximately 2.5~meV ($\sim 20.2$~cm$^{-1}$), corresponding to a zone-center magnon mode \cite{RN297,RN298}. Upon entering the ordered phase, this mode emerges sharply and displays a classic order-parameter-like temperature dependence. In addition, further studies observed two magnon modes at $\sim$2.0 and 2.4~meV, which exhibit distinct field-dependent behavior depending on the direction of polarization of the THz probe \cite{RN299,RN300}. As an external magnetic field is applied within the honeycomb plane, the two magnon resonances soften and mix, and their spectral weights redistribute nonlinearly. Around the critical field $B_c \sim 7$ T, where the zigzag order collapses, these sharp magnon modes give way to a broad magnetic continuum, which has been interpreted as a signature of a field-induced quantum disordered phase. This magnon-to-continuum crossover illustrates the breakdown of conventional spin-wave excitations near quantum criticality. Theoretical modeling using linear spin-wave theory (LSWT) and extensions of the Heisenberg-Kitaev-$\Gamma$ model have shown that an off-diagonal symmetric exchange ($\Gamma$ term) is essential to reproduce the observed mode splitting, selection rules, and field evolution. These features highlight the strong spin-orbit coupling and magnetic anisotropy in RuCl$_3$, which shape its magnon spectra \cite{RN299,RN300}. Overall, the terahertz magnon response of RuCl$_3$ reflects not only the underlying zigzag order but also the proximity to a quantum spin liquid regime, making it an exemplary system for studying the collapse and evolution of coherent spin-wave excitations.

\subsection{Optical Absorption}

\subsubsection{Exciton Dynamics in the Near Infrared-Visible-Ultraviolet Regions}

Exciton phenomena in magnetic insulators are in great agreement with the original description of excitons in crystals as ``excitation waves ”as conceived by Frenkel \cite{RN301}. In contrast to Wannier-Mott excitons commonly encountered in semiconductors, delocalized over length scales greater than unit cell sizes, excitons in magnetic insulators are usually Frenkel excitons that are understood to be highly localized to within interatomic distances. To visualize the characteristics of such excitons, we consider a single transition-metal ion embedded in a crystal. Its energy levels, highly degenerate in their free-ion state, are partially split inside a crystal because of internal magnetic fields associated with magnetic ordering and crystal electric fields generated by ligand ions. A $d$-$d$ transition in this single ion is bound to disturb, albeit very weakly, the neighboring ions because of exchange interactions present in the system. If the $d$-$d$ transition is spin-conserving, a dipole-dipole interaction can transfer the change in the original magnetic ion to its nearest-neighbor magnetic ions. A spin-violating transition can also occur by being assisted by another spin-violating transition nearby, in this case via spin-dependent (possibly off-diagonal) exchange. Therefore, internal transitions in each magnetic ion are quantum-mechanically coupled with those in other magnetic ions throughout the crystal. In this way, we conceive a magnetic exciton with its own dispersion (energy-momentum relation) and dynamics (coherent motion through a crystal).

The optical transitions associated with such magnetic excitons are typically found in the near-infrared-visible-ultraviolet regions of the electromagnetic spectrum. Close to such $d$-$d$ excitons, optical transitions associated with charge transfer also occur. An electron is transferred from ligand atoms to transition-metal ions or vice versa. However, in many cases, such transfers are partial in the sense that the transition metal d states are hybridized with ligand atom p states, expanding the Hilbert space of the relevant orbitals. Historically, in the field of high-temperature superconductivity, the so-called Zhang-Rice singlet state is known to occur when a hole in an O 2p orbital hybridizes with a hole in a Cu 3d orbital, forming a singlet that mimics a Cooper pair \cite{RN193}. A similar orbital-spin-entangled state has been reported in the 2D vdW antiferromagnet NiPS$_3$ \cite{RN80} as described in the following.

We also note that double transfers or exciton-magnon interactions have been studied extensively. Such an interaction mechanism involves the emergence of magnon sidebands on top of exciton absorption peaks. Both magnon-emitting and magnon-absorbing transitions were reported. In the former case, a photon is absorbed and produces an exciton plus a magnon, and the magnon sideband occurs at an energy higher than that of the exciton. In the latter case, a photon is assisted by a magnon to produce an exciton, so the magnon side band is placed at an energy lower than that of the exciton. Obviously, the latter mechanism cannot occur at very low temperatures because of the absence of magnons, whereas the former case can occur at low temperatures. The magnon sideband actually reflects the magnon density of states rather than magnon excitation peaks. Due to spin conservation, two-magnon sidebands are more commonly observed, as is true for 2D vdW magnets.

Exciton dynamics can be captured by optical absorption and Raman spectroscopy and resonant inelastic X-ray spectroscopy (RIXS). RIXS can even reveal the energy-momentum dispersion of the exciton in reciprocal space where localized Frenkel excitons typically exhibit rather flat dispersions. For 2D vdW magnets, much effort has been put into the realization of cavity-based modulation of excitons. Inside a carefully designed photonic-crystal-based cavity, the cavity modes hybridise with otherwise nearly flat excitonic modes of the crystal. A strong-coupling regime can be easily reached by this method, and many effects unique to the domain of atomic physics can be demonstrated as well. All of these new topics should stimulate further research in the near future.

\subsubsection{\texorpdfstring{NiPS$_3$}{NiPS3}}

\begin{figure}
    \includegraphics[width=\linewidth]{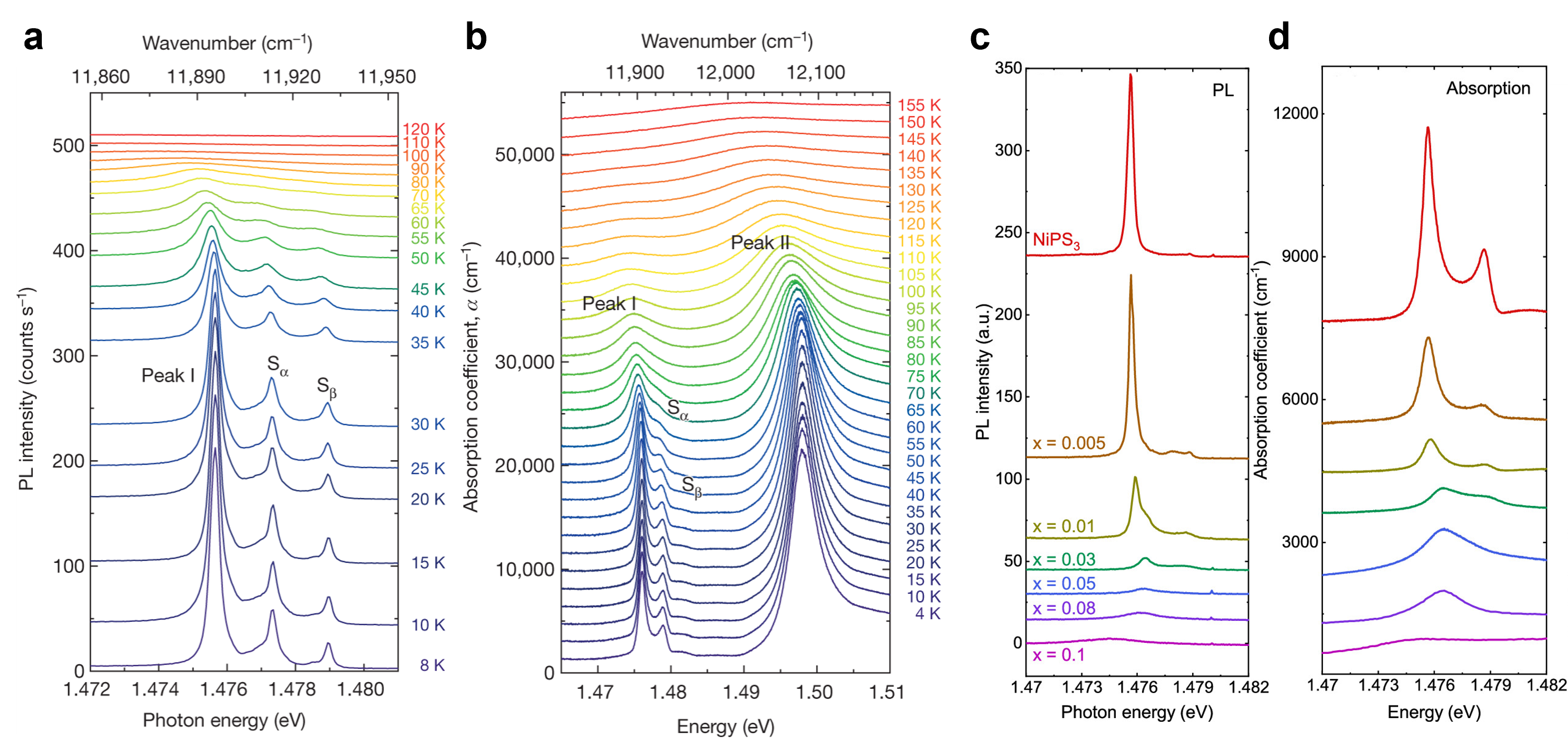}
    \caption{\label{fig:14}(a) PL spectra of NiPS$_3$ (excited by a 1.96 eV (632.8 nm) laser) as a function of temperature from 8 to 120 K (curves are vertically shifted for clarity). (b) Optical absorption spectra corresponding to the PL spectra in (a). From \citet{RN80}. (c) PL spectra as a function of Cd concentration $x$ at 8 K. (d) Optical absorption spectra as a function of the same set of Cd concentration $x$ at 4 K. From \citet{RN304}.}
\end{figure}

The most dramatic case of exciton dynamics is probably found in NiPS$_3$. In marked contrast to the Wannier-Mott excitons formed from extended Bloch states, the excitons in NiPS$_3$ are Frenkel excitons of highly localized character. Interestingly, the exciton located at 1.476 eV in NiPS$_3$ is closely related to the background AFM order and emerges together with the onset of AFM below 155 K. The many-body nature of this ultra-sharp exciton at 1.476 eV found at low temperatures deep in the AFM state was studied by PL, optical absorption, and RIXS measurements in combination with configuration interaction calculations \cite{RN299} (Fig.~\ref{fig:14}~(a) and (b)). Ultimately, the exciton was assigned to a Zhang-Rice triplet-to-Zhang Rice singlet transition. Moreover, the PL linewidth of this exciton mode reached almost 0.4 meV below 50 K, as if the system entered some condensation or super-radiance state. Later investigations confirmed the presence of ultra-narrow excitonic excitations and explored the magnetic field dependence \cite{RN302,RN303} and the stability under non-magnetic doping and the GPa-level pressure. It turned out that the exciton mode is quite sensitive to a small amount of Cd doping and also to pressure, suggesting that this Zhang-Rice mode is the result of fine-tuned balance between many competing processes \cite{RN304} (Fig.~\ref{fig:14}~(c) and (d)). More recently, however, W.~He \textit{et al.} \cite{RN190} proposed an alternative interpretation based on ultrahigh resolution RIXS measurements: the exciton was reclassified as a Hund's exciton, driven primarily by intraatomic Hund's exchange rather than in Zhang-Rice mode. This reinterpretation highlights a different underlying mechanism and suggests that the observed exciton corresponds to a singlet excitation from a subtle, yet high-spin d$^8$ ground state, primarily stabilized by intra-atomic Hund's coupling.

\subsubsection{\texorpdfstring{NiI$_2$}{NiI2}}

Another case of a Zhang–Rice exciton mode was discovered in NiI$_2$, an antiferromagnetic vdW insulator with improper helimagnetic order in its ground state, in which the material also exhibits spontaneous ferroelectric order, thereby establishing itself as a multiferroic vdW antiferromagnet \cite{RN187}. The broken inversion symmetry here induces the formation of magnetic excitons. Combining optical absorption, RIXS, and configuration interaction calculations, it was confirmed that the exciton type was again a transition between a Zhang–Rice-triplet state and a Zhang–Rice-singlet state, a manifestation of fundamentally quantum-mechanical entanglement of spin and orbital degrees of freedom. An ultrasharp exciton peak at 1.384 eV with a linewidth of 5 meV was reported.

\subsubsection{CrSBr}

\begin{figure}
    \includegraphics[width=\linewidth]{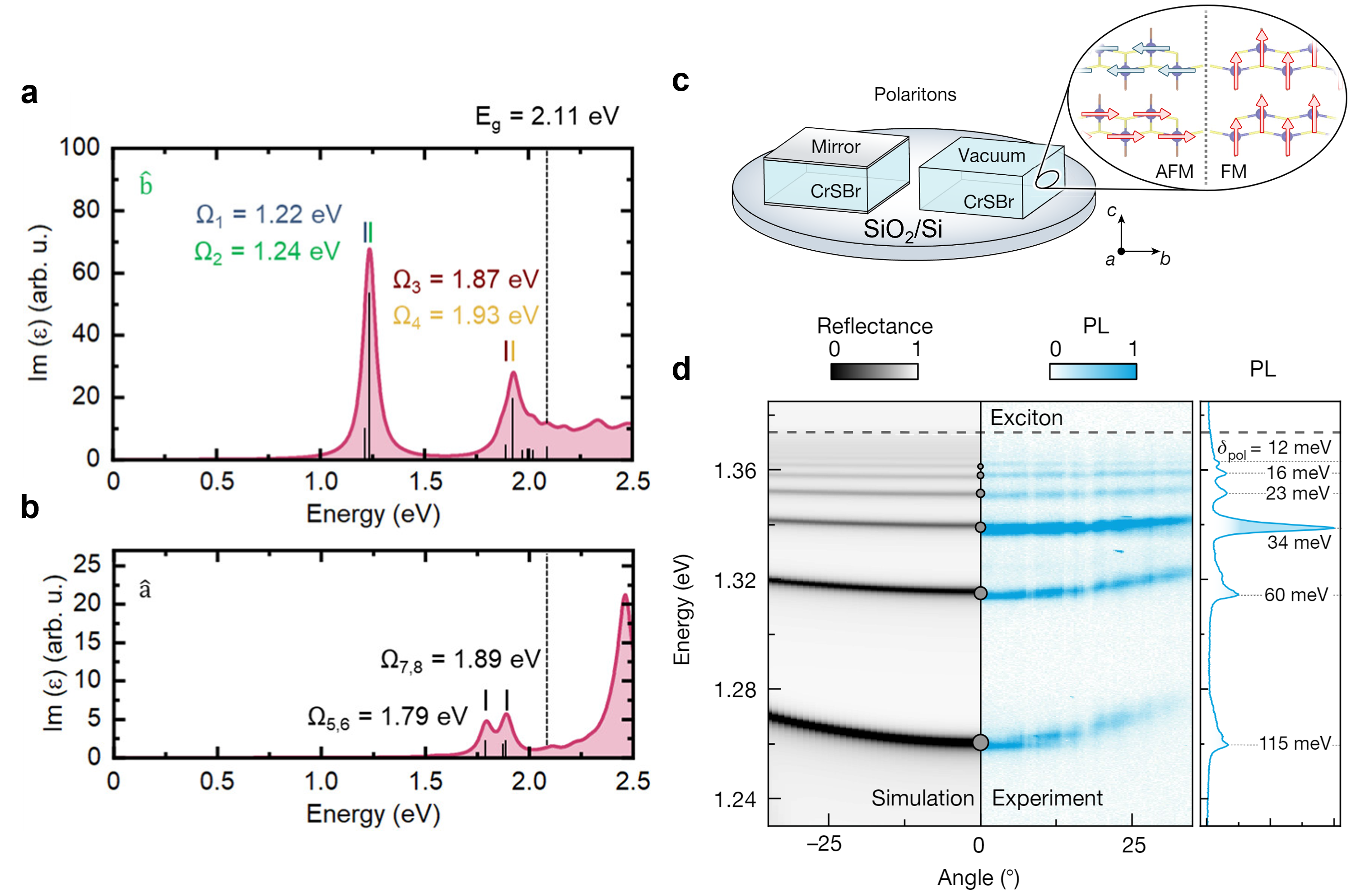}
    \caption{\label{fig:15}(a) Calculated optical absorption spectrum of monolayer (1L) CrSBr with an electric field polarized along the $b$ direction. (b) Calculated optical absorption spectrum obtained from \textit{ab initio} Bethe–Salpeter Equation (BSE) calculations of 1L CrSBr for an electric field polarized along the $a$ direction. From \citet{RN305}. (c) Schematic illustration of two experimental approaches to support exciton-polaritons in mesoscopic CrSBr. The magnified view shows antiferromagnetic (AFM) and ferromagnetic (FM) order with external magnetic field $B_{\text{ext}} \parallel c$. Magnetic axes correspond to the crystal $b$ (easy), $a$ (intermediate), and $c$ (hard) directions. (d) The right panel shows angle-resolved and angle-integrated photoluminescence (PL) emission of a nominally 580-nm-thick CrSBr sample recorded at $T = 1.6$~K. The crystal was rotated by $45^\circ$ with respect to the entrance slit of the spectrometer, and the PL was analyzed with a polarizer aligned along the $b$-axis. The left panel shows the simulated reflectance map for conditions matching the PL measurement. From \citet{RN306}.}
\end{figure}

The special feature of excitons in CrSBr is the anisotropy. Band calculations suggest that the material is quasi-one-dimensional (Q1D) with the $b$-axis as the high-mobility direction \cite{RN305} (Fig.~\ref{fig:15}~(a) and (b)). As such, excitons exhibit an electron density that spans nearly 20 unit cells along the $b$-axis, while being less delocalized along the other axis. As mentioned previously, CrSBr develops in-plane ferromagnetism and interlayer antiferromagnetism, that is, A-type AFM overall. This particular magnetic structure gives rise to a spin-polarized conduction band minimum and a valence band maximum. The excitons then inherit this spin polarization. For example, the bilayer case will have an antiparallel spin configuration in the two layers, and the excitons will be confined to individual layers because the interlayer coupling of excitons is spin-forbidden. With a strong magnetic field, spins in both layers are polarized along the same direction, and this now allows hybridization between excitons in the two layers, promoting interlayer hopping of electrons and holes. Consequently, the optical properties strongly depend on external magnetic fields, prompting promising optomagnetic applications. Another notable aspect of this spin-exciton coupling effect is the strong interaction between magnons and excitons \cite{RN306} (Fig.~\ref{fig:15}~(c) and (d)). The exciton energy can be modulated by up to 4 meV by driving magnons at a 0.1 meV energy scale.

\subsubsection{\texorpdfstring{CrI$_3$}{CrI3}}

Despite intensive searches and high-precision band theory calculations, no report was made of clear evidence of Wannier-Mott type excitons in CrI$_3$. Although the complete relativistic calculations of one group \cite{RN307} stand alone as the definitive prediction of these excitons in CrI$_3$, no experiment has reported the presence of these modes to the best knowledge of the authors. However, two dark excitons at 1.75 and 1.85 eV were identified using RIXS \cite{RN308} (Fig. 4.10). These transitions are optically forbidden, so that standard optical absorption measurements would not capture them. These excitons are quite sharp in linewidth and weakly dispersive. These excitons are essentially spin flips, strongly tied to the magnetic structure of the material through the Hund's coupling mechanism \cite{RN308}. What can be potentially associated with charge transfer excitons was also reported \cite{RN309,RN123}. However, the nature of these exciton modes is not clearly understood at this point.

\begin{figure}
    \includegraphics[width=\linewidth]{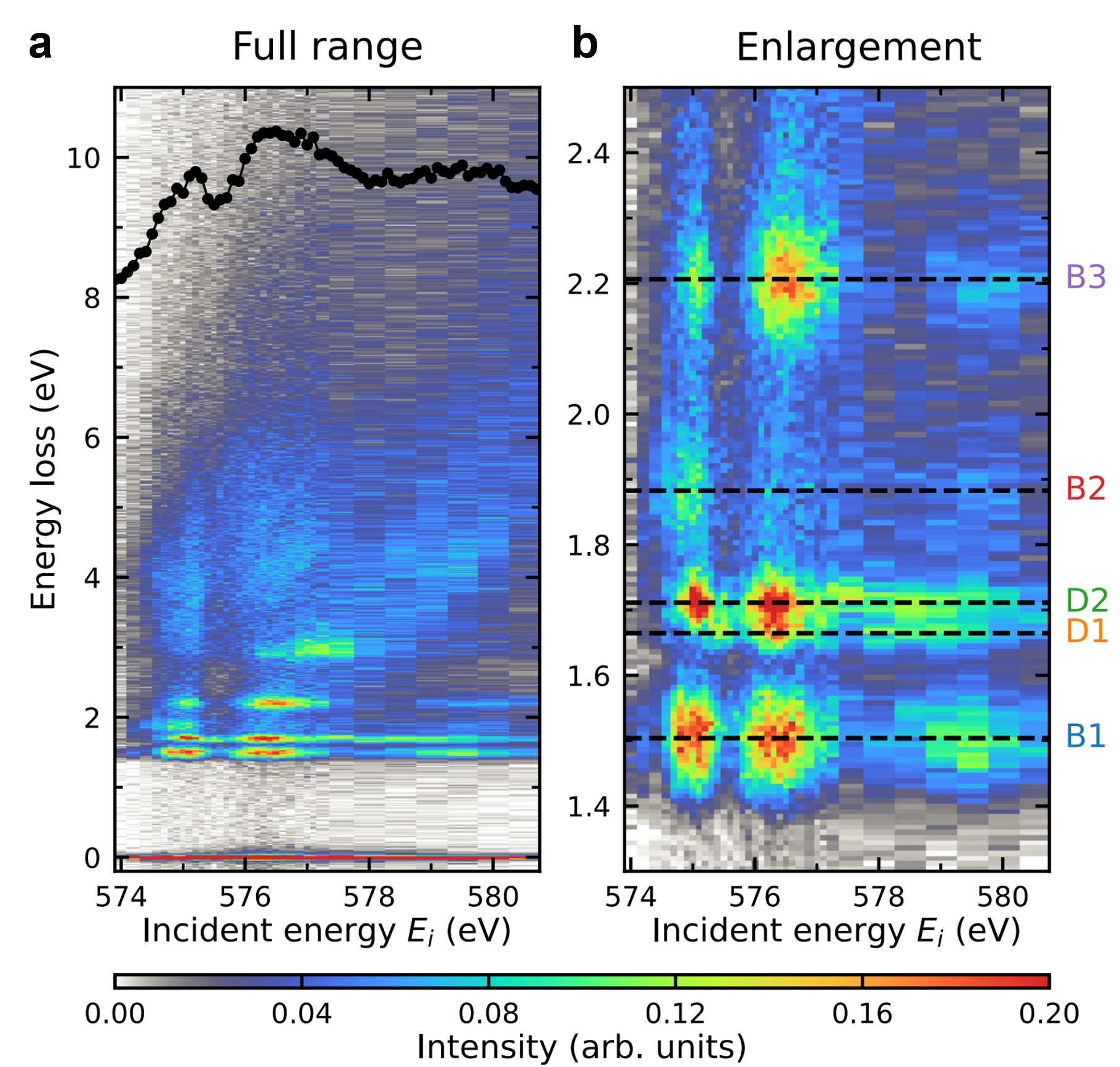}
    \caption{\label{fig:16}Resonance behavior of dark excitons in CrI$_3$. From \citet{RN190}. (a) Cr $L_3$-edge RIXS incident energy map recorded at $T = 30$~K. (b) Zoomed-in view of the excitonic resonances. Two peaks near 1.7~eV are identified as dark excitons and labeled as D1 and D2. The other three peaks are bright excitons previously observed in optical measurements and therefore denoted as B1-B3. From \citet{RN309,RN123}.}
\end{figure}

\subsection{Magnetooptical Studies on vdW Magnetic Materials}

Strong modulation of the optical properties of solids by external magnetic fields has been an active area of research for a long time. The field of magnetooptics has been reinvigorated by recent advances in vdW magnetic materials. Remarkably, the vdW antiferromagnet CrSBr exhibits spin‐polarized excitonic states \cite{RN306}. Consequently, external magnetic fields can effectively modify the exciton behavior, thereby strongly modulating the optical properties in the vicinity of exciton formation. This coupling mechanism is greatly enhanced by the formation of exciton–photon hybridization, i.e., exciton polaritons in a few nanometer-thick mesoscopic crystals. When an external out‐of‐plane magnetic field switches the spin polarizations from in‐plane AFM to interplane FM, a large modulation of optical absorption due to spin‐polarized excitons leads to strong magnetooptic effects. Several polariton branches form and appear even more than 0.1 eV below the original exciton level. A sensitive angular modulation via rotation of the external magnetic field is also possible. 

Similar exciton polariton studies were performed for NiPS$_3$ using a photonic cavity \cite{RN81}. Here, strong coupling between spin‐correlated excitons and photons within a microcavity was observed in the reflectance spectra in the vicinity of the ultranarrow Zhang-Rice exciton mode located at 1.476 eV \cite{RN80}. The resulting hybrid modes combine aspects of excitons, photons, and spins all together, promising novel applications in the field of optospintronics and polariton Bose–Einstein condensation. Incidentally, in NiPS$_3$, the aforementioned exciton displays strong optical anisotropy with the anisotropy axes themselves locked into the background magnetic zigzag ordering \cite{RN81}. This immediately led to magnetic linear dichroism (MLD) signals detected below the antiferromagnetic transition temperature \cite{RN310}. Therefore, for the case of NiPS$_3$ and some related materials in the family of MPX$_3$ (M = Fe, Mn, Ni and X = S, Se), the MLD measurements can capture the formation of magnetic order down to the atomic limit (up to bilayers for NiPS3, as the monolayer loses AFM order in this material).

This MLD technique is useful when the material under investigation is AFM because the magneto‐optical Kerr effect (MOKE), which is commonly adopted as a sensitive tool for detecting FM, vanishes in the first order at the atomic limit \cite{RN310}. In addition to MLD, there are other second‐order magnetooptic effects such as the magnetooptic Voigt effect (MOVE) and the magnetooptic Schäfer–Huber effect (MOSHE) \cite{RN122}. Being second order, these three techniques typically produce weak signals and were rarely used to study the AFM ordering of 3D solids. However, in the MPS$_3$ family of 2D vdW materials, giant MLD signals were detected in FePS$_3$ in zigzag order \cite{RN310}. Naturally, the effect comes from the electronic anisotropy associated with the background zigzag spin direction in the AFM ground state. Even the monolayer case yielded a detectable signal, and this offers ample opportunities in the field of vdW antiferromagnetic spintronics in detecting, characterizing, and manipulating AFM order in ultrathin antiferromagnets.

The MOSHE technique was also recently demonstrated for the case of MPS$_3$ family vdW antiferromagnets \cite{RN311}. While the MOVE technique is implemented in the transmission mode, the MOSHE technique is implemented in the reflection mode. The rotation of polarization upon magnetization is similar to that of MOKE, but the signal is proportional to the square of magnetization. The theoretical calculation performed by Yang \textit{et al.} \cite{RN311} shows that the MOSHE signal, again, originates from the in‐plane optical anisotropy associated with the magnetic ordering and depends sensitively on the relative direction between the Néel vector and the input polarization of light. The combination of rotation angle and ellipticity (i.e., complex rotation angle) is directly proportional to the anisotropy in in-plane optical conductivity. It turns out that the rotation of the polarization is insensitive to the orientation of the Néel vector. Still, the ellipticity, the accompanying measure of the optical activity, is sensitive to the interlayer magnetic ordering. A distinct advantage of this method is that full DFT calculations can be easily performed to yield the theoretical values of the optical activity parameters, i.e., optical rotation and ellipticity, and can be compared with experiment.

\section{\label{sec:IV}Light-induced out-of-equilibrium dynamics}
\subsection{Coherent phonon excitation}

\subsubsection{Coherent phonon generation mechanisms}
Upon impulsive optical excitation, collective lattice vibrations (phonons) can be coherently driven, resulting in phase-locked oscillations. The generation mechanisms for coherent phonons depend on their type (acoustic or optical) and symmetry (infrared-active or Raman-active) and can be broadly categorized based on how the excitation source interacts with the material. 

Coherent acoustic phonons can be excited through four major mechanisms, depending on the properties of the material \cite{RN312}. In metals, where carrier recombination is typically fast, optical excitation rapidly transfers energy from the electronic to the lattice degrees of freedom. This increases the lattice temperature of the material's properties, generating propagating strain waves through the thermoelastic effect. In contrast, carrier excitation alters the electronic landscape in gapped systems, which persists for some time. The deformation potential subsequently modifies the interatomic forces, thus launching strain waves. An applied electric field pulse can induce internal strain through the inverse piezoelectric effect in systems lacking inversion symmetry. Finally, coherent acoustic phonons can be excited in centrosymmetric materials through electrostriction, a second-order piezoelectric effect, provided the medium is transparent to pump light.

For optical phonons, without loss of generality, we consider systems with inversion symmetry, a common feature in vdW magnets. They can be classified as infrared (IR)-active or Raman-active, which are odd and even under spatial inversion. An electromagnetic field can resonantly drive coherent IR-active phonons, typically in the terahertz (THz) or mid-infrared (MIR) spectral range. However, coherent Raman-active phonons, which are frequently probed in ultrafast spectroscopy, can be excited through multiple pathways depending on the drive frequency. In the optical range, two primary pathways dominate. The first mechanism is displacive excitation of coherent phonons (DECP), where the creation of photo-carriers alters the potential energy surface of the lattice, shifting the quasi-equilibrium free energy minimum \cite{RN313,RN314}. Thus, an effective displacive force is exerted on the lattice, initiating coherent oscillations of fully symmetric $A_{1g}$ phonons. DECP is prominent in vdW magnets when the pump photon energy is resonant with ligand-to-metal charge-transfer transitions. The other well-known mechanism is impulsive stimulated Raman scattering (ISRS), where a pulse with a duration of time shorter than the phonon period impulsively initiates a stimulated Raman scattering process that excites all possible Raman-active phonons \cite{RN315}. In vdW magnets, ISRS can be triggered by a pump with frequencies resonant with $d$-$d$ transitions of the transition metals or completely off-resonant with any transitions. Alternatively, if the pump energy lies in the THz range, close to the frequencies of Raman-active phonons, various nonlinear pathways become accessible, including two-photon sum or differential frequency generation \cite{RN316}, one-photon-one-IR-phonon infrared resonant Raman scattering (IRRS) \cite{RN317}, two-IR-phonon ionic Raman scattering (IRS) \cite{RN318}, and other pathways involving additional collective modes such as magnons \cite{RN319}. These nonlinear channels can be effectively activated in vdW magnets upon intense THz electromagnetic stimuli \cite{RN320,RN321,RN322}.

\begin{figure*}
    \includegraphics[width=\linewidth]{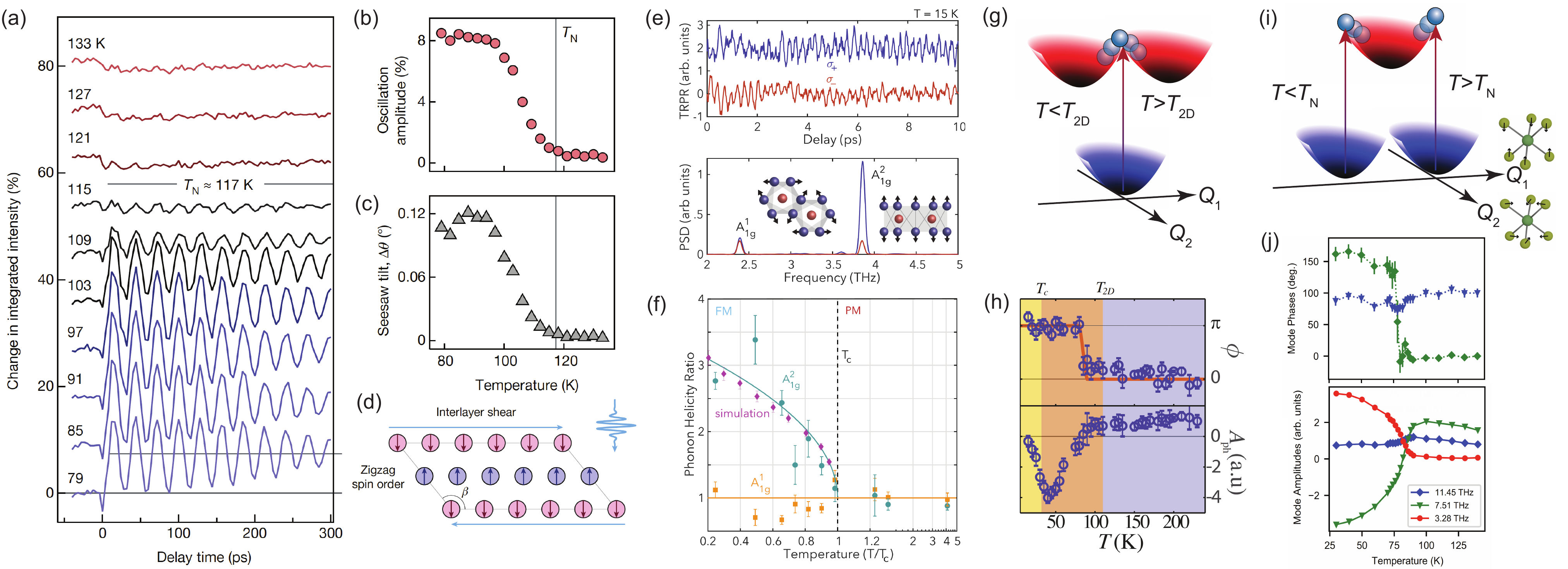}
    \caption{\label{fig:17}Coherent phonon excitation across various magnetic transitions. (a) Normalized time traces for differential Bragg peak (-3 3 1) intensity at various temperatures across $T_{\rm{N}}$ in FePS$_3$ following 4.76 eV photoexcitation. (b) Temperature dependence of the oscillation amplitude extracted from panel (a). (c) Temperature dependence of the seesaw tilt angle resulting from the acoustic phonon. (d) Schematic illustration of the interlayer shear motion of the antiferromagnetic state with a variation of the monoclinic angle $\beta$. (e) Time-resolved polarization rotation change with demagnetization background subtracted upon a 1.55 eV pump under left-handed $\sigma^+$ (blue) and right-handed $\sigma^-$ (red) polarizations in CrI$_3$, and their corresponding Fourier transform power spectral densities. The inset shows the schematic of the eigenvectors of two observed phonon modes. (f) $\sigma^+/\sigma^-$ ratio of the integrated Fourier transform peak intensities of the two modes shown in (e), overlaped with the simulated result for $A_{1g}^2$ and a fit using an order-parameter-like function $\sqrt{T_c-T}$ (green line). (g) Schematic of the potential energy surface change as temperature decreases across $T_{2D}$ as a function of two eigenmodes in CrSiTe$_3$. (h) Temperature dependence of the phase and amplitude of the $A_g$ mode extracted from transient reflectivity data upon a 1 eV pump in CrSiTe$_3$. (i) Schematic of the potential energy surface change as temperature decreases across $T_{\rm{N}}$ in FePS$_3$ as a function of two eigenmodes with eigenvectors displayed. (j) Temperature dependence of the phase and amplitude of several modes extracted from transient reflectivity data upon a 1.63 eV pump in FePS$_3$. Panels (a)-(d) are adapted from \citet{RN67}. Panels (e) and (f) are adapted from \citet{RN323}. Panel (h) is adapted from \citet{RN324}. Panel (j) and the inset of panel (i) are adapted from \citet{RN325}.}
\end{figure*}

\subsubsection{Magnetism-mediated phonon dynamics}

A hallmark of vdW magnets is their intricate spin-lattice coupling, which leads to unconventional changes in the amplitude, frequency, and phase of certain phonons across magnetic transition temperatures, which are often undetectable in materials with weakly coupled spin and lattice degrees of freedom. The most pronounced manifestation of this coupling is the appearance of new phonon modes, or significant enhancements in the phonon amplitudes, below the magnetic transition temperatures. This phenomenon is ubiquitous in vdW antiferromagnets, where spin ordering doubles the unit cell and folds the Brillouin zone. Consequently, phonons at finite momenta are folded to the center of the Brillouin zone, making them optically active. Such zone-folded optical phonons are not only observed with static probes, such as Raman spectroscopy \cite{RN7,RN326,RN327,RN76,RN72,RN274}, but can also be coherently populated by an external drive \cite{RN320,RN321,RN322,RN328}. For example, coherent breathing modes that modulate interplane exchange interactions emerge across the magnetic transition in the topological antiferromagnet series MnBi$_{2n}$Te$_{3n+1}$ \cite{RN61,RN329}. These $A_{1g}$ modes, argued to be zone-boundary phonons, become optically bright due to scattering with the magnetic wavevector. Their amplitudes change as antiferromagnetism melts or switches to ferromagnetism under a high external magnetic field. 

Optical phonons that modulate spin-exchange interactions not only exhibit enhancements across the magnetic ordering temperature but also show intimate correlations with specific electronic transitions. In the zigzag vdW antiferromagnet CoPS$_3$, polarization rotation measurements show that a coherent $B_g$ phonon that involves the in-plane motion of Co$^{2+}$ is significantly enhanced below the N\'{e}el temperature $T_{\rm{N}}$. This amplification is most efficiently driven through ISRS when the pump photon energy is resonant with the $d$-$d$ transition of Co \cite{RN330}. Similarly, in the XY-type zigzag antiferromagnet NiPS$_3$, an $A_g$ phonon that generates equally spaced sidebands in the electronic structure becomes visible below $T_{\rm{N}}$ \cite{RN87}. This mode, which modulates the interatomic Ni-S bond length, is strongly correlated with localized $d$-$d$ transitions due to the local inversion symmetry breaking at Ni sites below $T_{\rm{N}}$. However, in the Ising-type zigzag antiferromagnet FePS$_3$, polarization rotation measurements reveal that the zone-folded $A_g$ mode, though also detectable only in the presence of magnetic order, is effectively excited when the pump photon energy is resonant with $d$-$d$ transitions or above the charge-transfer bandgap \cite{RN328}. Because the phonon amplitude follows the trend of the absorption spectrum, DECP likely plays the dominant role in the initiation of this fully symmetric mode. 

Spin-lattice coupling also dramatically affects the excitation of acoustic modes. A striking example can be found in antiferromagnet FePS$_3$, where a gigahertz acoustic phonon undergoes an order of magnitude amplification as the temperature drops below $T_{\rm{N}}$ [Figs. \ref{fig:17}(a),(b)] \cite{RN67}. Using a suite of ultrafast diffraction techniques, Zong $et$ $al$. identify this coherent acoustic phonon as a seesaw-like interlayer shear mode that modulates the angle of the monoclinic structure $\beta$, with frequencies inversely proportional to the sample thickness [Figs. \ref{fig:17}(c),(d)] \cite{RN67,RN331}. By probing the magnetization dynamics through transient optical rotation measurements, they show that the enhancement of this interlayer shear correlates with ultrafast demagnetization, which serves to release the elastic energy coming from local strains. Moreover, ultrafast microscopy directly reveals multiple odd and even harmonics of this acoustic phonon below $T_{\rm{N}}$, suggesting the emergence of standing waves in the acoustic cavity confined by the two surfaces of the FePS$_3$ thin film \cite{RN332}. Above $T_{\rm{N}}$, however, the dominant acoustic response shifts to in-plane propagating modes, suggesting that the development of magnetism can profoundly influence the phonon excitation channels.

In ferromagnets, net magnetization can induce a dichroic response to left- and right-handed circularly polarized light. A notable example is bulk CrI$_3$, where coherent phonons exhibit a clear dependence on driving light helicity. Resonant pumping with the $d$-$d$ transition ($t_{2g}$-$e_g$, which is Jahn-Teller active), launches two $A_{1g}$ modes through ISRS \cite{RN323}. Whereas the mode that involves in-plane motion of I ions exhibits no particular pump-helicity dependence across all sampled temperatures, the mode associated with trigonal breathing of I shows an unambiguous dependence on pump helicity [Figs. \ref{fig:17}(e),(f)]. The amplitude ratio of this $A_{1g}^2$ phonon upon left versus right-handed circularly polarized pulses exhibits an order-parameter-like onset at the Curie temperature $T_{\rm{C}}$. Such circular dichroism arises because the out-of-plane phonon modulates the exchange interaction, leading to intertwined lattice and spin oscillation assisted by local single-ion moment changes. These modes are also excited when pumping resonant with the charge-transfer transition, with DECP being the predominant mechanism \cite{RN333}.

The frequency of coherent phonons can also show an abrupt change upon onset of magnetism, a feature that can be monitored with conventional Raman spectroscopy, as has been shown in various antiferromagnets \cite{RN7,RN327,RN76,RN72,RN274}. Time-domain techniques have demonstrated particularly salient changes in phonon frequencies that are coupled to magnons in vdW antiferromagnets. In CoPS$_3$, FePS$_3$, and FePSe$_3$, optical phonons with $A_g$ or $B_g$ symmetries that couple to magnons with identical symmetries exhibit an unconventional softening as the temperature approaches $T_{\rm{N}}$ from below, which is absent for modes that exhibit negligible coupling to magnons \cite{RN320,RN321,RN330}. Moreover, in the itinerant vdW ferromagnet Fe$_3$GeTe$_2$, the frequency of an out-of-plane $A_{1g}$ mode can be manipulated by adjusting the pump light polarization, as it modulates the magnitude of the restoring forces \cite{RN334,RN335}. These modes can further mediate the ultrafast demagnetization process, as demonstrated by simulations of the time-dependent density functional theory (rt-TDDFT) in real time \cite{RN336,RN337}.

Time-domain coherent phonon spectroscopy also offers unique insights into the phase of coherent lattice vibrations that cannot be discerned from conventional Raman spectroscopy. In vdW magnets, the oscillation phase of coherent phonons can undergo dramatic changes across magnetic transitions, revealing alterations in the excitation mechanism as magnetism develops. For instance, in the ferromagnet CrSiTe$_3$, exciting the charge-transfer transition reduces the electrostatic interaction between Te and Cr, creating a repulsive force that expands the distance between the two layers of Te sandwiching Cr. Thus, a $A_{1g}$ phonon will be generated using conventional DECP. However, as short-range magnetic correlations develop, another competing mechanism dubbed spin-DECP starts to participate, producing an attractive force that leads to a higher exchange interaction value [Fig.~\ref{fig:17}(g)] \cite{RN324}. Therefore, the sign of initial displacement along the relevant phonon coordinate will be reversed when spin-DECP dominates over conventional DECP, resulting in a phase flip of $\pi$. Note that such a phase flip occurs at the elevated temperature $T_{\rm{2D}}$ where short-range correlations occur, higher than the 3D magnetic ordering temperature $T_{\rm{C}}$ [Fig.~\ref{fig:17}(h)]. If the driving light frequency is red-detuned from the charge-transfer resonance, the spin-DECP channel is shut off, and the phase flip disappears. In particular, such a phase change is absent for the symmetry-breaking $E_g$ mode. Similarly, in the antiferromagnet FePS$_3$, a 7.41 THz $A_g$ mode involving trigonal distortions of the Fe-S octahedra shows a decrease in amplitude and a simultaneous phase shift of $\pi$ around $T_{\rm{N}}$ [Fig.~\ref{fig:17}(j)] \cite{RN325}. This behavior is absent in another 11.45 THz $A_g$ mode, which only involves in-plane S distortions. In contrast to the previous case, which can be understood as a change in the photoexcited potential energy surface across the transition temperature, the phase flip here arises from the magnetoelastic coupling-induced displacement of the equilibrium potential energy surface mainly along the trigonal distortion coordinate [Fig.~\ref{fig:17}(i)].

\subsection{Ultrafast exciton dynamics}

\begin{figure*}
    \includegraphics[width=\linewidth]{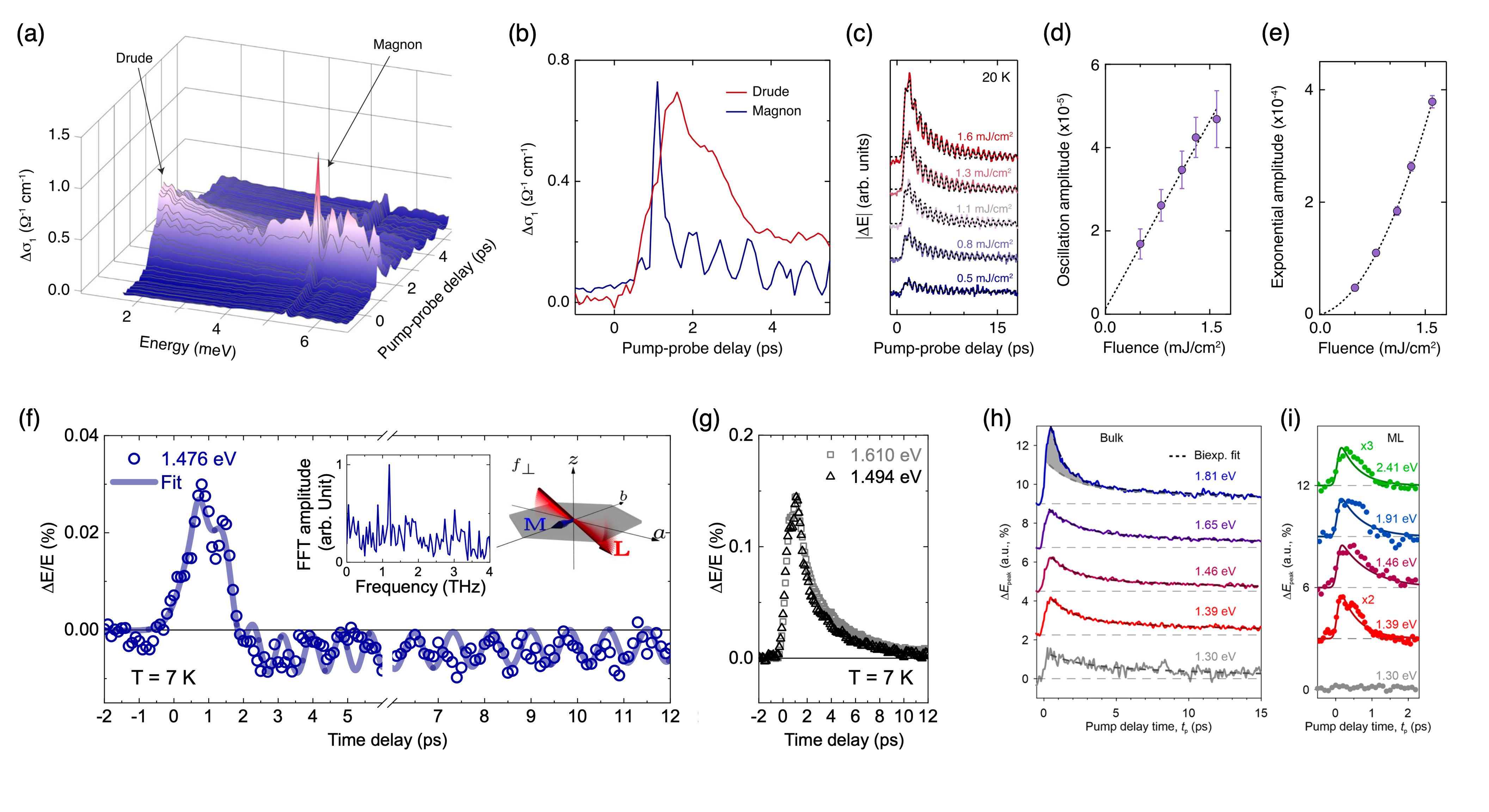}
    \caption{\label{fig:18}Ultrafast exciton dynamics. (a) Spectro-temporal evolution of the pump-induced change in the real part of the THz conductivity ($\Delta\sigma_1$) of NiPS$_3$ at 20~K. (b) Temporal evolution of $\Delta\sigma_1$ at the energies of the Drude peak and the magnon peak taken as line cuts from the spectrum in (a). (c) Spectrally-integrated pump-induced change of the THz electric field ($\Delta E$) as a function of time delay. The black dashed lines are fits to the sum of a damped oscillation and an exponential background. (d),(e) Amplitude of the oscillation (d) and the Drude exponential decay (e) extracted from the fits in (c). (f) Pump-induced change in the THz electric field ($\Delta E/E$) of NiPS$_3$ at 7~K for a pump photon energy of 1.476~eV. $\Delta E/E$ remains negative for tens of picoseconds. The oscillations correspond to a magnon mode. (g) $\Delta E/E$ for pump photon energies of 1.476~eV and 1.610~eV (absorption onset) in which the signal is short lived and positive. (h),(i) Pump-induced change of the THz peak electric field of the near-field response of bulk (e) and monolayer (f) CrSBr as a function of pump-probe time delay for various pump photon energies. Panels (a)-(e) are from \citet{RN86}. Panels (f) and (g) are from \citet{RN338}. Panels (h) and (i) are from \citet{RN339}.}
\end{figure*}

vdW magnets have been found to host unique types of excitons coupled to the magnetic order that are distinct from other conventional types of excitons, such as Frenkel and Wannier excitons, which form due to the Coulomb interaction between electrons and holes \cite{RN340}. For example, in the vdW antiferromagnet NiPS$_3$, a very sharp exciton was observed only below the N\'eel temperature using resonant inelastic X-ray scattering (RIXS), photoluminescence, and optical absorption measurements \cite{RN80}. This exciton was initially characterized as a transition from a Zhang-Rice triplet to a Zhang-Rice singlet. However, a later study using ultra-high-resolution RIXS clarified that the exciton in NiPS$_3$ is a Hund's exciton, i.e. its formation is due to Hund's exchange interactions \cite{RN190}. By studying the dispersion of the Hund's exciton, they revealed that its propagation is similar to a double-magnon. Other works have highlighted the anisotropic nature of the exciton and the coupling of its polarization to the magnetic order \cite{RN81,RN82}.

Given the unique magnetic coupling of the exciton in NiPS$_3$ at equilibrium, it holds promise for unconventional phenomena when driven out of equilibrium. An experimental study \cite{RN86} investigated the low-energy response of NiPS$_3$ when excitons are photoexcited. They mapped the THz spectrum as a function of time after photoexcitation and revealed both coherent magnon oscillations and a Drude response at low frequencies, as shown in Fig.~\ref{fig:18}(a), indicating a coexistence of itinerant conductivity and antiferromagnetic order in the photoinduced state. The Drude response was found to have a quadratic dependence on pump fluence [Fig.~\ref{fig:18}(d)], indicating that exciton dissociation is responsible for the production of itinerant carriers. The coherent magnon response is discussed in more detail in the next section. Another ultrafast THz study on NiPS$_3$ \cite{RN338} studied the effect of tuning the pump photon energy around the exciton at 1.476~eV. They observed a long-lived state with negative photoconductivity, indicating population inversion, as shown in Fig.~\ref{fig:18}(f).

Another compound with interesting exciton properties is the vdW antiferromagnetic semiconductor CrSBr. This compound is an A-type antiferromagnet, in which each layer orders ferromagnetically and has antiferromagnetic coupling between adjacent layers. This material hosts Wannier exciton transitions that are sensitive to interlayer electronic coupling. The Wannier excitons have a large binding energy of 0.5~eV for a monolayer and 0.25~eV for bulk CrSBr, and these transitions are anisotropic as shown in polarization-resolved optical spectroscopy, in which the exciton transition at 1.34~eV only visible when light is polarized along the crystallographic $b$-axis \cite{RN177}. Photoluminescence measurements under an applied magnetic field demonstrate that the exciton is coupled to the magnetic order. Time-resolved measurements \cite{RN341} have also demonstrated that above-gap photoexcitation generates coherent magnons that modulate exciton energies (see the next section for more details). This magneto-excitonic coupling is found to originate from interlayer electronic interaction since the monolayer CrSBr shows no change in the exciton energy while in the bilayer the exciton energy redshifts abruptly at a critical field of 0.134~T since it undergoes a spin flip transition from antiferromagnetic to ferromagnetic ordering. This effect is attributed to spin-allowed interlayer hybridization of electron and hole orbitals as shown by Green's function-Bethe-Salpeter equation (GW-BSE) calculations. Such a coupling between excitons and magnetism in CrSBr offers an efficient way to monitor and manipulate the spin degrees of freedom using optical light tuned near the exciton resonance.

Using time-resolved polarization nanoscopy at THz frequencies, another experimental study \cite{RN339} investigated the ultrafast dynamics of excitons in CrSBr. They excite the sample with ultrashort optical pulses and probe the near-field THz response. In bulk CrSBr, they observe relaxation dynamics due to scattering with phonons, defects, or paramagnons, with an exciton lifetime of 15~ps [Fig.~\ref{fig:18}(h)]. In a monolayer, the recombination dynamics occurs on the very fast timescale of 0.5~ps [Fig.~\ref{fig:18}(i)], indicating the formation of 1$s$ excitons, as excitons formed by energetically distant bands or unbound electron-hole pairs become quenched by the Coulomb interaction.

\subsection{Coherent magnon generation}

The optical excitation of coherent magnons has been observed in various magnetic materials over the last couple of decades (see \cite{RN342,RN343} for excellent reviews covering this topic). Here, we focus specifically on vdW magnets and survey the mechanisms for coherent magnon generation recently reported in these systems. Although some mechanisms are the same as those for 3D magnetic systems, certain features resulting from the quasi-2D nature of vdW magnets allow for unique optical excitation pathways.

\subsubsection{vdW antiferromagnets}

\begin{figure*}
    \includegraphics[width=\linewidth]{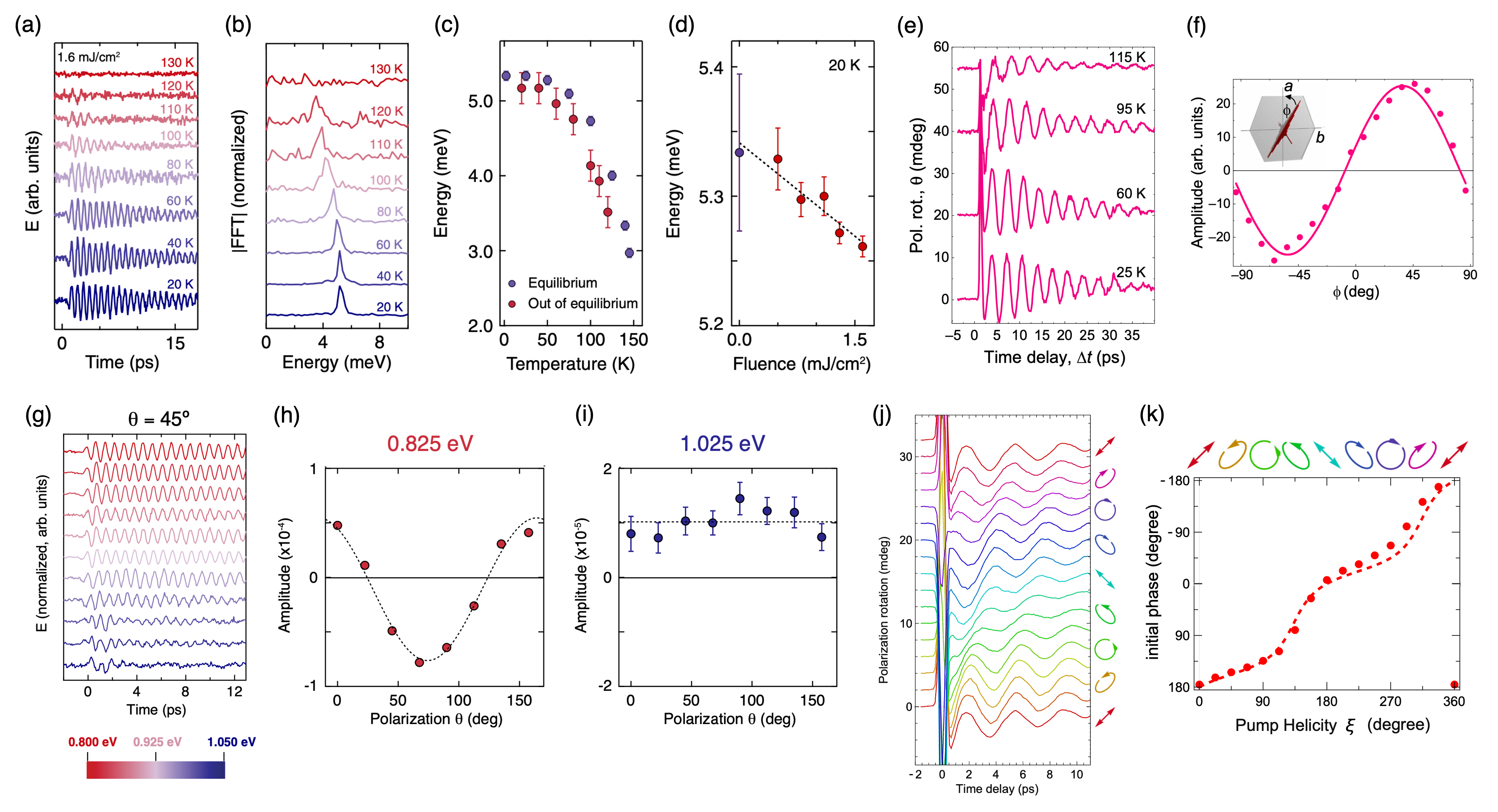}
    \caption{\label{fig:19}Coherent magnon generation in NiPS$_3$. (a) Temporal evolution of the emitted THz electric field ($E$) from NiPS$_3$ at various temperatures showing coherent magnon oscillations. (b) Fourier transform of the time traces in (a). The magnon energy softens as the temperature approaches $T_N$. (c) Comparison of the magnon energies as a function of temperature at equilibrium and out of equilibrium from the THz emission, revealing that the magnon energy of the driven system is redshifted. (d) Absorbed pump fluence dependence of the energy of the magnon oscillations at 20~K obtained from fits to the THz emission data. The magnon energy decreases linearly in both cases. (e) Temperature dependence of the time-resolved Faraday rotation after photoexcitation with a photon energy of 1.08~eV (resonant with the $d$-$d$ transition). (f) Pump polarization dependence of the magnon amplitude. The angle $\phi$ is the angle between the linear pump polarization and the crystallographic $a$-axis. (g) Normalized THz emission at 15~K as a function of time after the pump pulse for a range of pump photon energies (0.80--1.05~eV). (h),(i) Pump polarization dependence of the magnon amplitude extracted from fits to the THz emission for pump photon energies 0.825~eV (transparency) and 1.025~eV (resonant with the $d$-$d$ transition), respectively. For the transparent region (h), the magnon amplitude follows a cos(2$\theta$) dependence, while for the resonant case (i), the amplitude is nearly independent of the pump polarization angle. (j) Pump-induced change in the Faraday rotation angle as a function of time at 12~K for varying polarizations of the pump (depicted by the arrows on the right). (k) Pump polarization dependence of the initial phase of the magnon oscillations extracted from fits to the time traces in (j). Panels (a)-(d) are from \citet{RN86}. Panels (e) and (f) are from \citet{RN344}. Panels (g)-(i) are from \citet{RN345}. Panels (j) and (k) are from \citet{RN346}.}
\end{figure*}

In the previous section, we saw that the vdW antiferromagnet NiPS$_3$ possesses a unique type of magnetically coupled exciton. When this exciton transition is resonantly photoexcited, it leads to the generation of a coherent magnon \cite{RN86}. This experiment used ultrafast THz emission spectroscopy, which detects THz radiation emitted by the sample following an optical pump pulse. Such a signal is a direct measure of coherent magnons, since dipoles in the sample oscillating perpendicular to the light propagation direction will emit radiation. Figure~\ref{fig:19}(a) shows the THz emission response of NiPS$_3$ when the exciton is resonantly excited. The signal consists purely of a single damped sinusoidal oscillation at the frequency of a magnon around 1.3~THz. The temperature dependence of the THz emission reveals a softening toward the N\'eel temperature ($T_N=157$~K), confirming its magnon character [Fig.~\ref{fig:19}(b)]. However, the magnon frequency is slightly redshifted compared to the equilibrium frequency, as shown in Fig.~\ref{fig:19}(c), and the amount of redshift is linearly proportional to the density of photogenerated excitons. It cannot be explained by heating [Fig.~\ref{fig:19}(d)]. A simple theoretical model demonstrates that the change in spin of the exciton transition leads to a reduction in the magnon frequency. Therefore, this unique magnetically coupled exciton is responsible for the coherent generation of the magnon.

Additional ultrafast studies on NiPS$_3$ have uncovered even more mechanisms for coherent magnon generation when the system is pumped in other regions of the optical absorption spectrum, in particular, in the vicinity of a lower energy spin-allowed ($\Delta S=0$) $d$-$d$ transition. One study \cite{RN344} found that resonantly exciting this $d$-$d$ transition results in the coherent excitation of a magnon at 0.3~THz. This experiment used time-resolved Faraday rotation and is therefore sensitive to magnons with out-of-plane magnetization oscillations (compared to THz emission which is sensitive to in-plane oscillations). The Faraday rotation signal yields sinusoidal oscillations at 0.3~THz, as shown in Fig.~\ref{fig:19}(e), whose temperature dependence shows a frequency softening that matches the expected critical scaling. The dependence of the oscillation amplitude on the linear polarization angle of the pump reveals a cos(2$\theta$) behavior [Fig.~\ref{fig:19}(f)]. The mechanism of the coherent generation of this magnon can be attributed to an ultrafast light-induced change in the magnetocrystalline anisotropy.

A subsequent theoretical study \cite{RN347} proposed two microscopic mechanisms for the optical excitation of coherent magnons in vdW magnets: the resonant pumping of atomic orbital excitations and a light-induced Floquet spin Hamiltonian. The former was shown to agree with the experiment of \cite{RN344}, while the latter provides a microscopic framework for understanding magneto-optical effects (e.g., inverse Faraday effect and inverse Cotton-Mouton effect; also described phenomenologically as impulsive stimulated Raman scattering) that have been reported in many magnetic systems. This theoretical framework can also be applied to 2D magnets with XY spin Hamiltonians (such as NiPS$_3$ in the monolayer limit) and therefore has implications for the study of excitations of a Berezinskii-Kosterlitz-Thouless phase.

Coherent magnon dynamics in NiPS$_3$ has also been measured using ultrafast THz emission spectroscopy \cite{RN345}. By pumping the system in the same spectral region as the ultrafast Faraday rotation study \cite{RN344}, another generation mechanism was uncovered. The experiment found that when the pump is on resonance with the spin-allowed $d$-$d$ transition, the magnon at 1.3~THz is coherently excited through this new mechanism, while when the pump photon energy is tuned to a transparent region below this transition, the magnon is instead launched coherently via a magneto-optical effect. Figures~\ref{fig:19}(g)-(i) display the striking change in the pump polarization dependence across these two different photoexcitation conditions, indicating that distinct microscopic mechanisms are responsible for coherent magnon generation. The mechanism of resonant excitation, which does not yield dependence on the pump polarization angle $\theta$, involves dissipative dynamics as a key component, such that the system must be modeled as an open quantum system using a Lindbladian picture. The decay process from the excited state back to the ground state manifold provides the initial conditions for launching the coherent spin precession. We note that this dissipation-based mechanism can also explain the coherent generation of the 0.3~THz magnon mode but since this mode is less sensitive to dissipation, the earlier theory of \cite{RN347} is sufficient to explain the experimentally observed results in \cite{RN344}. In the transparent region of photoexcitation, the coherent magnon amplitude follows the expected cos(2$\theta$) dependence for magneto-optical effects.

\begin{figure*}
    \includegraphics[width=\linewidth]{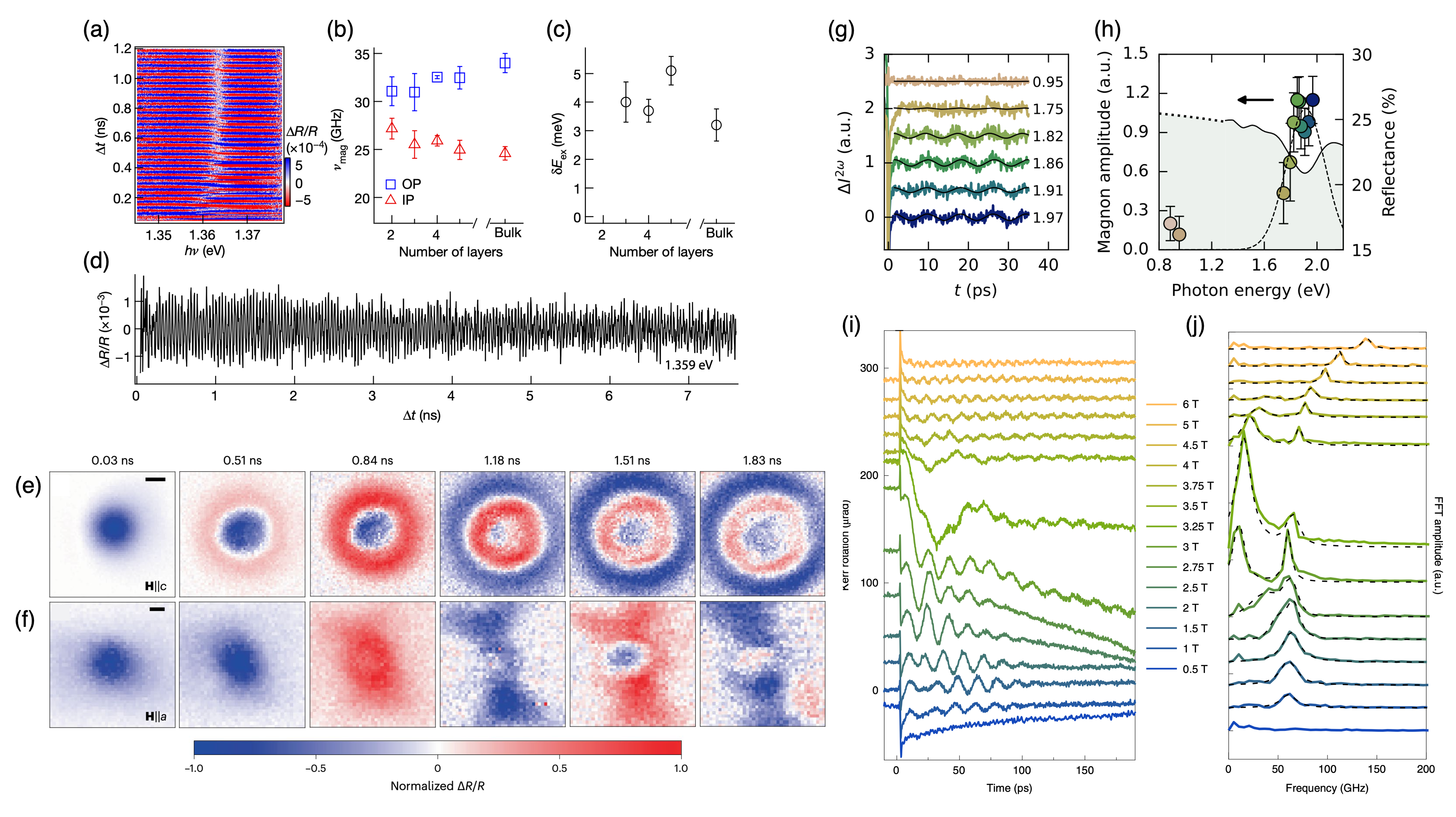}
    \caption{\label{fig:20}Coherent magnon generation in other vdW antiferromagnets. (a) Transient reflectivity spectra $\Delta R/R$ as a function of pump-probe delay time $\Delta t$ and probe energy $h\nu$ for five layer CrSBr. (b) Frequency of the in-phase (IN) and out-of-phase (OP) magnon modes as a function of the number of layers in the sample. (c) Magnitude of the modulation of the exciton energy $\delta E_{\text{ex}}$ as a function of the number of layers. (d) Vertical cut of the spectrum in (a) at $h\nu=1.359$~eV. (e) Snapshots of spatially resolved transient reflectivity measurements with a magnetic field of 1~T applied along the $c$ direction. (f) Snapshots of spatially resolved transient reflectivity measurements with a magnetic field of 0.35~T applied along the $a$ direction. The images in (e) and (f) were obtained by rastering the pump across the sample and measuring the reflectivity at fixed time delays between the pump and probe. (g) Pump-induced change in the SHG intensity of MnPS$_3$ for different pump photon energies ranging from 0.88 to 1.97~eV. (h) Magnon oscillation amplitude obtained from fits to the time traces in (g) plotted with reflectance showing the orbital resonance. (i) Pump-induced Kerr rotation as a function of pump-probe delay time in bilayer CrI$_3$ for different in-plane magnetic field strengths. (j) FFT amplitude of the time traces in (i) after subtraction of the exponential background due to demagnetization. Panels (a)-(d) are from \citet{RN341}. Panels (e) and (f) are from \citet{RN348}. Panels (g) and (h) are from \citet{RN349}. Panels (i) and (j) are from \citet{RN350}.}
\end{figure*}

Beyond discovering new optical generation mechanisms for coherent magnons, there is also a desire to control the properties of the excited magnons. Another time-resolved Faraday rotation study on NiPS$_3$ \cite{RN346} demonstrated control over the phase of coherent magnons using both circularly polarized and linearly polarized light. This work used the same resonant photoexcitation conditions as \cite{RN344} to generate the 0.3~THz magnon coherently. They found that the magnon phase changes by 180$^{\circ}$ for opposite helicities of a circularly polarized pump, and for orthogonal directions of a linearly polarized pump, which were attributed to non-thermal magneto-optical effects. Moreover, they were able to achieve a continuous tuning of the coherent magnon phase by varying the ellipticity of the pump polarization between linear and circular, as shown in Figs.~\ref{fig:19}(j),(k).

Therefore, we see that the vdW antiferromagnet NiPS$_3$ enables the optical generation of coherent magnons via various microscopic mechanisms depending on the photon energy of the pump relative to electronic transitions in the material, as well as the symmetry of the magnon modes that can be excited and detected. These works also lay the groundwork for the control of magnon properties and dynamics via selective optical excitation, which could have applications in spin-based technological devices.

Coherent magnon generation has also been reported in another vdW antiferromagnet of the same family, MnPS$_3$. A study of MnPS$_3$ using time-resolved SHG \cite{RN349} demonstrated the launch of a coherent magnon when pumping an orbital resonance. Since the magnetic point group of MnPS$_3$ breaks the inversion symmetry, the magnetic order can be detected through SHG. They pump the system with a photon energy in resonance with a transition from the ground state of Mn$^{2+}$ ions $^6A_{1g}$ ($t^3_{2g}e^2_g$) which is an orbital singlet with $S=5/2$ to an excited state $^6T_{1g}$ ($t^4_{2g}e^1_g$) which is an orbital triplet with $S=3/2$ at an energy of 1.92~eV. As shown in Fig.~\ref{fig:20}(g), the pump-induced change in SHG intensity shows oscillations at a frequency of 119~GHz, which corresponds to the magnon energy gap of MnPS$_3$. The coherent magnon oscillations are only observed when the pump is resonant with this transition and are absent for pump photon energies below the orbital resonance [Fig.~\ref{fig:20}(h)]. The initial phase of the magnon oscillations is near zero, indicating that the spins precess due to an impulsive force rather than a thermal displacive excitation. The microscopic mechanism of coherent magnon generation can be understood as follows: the pump excites electrons into an orbital-angular-momentum-carrying excited state that enables a transient coupling of spin and orbital angular momentum. This in turn reorients the magnetic anisotropy direction during the excited state lifetime and exerts an impulsive torque that initiates the spin precession (similar to that observed in NiPS$_3$ \cite{RN344}).

Coherent magnons can also be generated via the magnetic dipolar process by applying an AC magnetic field that is resonant with their energies \cite{RN351}. In this case, the magnons are linearly driven, with their amplitude scaling proportionally to the driving field strength. In the Ising-type zigzag antiferromagnets FePS$_3$ and FePSe$_3$, the strong onsite magnetic anisotropy gives rise to a sizable magnon gap in the THz range, enabling the excitation of magnons with intense THz pulses. Recent experiments have confirmed the coherent linear excitation of magnons with $A_g$ and $B_g$ symmetries (or a combination of both, see the Magnon-Phonon Coupling section) when subjected to a high-field THz pulse with a bandwidth encompassing their energies \cite{RN321,RN322}.

The vdW semiconductor CrSBr has also attracted significant attention in the context of coherent magnon generation. This material also features excitons coupled to magnons, as we saw in the previous section. An experimental work \cite{RN341} used photoexcitation above the gap to launch coherent magnons and probed their oscillations using broadband reflectivity. The spin-dependent interlayer electron-exchange interaction in this compound couples the Wannier excitons to the interlayer magnetic order. When coherent magnons are generated, the exciton energy changes due to changes in this interlayer electron-exchange interaction, and therefore the magnon oscillations can be detected as coherent oscillations in the exciton energy. Two coherent magnons are observed, at frequencies of 24 and 34~GHz, corresponding to in-phase and out-of-phase spin precessions, respectively. Additionally, these measurements can be extended to the 2D limit of CrSBr. Figure~\ref{fig:20}(b) shows that the coherent magnon oscillations are observed in the transient reflectivity spectra of the bulk CrSBr down to the two-layer CrSBr. The magnon oscillations are very long lived, with a coherence time of 5~ns [Fig.~\ref{fig:20}(d)].

A subsequent study on CrSBr \cite{RN348} investigated the propagation of coherent magnon wavepackets following optical excitation. They found that magnon propagation is mediated by long-range magnetic dipole-dipole interactions rather than short-range exchange interactions. Figures~\ref{fig:20}(e),(f) show the magnon wavepacket propagation for two different low-frequency magnon modes at 21 and 32~GHz, respectively. The two magnon branches have very different propagation dynamics: wave packets of the lower frequency band propagate with an isotropic group velocity, while those of the higher frequency band show unidirectional propagation with twice the group velocity of the lower band. These features, along with their observation that the group velocities of the magnon wave packets depend on the sample thickness, implies that long-range magnetic dipole coupling is responsible for the propagation dynamics. Quantitative agreement is found between the experimental data and a theory of dipolar spin waves for easy-plane antiferromagnets.

Coherently generated magnons have also been observed in other families of atomically thin vdW antiferromagnets. A study on antiferromagnetic CrI$_3$ bilayers \cite{RN350} revealed optically excited coherent magnons using the time-resolved magneto-optical Kerr effect (MOKE). The sample used was a heterostructure of bilayer CrI$_3$ on monolayer WSe$_2$. Figure~\ref{fig:20}(i) shows oscillations in the Kerr rotation signal as a function of time corresponding to magnon modes. The proposed mechanism for the generation of coherent magnons is that the optical pump excites excitons in the WSe$_2$ layer, which then dissociate and lead to electron transfer at the CrI$_3$-WSe$_2$ interface and an impulsive perturbation of the magnetic interactions in CrI$_3$ by these hot carriers. Additionally, they demonstrated that the magnon frequencies can be tuned by tens of gigahertz by varying the gate voltage, changing the system from hole doping to electron doping.

\begin{figure*}
    \includegraphics[width=\linewidth]{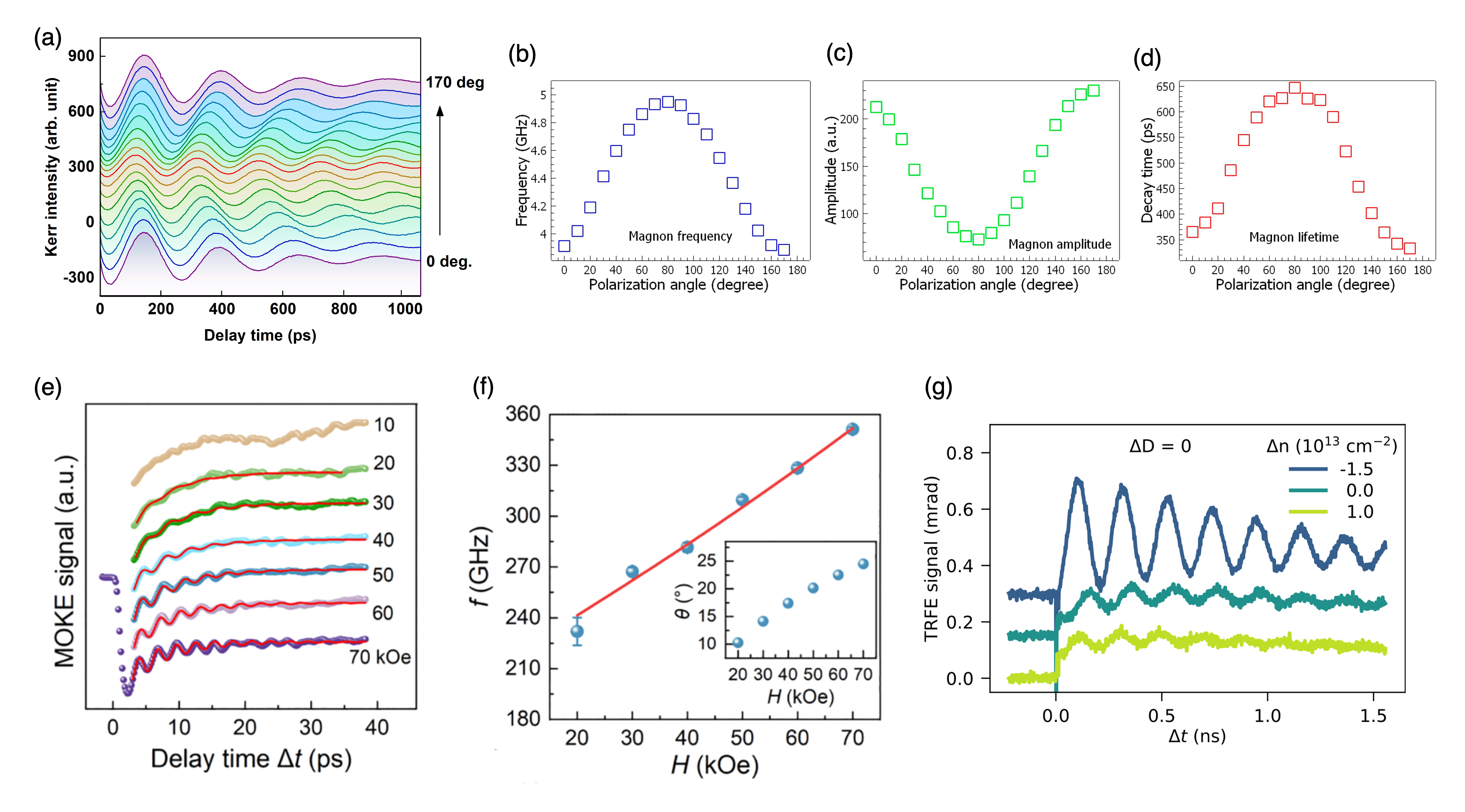}
    \caption{\label{fig:21}Coherent magnon generation in vdW ferromagnets. (a) Pump-induced Kerr rotation of Fe$_{3.6}$Co$_{1.4}$GeTe$_2$ as a function of pump-probe delay time for different pump polarization angles. (b)-(d) Magnon frequency (b), amplitude (c), and decay time (d) as a function of pump polarization angle extracted from fits to the time traces. (e) Pump-induced MOKE signal of Fe$_3$GaTe$_2$ at 10~K under different external magnetic field strengths. (f) Frequency of the spin precession extracted from fits to the time traces in (e). (g) Time-resolved Faraday ellipticity of Cr$_2$Ge$_2$Te$_6$ with an applied magnetic field of 100~mT for three different values of charge carrier density $\Delta n$. Panels (a)-(d) are from \citet{RN352}. Panels (e) and (f) are from \citet{RN353}. Panel (g) is from \citet{RN354}.}
\end{figure*}

\subsubsection{vdW ferromagnets}

Now, we turn to coherent magnon generation in ferromagnets. An experimental study on the ferromagnetic vdW compound Fe$_{3.6}$Co$_{1.4}$GeTe$_2$ using time-resolved MOKE \cite{RN352} found that coherent magnons can be launched and controlled with ultrafast optical excitation. As shown in Fig.~\ref{fig:21}(a), oscillations are observed in the pump-induced Kerr rotation as a time delay function, corresponding to a coherent magnon. By tuning the pump laser polarization, they observed a substantial modulation of the magnon frequency, amplitude, and lifetime [Figs.~\ref{fig:21}(b)-(d)]. First-principles calculations demonstrate anisotropic optical absorption of different crystal orientations, suggesting that the coherent generation of magnons is due to the modification of the effective demagnetization field and magnetic anisotropy caused by the anisotropic optical absorption for different pump polarizations.

Similar work was performed on Fe$_3$GaTe$_2$ \cite{RN353}. This study used time-resolved MOKE to observe coherent magnons as a function of applied magnetic field [Fig.~\ref{fig:21}(e)]. They observe an ultrafast demagnetization upon laser excitation followed by magnetization precession and relaxation. The spin precession frequency reaches up to 351.2~GHz, as shown in Fig.~\ref{fig:21}(f), which is the highest reported for any 2D ferromagnetic material and is attributed to the strong perpendicular magnetic anisotropy in Fe$_3$GaTe$_2$ owing to the large spin-orbit coupling between the Fe $d$ orbitals and the Te and Ga $p$ orbitals. Such high spin frequencies are promising for spintronics applications.

Another work investigated the magnetization dynamics in the vdW ferromagnetic semiconductor Cr$_2$Ge$_2$Te$_6$ \cite{RN354}. In contrast to purely thermal mechanisms reported for other 2D ferromagnets, they find that coherent opto-magnetic phenomena play an important role in exciting magnetization dynamics. They observe laser-induced oscillations of the magnetization in the time-resolved Faraday ellipticity as shown in Fig.~\ref{fig:21}(g). The results can be explained by an interplay of the inverse Cotton-Mouton effect and photo-induced magnetic anisotropy, as well as a more conventional thermal mechanism. This work demonstrates that the magnetization precession amplitude can be controlled by tuning the charge carrier density through electrostatic gating. A study on Cr$_2$Ge$_2$Te$_6$ using time-resolved Faraday rotation \cite{RN355} also observed pump-induced precession of magnetization. Still, it was attributed to laser-induced heating of the sample, causing instantaneous demagnetization that triggers the precession.

\subsubsection{vdW multiferroics}

Lastly, we discuss the coherent generation of electromagnons in vdW multiferroics. Electromagnons are electric dipole active spin waves and are soft modes of a multiferroic phase. The optical generation of coherent electromagnons has been demonstrated in the vdW multiferroic NiI$_2$. An experimental study on NiI$_2$ \cite{RN188} using time-resolved SHG and Kerr rotation microscopy revealed coherent oscillations of electromagnon modes. They extract the dynamical magnetoelectric coupling constant and find it to be very large, surpassing that of other known multiferroics. The reason for this large value can be understood by considering that the electromagnons in NiI$_2$ originate from the inverse Dzyaloshinskii-Moriya interaction, rather than phonon-mediated mechanisms in other multiferroics, and this electronic nature combined with strong $d$-$p$ hybridization between the nickel and iodine atoms gives rise to a giant electric polarization and magnetoelectric coupling strength. Coherent electromagnons in NiI$_2$ have also been observed using ultrafast THz spectroscopy \cite{RN356}. Investigations of the generation of coherent electromagnons are still in their infancy, and the microscopic mechanisms for such processes remain an open question.

\subsection{Coupling of collective modes}

Emerging from the interplay between charge, spin, and lattice sectors, collective bosonic excitations, namely excitons, magnons, and phonons, exhibit intricate interactions within and between sectors. These couplings can generate new hybrid quasiparticles that combine the characteristics of their parent modes, paving the way for optically tunable intertwined phononic, magnonic, and spintronic applications. 

\begin{figure*}
    \includegraphics[width=\linewidth]{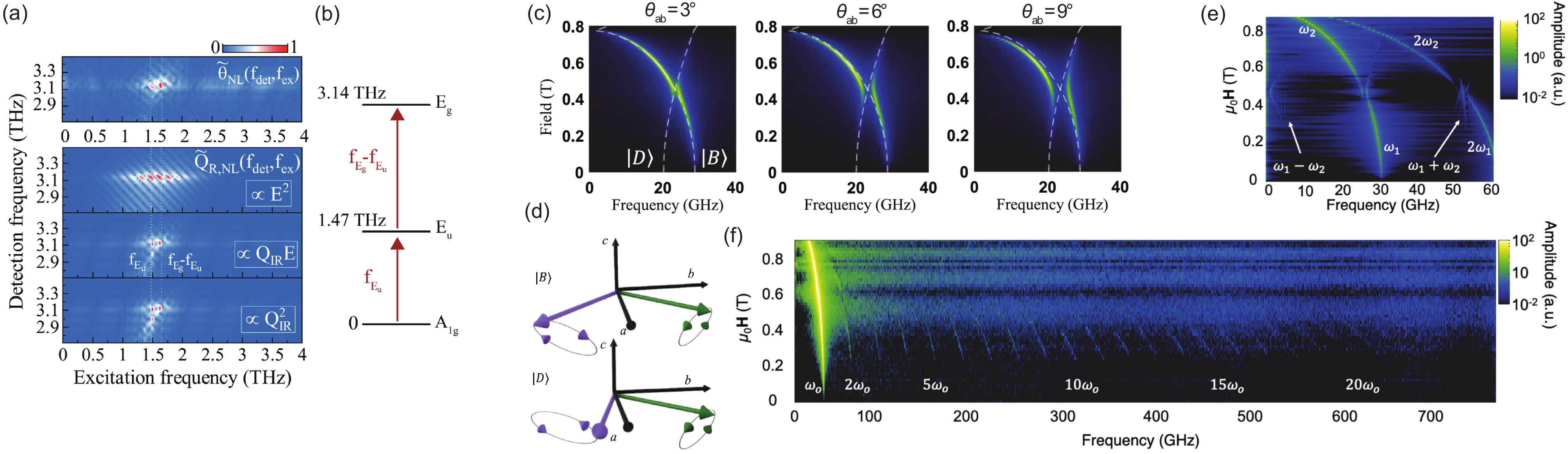}
    \caption{\label{fig:22}Nonlinear phonon and magnon coupling. (a) Upper: experimental two-dimensional nonlinear THz spectroscopy of MnBi$_2$Te$_4$. Lower: simulated nonlinear THz spectroscopy with two-photon ($\propto E^2$), one-photon-one-IR-phonon ($\propto Q_{\rm{IR}}E$), and two-IR-phonon ($\propto Q_{\rm{IR}}^2$) as the driving force. (b) Schematic illustration of the one-photon-one-IR-phonon excitation mechanism. (c) Magnon spectra of CrSBr measured at three selected magnetic field angles $\theta_{ab}$ with the dispersion of the bright ($\ket{B}$) and dark ($\ket{D}$) magnons highlighted by white dashed lines. (d) Schematic illustration of the eigenvector of the bright and dark magnons. (e) Magnon spectrum at $\theta_{ab}=2^\circ$ with all the second-order nonlinear components denoted. SHG: $2\omega_{1,2}$; DFG: $\omega_1-\omega_2$; SFG: $\omega_1+\omega_2$. (f) Magnon spectrum with high harmonics. Panels (a) and (b) are adapted from \citet{RN357}. Panels (c) and (d) are adapted from \citet{RN358}. Panels (e) and (f) are adapted from \citet{RN359}.}
\end{figure*}

\subsubsection{Phonon-phonon coupling}
Focusing on interactions exclusively within the lattice degrees of freedom, anharmonic coupling between different phonons arises from the symmetry properties of crystal structures and phonon modes. When an IR-active phonon is optically driven to large amplitudes, it can nonlinearly couple to other Raman-active modes in the lattice under specific symmetry constraints \cite{RN318}. Nonlinear phononics has opened a multitude of novel pathways for engineering quantum phases with mode-selective control, such as enhancing or inducing superconductivity \cite{RN360,RN361,RN362}, reversing or stabilizing ferroelectricity \cite{RN363,RN364,RN365}, driving structural phase transitions \cite{RN366}, modulating exchange interactions \cite{RN367}, flipping magnetism \cite{RN368}, polarizing ferrimagnetism through the piezomagnetic effect \cite{RN369}, and stabilizing fluctuating ferromagnetism \cite{RN370}. Given the strong magnetoelastic coupling in vdW magnets, exploiting the nonlinearity among phonons can potentially lead to light-driven magnetic (or even topological) phase transitions, enabling new controllable opto-mechanical-spintronic functionalities \cite{RN371}. 

A state-of-the-art optical technique for probing nonlinear couplings between different modes is two-dimensional (2D) nonlinear spectroscopy, particularly in the THz range, which is resonant or near-resonant with many targeted excitations \cite{RN372,RN373}. Technically, this method introduces an additional pump with a tunable time delay $\tau$ along with the delay of the primary pump probe $t$. A 2D Fourier transform of signals as a function of $t$ and $\tau$ produces a 2D THz spectrum as a function of the detection frequency $\omega_t$ and the excitation frequency $\omega_{\tau}$. If an excited mode at $\omega_{1}$ is coupled to another mode at $\omega_{2}$, a resonance peak appears at $(\omega_{\tau}=\omega_{1},\omega_{t}=\omega_{2})$. Furthermore, this technique can distinguish between the various aforementioned mechanisms for exciting Raman-active phonons, such as two-photon sum or differential frequency generation, one-photon-one-phonon IRRS, or two-phonon IRS, by analyzing the detailed spectral structure [Fig.~\ref{fig:22}(b)]. As an example, in the vdW topological antiferromagnet MnBi$_2$Te$_4$, 2D THz spectroscopy revealed that the interaction between a coherently excited IR-active $E_u$ phonon and a photon drives coherent oscillations of a Raman-active $E_g$ phonon through IRRS [Figs.~\ref{fig:22}(a),(b)] \cite{RN357}. Similarly, in the zigzag antiferromagnet FePS$_3$, coherent oscillations of $A_g$ and $B_g$ phonons are launched by $A_u$ and $B_u$ IR-active phonons through IRRS \cite{RN320}. Since the lattice structure is tied to topological and magnetic properties, uncovering nonlinear phononic channels raises a plethora of opportunities for ultrafast engineering of magnetism.

\subsubsection{Magnon-magnon coupling}
In analogy to the lattice sector, coupling between magnons gives rise to various nonlinear magnon-magnon mixed states, forming the foundation of nonlinear magnonics \cite{RN374,RN375}. The vdW antiferromagnet CrSBr exhibits a rich variety of states arising from such nonlinear magnonic interactions. In CrSBr, the antiferromagnetically ordered spins align collinearly along the crystallographic $b$-axis. A magnetic field applied along the $a$-axis slightly cants the spins, resulting in two magnon eigenmodes. The acoustic magnon, involving in-phase precession of the two sublattice spins ($S_1$,$S_2$) with a fixed angle between them, is optically invisible and forms a dark state $\ket{D}$. In contrast, the optical magnon, involving synchronized counter-rotations of $S_1$ and $S_2$ with a time-dependent net magnetization and varying intersection angle, is optically bright ($\ket{B}$) [Fig.~\ref{fig:22}(d)]. When the magnetic field is rotated away from the $a$-axis by an angle $\theta_{ab}$ in the $ab$-plane, the frequencies of both modes can be continuously tuned, and optical detection of the dark state becomes possible. By varying the magnitude of the external magnetic field, an avoided crossing between $\ket{B}$ and $\ket{D}$ is observed, proving their hybridization [Fig.~\ref{fig:22}(c)] \cite{RN358}. As $\theta_{ab}$ increases, the splitting grows linearly, effectively brightening the dark state. Further application of uniaxial strain can enhance the coupling between $\ket{B}$ and $\ket{D}$ and alter their dispersion curvature. Recent experiments with improved signal-to-noise ratio have further revealed the existence of many nonlinear magnonic phenomena in this system: when $\ket{D}$ is brightened due to magnetic-field-induced symmetry breaking, sum frequency generation (SFG) and differential frequency generation (DFG) of $\ket{B}$ and $\ket{D}$ also emerge [Fig.~\ref{fig:22}(e)] \cite{RN359}. Notably, since the energy of DFG, as well as that of the two individual magnons, can be tuned by adjusting the angle and magnitude of the external magnetic field, the DFG peak can be tuned into resonance with one magnon, giving rise to the parametric amplification of this mode. Additionally, high harmonic generation (HHG) of $\ket{B}$ up to at least the 20$^{th}$ order has been observed, exhibiting a non-perturbative nature and underscoring the extreme magnonic nonlinearity of this system [Fig.~\ref{fig:22}(f)]. This unique optical access to nonlinear magnonics in the microwave range offers the potential to use CrSBr as a quantum transducer, while different nonlinear processes pave the foundation for generating entangled magnons and parametric amplification of weak signals.

\subsubsection{Magnon-phonon coupling}
The interplay can occur between the lattice and magnetic degrees of freedom, resulting in the emergence of hybridized quasiparticles known as magnon polarons (MPs). Conventionally, these modes arise when the energies of the two quanta come into proximity. Historically, previous research has focused extensively on MPs formed by acoustic phonons and magnons in ferromagnetic materials, where their hybridization occurs at finite momenta in the gigahertz range \cite{RN376}. However, recent static Raman and IR spectroscopy studies have revealed MPs originating from optical phonons and magnons with identical irreducible representations at zero momentum in various vdW antiferromagnets \cite{RN274,RN377,RN378,RN379}. In the Ising-type zigzag antiferromagnets FePS$_3$ and FePSe$_3$, the large onsite magnetic anisotropy produces a significant magnon gap, situating the acoustic magnons in the THz range. Under finite magnetic fields, magnon energies can approach various Raman-active phonon frequencies, generating avoided crossings between them, marking the formation of MPs \cite{RN326,RN327,RN271,RN380}. 

\begin{figure*}
    \includegraphics[width=\linewidth]{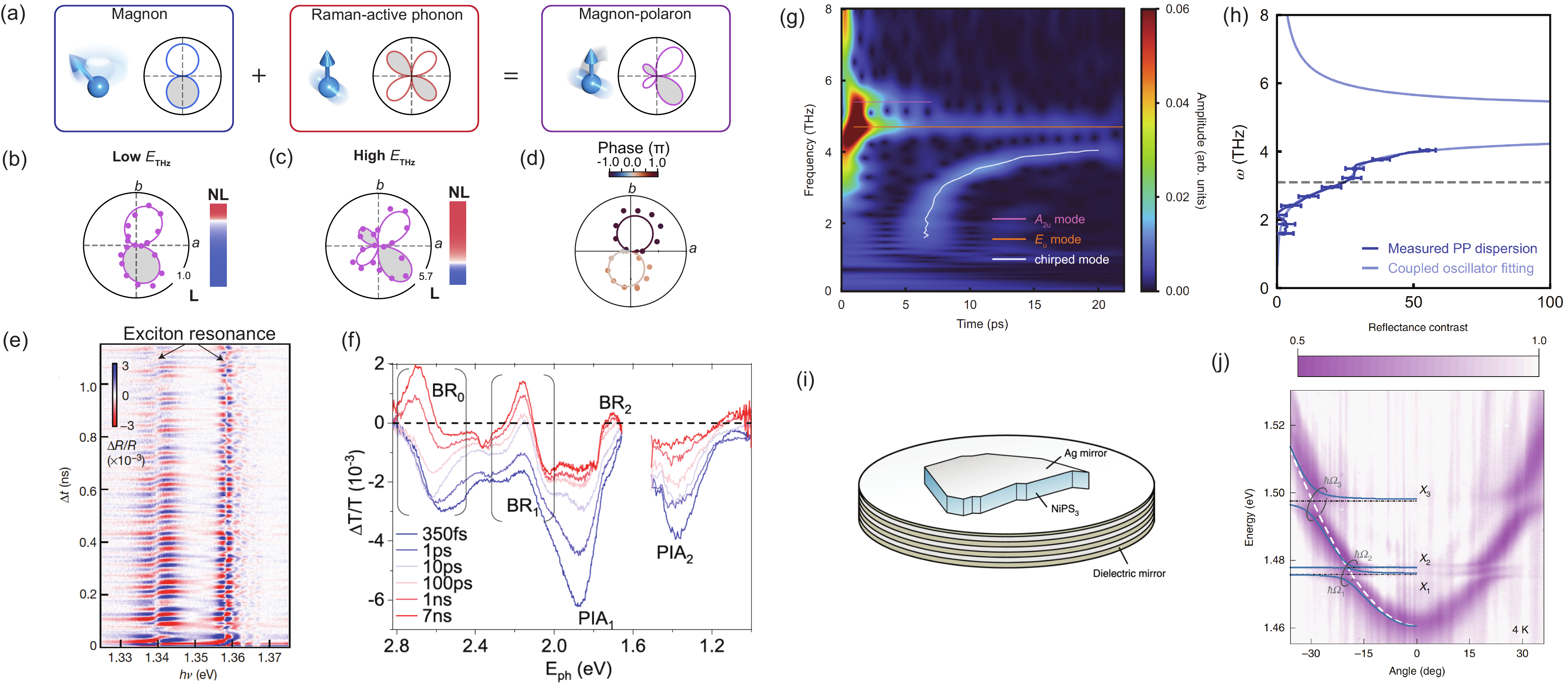}
    \caption{\label{fig:23}Coupling between phonons, magnons, excitons, and photons. (a) Schematic illustration and simulated driving field polarization dependence of the amplitudes of magnon, Raman-active phonon, and MP. Grey and white lobes correspond to opposite phases. (b),(c) Amplitudes as a function of the driving field polarization angle extracted from fitting to the broadband THz pump optical polarization ellipticity probe time traces at relatively low and high driving field strengths in FePS$_3$. The color bars on the right indicate the relative strengths of the nonlinear (phonon, red) versus linear (magnon, blue) components obtained by the fitting (solid lines). (d) Phase-resolved polarimetry of the chiral MP in FePSe$_3$ upon a broadband THz pump. (e) Transient reflectance spectra of CrSBr around its exciton resonances with the incoherent decay background subtracted upon an above-bandgap pump at 1.7 eV. (f) Transient absorption spectra of CrBr$_3$ at selected time delays under an above-gap pump at 2.95 eV. BR$_0$, BR$_1$, and BR$_2$ represent band renormalizations of charge transfer and excitonic transitions, respectively. PIA$_1$ and PIA$_2$ represent photo-induced absorption coming from exciton polarons. (g) Spectrogram of the THz-pump-SHG-probe time trace in NiI$_2$. (h) Reconstructed phonon polariton dispersion from (g) and fitting with a coupled Lorentz oscillator. (i) Angle-resolved simulated reflectance spectrum and experimental photoluminescence spectrum of CrSBr. (j) Angle-resolved reflectance contrast mapping of NiPS$_3$ overlaid with fitting by a coupled model between a single cavity mode and three exciton resonances $X_1$, $X_2$ and $X_3$. Panels (a)-(c) are adapted from \citet{RN322}. Panel (d) is adapted from \citet{RN321}. Panel (e) is adapted from \citet{RN341}. Panel (f) is adapted from \citet{RN381}. Panels (g) and (h) are adapted from \citet{RN382}. Panel (i) is adapted from \citet{RN306}. Panel (j) is adapted from \citet{RN383}.}
\end{figure*}

Recent ultrafast experiments further corroborate the persistence of MPs even at a zero external magnetic field in FePS$_3$ and FePSe$_3$. This is evident in the coexistence of distinct excitation pathways characteristic of both magnons and phonons. Specifically, in these inversion-symmetric compounds, magnons can be linearly excited by a THz magnetic field through the magnetic-dipole process, while Raman-active phonons can only be excited through second-order nonlinear pathways, as discussed previously. These channels exhibit different behaviors in polarimetry experiments: variation in the amplitude of the mode as a function of the driving THz field polarization angle [Fig.~\ref{fig:23}(a)]. For FePS$_3$ and FePSe$_3$, the amplitude of the linear channel is maximized when the driving magnetic field is aligned with the magnon magnetization, showing a two-petal pattern represented by $\cos(\phi)$ or $\sin(\phi)$. In contrast, the amplitude of the nonlinear channel follows a homogeneous quadratic polynomial of $\cos(\phi)$ and $\sin(\phi)$, exhibiting four petals or a ``peanut" like shape depending on the symmetry. For an MP, both channels are activated. Their coherent superposition will produce a polar pattern with ostensible symmetry breaking. Experimentally, using a broadband THz pump to activate both linear and nonlinear excitations at various polarization angles, various reported phonons manifest as coherent oscillations in the polarization ellipticity change of a transmitted optical pulse, which display drive-amplitude-dependent polarimetry patterns. For example, a $B_g$ phonon exhibits a distorted two-lobe pattern at low field strengths, which transforms into a four-petal pattern with unequal lobe sizes at high field strengths [Figs.~\ref{fig:23}(b),(c)] \cite{RN322}. This crossover can be quantitatively described by the combination of linear and nonlinear excitation channels, with the linear pathway dominating at low field strengths and the nonlinear pathway prevailing at high field strengths. These results highlight that the phonons determined by static Raman or IR spectroscopy acquire a magnetic dipole, signifying the emergence of MPs even at a zero external magnetic field. Furthermore, this tunability enables control of the MP dynamical symmetry, expanding the current scope of magno-phononic control of magnetism.

The magnon-phonon coupling can also induce the emergence of chiral quasiparticles. Chirality appears in crystals that break all improper rotational symmetries, which can host exotic electronic excitations such as Weyl fermions \cite{RN384}. Chiral structural excitations carrying angular momentum, with atoms circulating in elliptical trajectories, can even appear in achiral materials. Recent reports related to chiral phonons suggest that they may underlie various phenomena, including the induction of unconventional thermal Hall effects \cite{RN385,RN386,RN387}, the activation of the Spin Seebeck effect \cite{RN388}, mediating angular momentum transfer during demagnetization \cite{RN389,RN390}, the generation of giant magnetic moments \cite{RN391,RN392,RN393,RN394}, and engineering magnetic orders \cite{RN395,RN396}. Typically, these chiral modes are degenerate at the center of the Brillouin zone. Selective excitation of a single-handed enantiomer usually requires circularly polarized pulses in the THz or MIR range, or magnetic fields to lift the degeneracy \cite{RN327,RN369,RN393,RN395,RN396,RN397,RN398,RN399,RN400}. However, intrinsic magnon-phonon coupling provides a route to realize nondegenerate chiral phonons by transferring the intrinsic ellipticity of the spin procession to phonons, leading to chiral MPs. Chiral phonons were first theoretically proposed in the vdW ferromagnet CrI$_3$ and N\'{e}el antiferromagnet VPSe$_3$ \cite{RN401,RN402} and have been experimentally identified in the antiferromagnet FePSe$_3$ \cite{RN321}. In this material, two phonons of $A_g$ and $B_g$ symmetries and two magnons of $A_g$ and $B_g$ symmetries inherently reside at nearly equivalent energies, forming strongly coupled MPs. THz pump experiments have confirmed the dominant linear excitation channel. However, the observed polar pattern, a two-petal pattern rotated away from the crystallographic axes and lacking nodes, clearly deviates from conventional magnetic-dipole symmetry [Fig.~\ref{fig:23}(d)]. Quantitative fitting and modeling suggest that the eigenbases at zero magnetic field are no longer pure $A_g$ and $B_g$ symmetries but rather their combination with phase retardation, i.e. $A_g+re^{i\psi}B_g$, where $r$ is the ratio of the two compositions and $\psi$ is their relative phase difference. This highlights the spontaneous emergence of nondegenerate chiral phonons at the center of the Brillouin zone without external stimuli. 

\subsubsection{Exciton-magnon and exciton-phonon couplings}
Magnons and phonons can strongly interact with various charge-transfer, $d$-$d$, or excitonic transitions, as discussed in the previous section. Here, we focus on the coupling between magnons and phonons with excitons, charge-neutral electron-hole pairs, in vdW magnets. In the A-type antiferromagnet CrSBr, exciton-magnon coupling manifests itself as coherent magnons that are only detectable when the probe energy is near an exciton resonance \cite{RN341,RN348}. This interaction is strongly evidenced by the $\pi$-phase flip of magnon oscillations across excitonic resonances, a signature of mode-modulated optical transitions [Fig.~\ref{fig:23}(e)]. With respect to exciton-phonon interactions, strong electron-phonon coupling can lead to the formation of exciton-polarons, where excitons are dressed by lattice distortions. These composite quasiparticles appear as multiple evenly spaced polaronic bound states close to the excitonic peaks in the spectrum, reminiscent of polaron shake-off sidebands observed in other condensed matter systems \cite{RN403,RN404,RN405,RN406}. For example, in the XY-type zigzag antiferromagnet NiPS$_3$, equilibrium spectral measurements reveal evenly spaced sidebands due to coupling with $A_{1g}$ phonons on the high-energy side of excitonic peaks \cite{RN81}. Similarly, in the A-type antiferromagnet CrSBr, photoluminescence spectroscopy reveals various phonon side bands appearing on the low-energy end of the exciton peak \cite{RN407}. Time-resolved photoluminescence further shows that the exciton-phonon scattering lifetime increases as the temperature approaches $T_{\rm{N}}$ from below and then saturates. Alternatively, exciton polarons may form transiently upon light excitation, as observed in ferromagnets CrI$_3$ and CrBr$_3$ \cite{RN333,RN381,RN309}. In these systems, small localized exciton polarons emerge as photoinduced peaks near the excitonic peaks in the absorption spectra, accompanying, but distinct from, features from band renormalization of the excitonic or charge-transfer transitions [Fig.~\ref{fig:23}(f)] \cite{RN381}. They appear within picoseconds due to the spontaneous localization of photoexcited electron-hole pairs and decay in two steps within $\sim$nanoseconds via radiative recombination and nonradiative Auger recombination. Combined time-dependent transient absorption, photoluminescence, and first-principles simulations demonstrate that the formation mechanisms of these highly localized exciton polarons involve local Jahn-Teller distortions that break the symmetry \cite{RN333}. Interestingly, the phonons involved in polaron formation in all aforementioned cases correspond to out-of-plane vibrations of ligand atoms with $A_{1g}$ symmetry, potentially suggesting the universal but unconventional nature of this trigonal distortion in vdW magnets. 

\subsubsection{Polaritons}
In the presence of external light excitation, bosonic excitations with dipole moments such as the previously mentioned IR-active phonons and excitons can interact with quantized photons, forming polaritons \cite{RN408,RN409}. These new eigenmodes manifest as anticrossings when the dispersive photons coincide with the relatively dispersionless bosonic resonances at finite momentum and energy. The confinement of light possibly on subdiffraction-limited length scales in vdW materials establishes polaritonics as a promising pathway to harvest and manipulate light-matter interactions. 

Photoexcited coherent IR-active phonons radiate electromagnetic waves that reciprocally re-excite these dipolar vibrations, forming propagating light-lattice hybrids known as phonon polaritons (PPs). These quasiparticles further enable exploration of nonlinear phononics and modulation of material properties \cite{RN410,RN411,RN412}. Recent studies on vdW antiferromagnets that break inversion symmetry, such as NiI$_2$ and MnPS$_3$, demonstrate the resonant excitation of broadband PP using THz pulses and their ``time-of-flight" detection with time-resolved SHG \cite{RN382}. The spectrogram $\omega(t)$ of the time traces reveals a time-dependent chirped mode corresponding to the spread-out PP wavepacket, in addition to two IR phonons with $E_u$ and $A_{2u}$ symmetries [Fig.~\ref{fig:23}(g)]. The group velocity, the derivative of the dispersion relation $v_g=d\omega/dk$, can also be easily derived from $v_g= 2l/t(\omega)$, where $l$ represents the thickness of the sample.  PP dispersions can be reconstructed by integrating $v_g$, further revealing a hidden magnon-phonon bilinear coupling as a kink in the lower dispersion branch [Fig.~\ref{fig:23}(h)]. 

Exciton polaritons (EP) are commonly realized using Febry-P\'{e}rot microcavities with significant Rabi splitting. However, vdW materials can serve as intrinsic resonators due to their large dielectric mismatch with the vacuum, enabling confinement of photons and self-hybridization without a closed cavity structure \cite{RN413}. In antiferromagnetic CrSBr, they emerge as various unevenly distributed side bands below exciton resonances \cite{RN306}. Angle-dependent photoluminescence and reflectance spectra resolve the dispersion of up to seven mode branches, with increasing exciton-like character leading to reduced curvature and larger effective mass [Fig.~\ref{fig:15}(d)]. The strong coupling strength $g$ places bare CrSBr in the ultrastrong coupling regime defined by $g/E_X>0.1$, where $E_X$ is the bare exciton energy. In particular, the large anisotropy of CrSBr gives rise to permittivities of opposite sign along different crystallographic axes, producing hyperbolic EPs that facilitate light confinement at sub-diffraction-limited length scales \cite{RN413}. Moreover, the EPs in CrSBr also show strong magnetic field and temperature dependence, making them sensitive to both coherent magnon oscillations and incoherent thermal magnon populations \cite{RN306,RN414,RN415}. In another zigzag antiferromagnet, NiPS$_3$, spin-dressed many-body excitons interact with photons to form EPs within microcavities. These EPs manifest as anticrossing signatures in the reflectance spectra, which vanish upon heating to $T_{\rm{N}}$ [Fig.~\ref{fig:23}(j)] \cite{RN383}. Further photoluminescence results uncover an unconventional relaxation bottleneck characterized by suppression of exciton-exciton dissociation, further enriching the dynamics of these hybrid modes.

\subsection{Metastability and optical switching of magnetism}

\begin{figure*}
    \includegraphics[width=\linewidth]{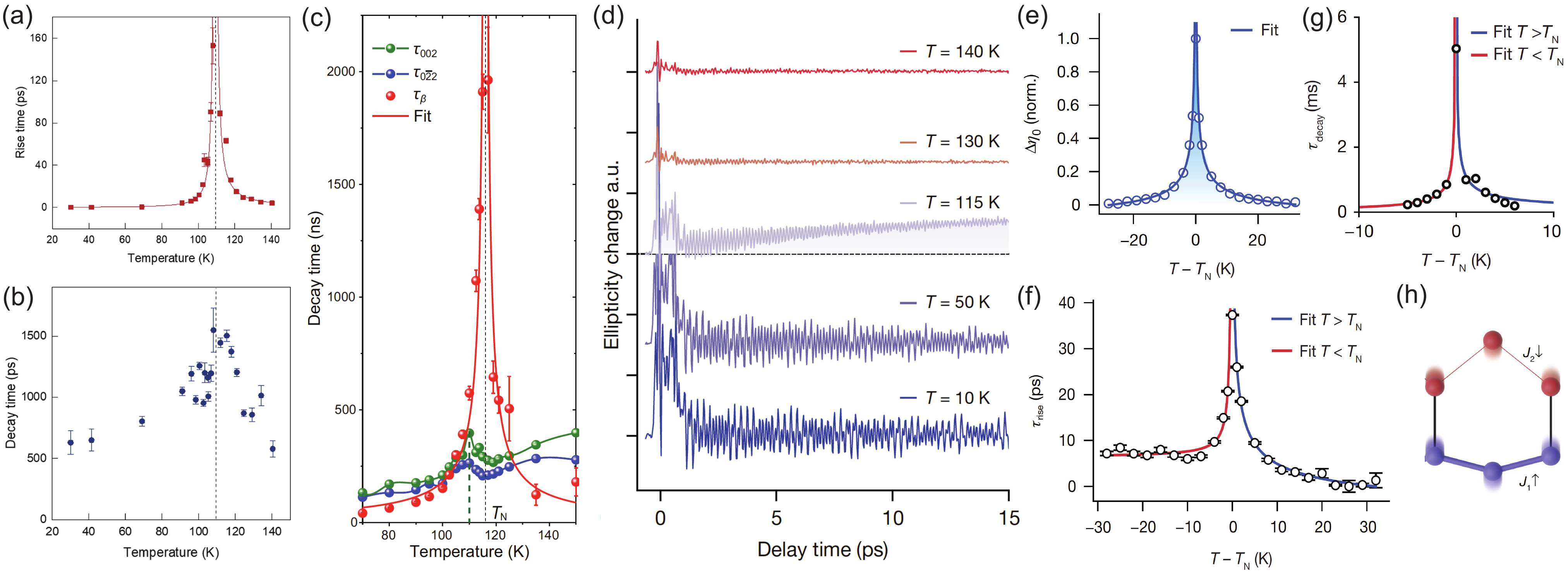}
    \caption{\label{fig:24}Critical behaviors in FePS$_3$. (a),(b) Fitted rise and decay time of polarization rotation dynamics upon a 3 eV pump as a function of temperature, with a divergence around $T_{\rm{N}}$. (c) Relaxation time of (0 0 2), (0 -2 2) peak shifts and monoclinic angle change upon a 3 eV pump as a function of temperature. (d) Polarization ellipticity change time traces upon a broadband THz pump at selected temperatures. (e)-(g) Fitted amplitude, rise time, and decay time of polarization ellipticity time traces in panel (d) as a function of temperature. (h) Schematic illustration of the eigenvector of the phonon that modulates exchange couplings and thus induces net magnetization. Bonds with enhanced and suppressed exchange interaction are shown by thick and thin lines, respectively. Red and blue spheres correspond to spin-up and spin-down Fe ions, respectively. All solid lines are power-law fit. Panels (a) and (b) are adapted from \citet{RN416}. Panel (c) is adapted from \citet{RN68}. Panels (d)-(h) are adapted from \citet{RN320}.}
\end{figure*}

The ability to coherently launch various collective modes using optical pulses, along with the complex interplay among these excitations and their interactions with light, has opened up a wealth of possibilities for driving nonthermal phase transitions. Intense optical stimuli can generate electronic, magnetic, or phononic excitations, steering the system into regions of its potential energy landscape that are otherwise inaccessible under equilibrium conditions \cite{RN342,RN417,RN418,RN419,RN420,RN421}. These advances have enabled a large portfolio of photoinduced phase transitions across structural \cite{RN422,RN423,RN424}, ferroelectric \cite{RN363,RN364}, magnetic \cite{RN426,RN427,RN428}, topological \cite{RN366,RN429,RN430,RN431}, charge density wave \cite{RN432,RN433,RN434,RN435}, superconducting \cite{RN360,RN361}, excitonic \cite{RN436,RN437}, orbital \cite{RN438,RN439,RN440}, and metal-insulator transitions \cite{RN441,RN442}, heralding a new frontier of dynamical and controllable quantum phase engineering. 

To this end, myriad light-induced dynamical phase transitions have been theoretically predicted and observed in a vast array of vdW magnets. For instance, photoexciting spin-orbit-entangled excitons in the antiferromagnetic insulator NiPS$_3$ can lead to a transient antiferromagnetic metallic phase \cite{RN86}. Transient magnetization accompanied by interorbital charge motion and weakening of antiferromagnetism is theoretically proposed and experimentally observed in the antiferromagnet RuCl$_3$ \cite{RN443,RN444,RN445}, with the possibility of accessing the proximate Kitaev quantum spin liquid phase \cite{RN446,RN447}. The topological antiferromagnet MnBi$_2$Te$_4$ was proposed to undergo a transition to a ferromagnet when nonlinearly driven by the displacement of a breathing $A_{1g}$ phonon \cite{RN371}. Furthermore, in heterostructures formed by an antiferromagnet and a ferromagnet or by two different ferromagnets with antiparallel magnetizations, rt-TDDFT simulations predict the development of transient ferrimagnetism during demagnetization processes via interlayer asymmetric spin or charge transfer \cite{RN448,RN449}.

A pressing challenge is to identify strategies to extend these nonequilibrium phases beyond a few picoseconds to metastable or permanently switched states, which have only been rarely observed \cite{RN362,RN365,RN369,RN370,RN440,RN442,RN450,RN451,RN452,RN453,RN454,RN455,RN456,RN457}. In this section, a few examples of optically induced metastable or permanent switching of magnetism in vdW magnets will be introduced.

\subsubsection{Metastable state}

A promising strategy to sustain a metastable phase involves operating near phase-transition temperatures where critical slowing down occurs. At these temperatures, the order parameters associated with competing phases remain unhardened, with significant fluctuations that allow external stimuli to effectively tip the balance between different energy scales in favor of a new ground state. In the Ising-type antiferromagnet FePS$_3$, pulses with high-energy photons that directly excite the electronic subsystem induce demagnetization through heating. Polarization rotation measurements, which directly track the dynamics of the magnetic order, reveal a power-law divergence in both the rise and decay times of the demagnetization at $T_{\rm{N}}$, extending to 100 ps and 1 ns, respectively [Figs.~\ref{fig:24}(a),(b)] \cite{RN416}. Similarly, time-resolved X-ray diffraction, which monitors the dynamics of lattice degree of freedom, shows that the recovery of interlayer shear --- the primary structural distortion coupled to magnetic ordering --- exhibits a critical slowing down at $T_{\rm{N}}$, reaching up to $\sim$microseconds [Fig.~\ref{fig:24}(c)] \cite{RN68}. Furthermore, recent optical characterization uncovers emerging ferromagnetism in FePS$_3$ interfaced with WS$_2$ layers due to a large interfacial exchange field \cite{RN458}. These results establish FePS$_3$ as an ideal platform for exploring potential metastable magnetism. Indeed, recent advances have demonstrated that pumping with intense broadband THz pulses can induce metastable magnetization, signified by circular dichroism emerging out of equilibrium [Figs.~\ref{fig:24}(d)] \cite{RN320}. Remarkably, this metastable state has lifetimes as long as $\sim$milliseconds. The amplitude and lifetime of the metastable state both achieve maxima at $T_{\rm{N}}$ with power-law divergence, suggesting the critical role of fluctuations in prolonging the lifetime [Figs.~\ref{fig:24}(e)-(g)]. Combined Monte Carlo and spin dynamics simulations further elucidate the microscopic mechanism underlying metastable magnetization: the displacement of a specific nonlinearly generated Raman-active phonon enhances the magnetic exchange interactions within one zigzag chain while weakening it in the neighboring one. This imbalance induces a finite out-of-plane net magnetization from the slightly perturbed antiferromagnetic motif [Fig.~\ref{fig:24}(h)]. These theoretical results confirm that the relaxation back to equilibrium follows critical slowing down close to the second-order phase transition, with the critical exponent consistent with a 3D Ising universality class. These findings provide a concrete benchmark vdW system to engineer magnetic phases with THz nonlinear phononics and establish regions near critical points with significant order parameter fluctuations as promising platforms for searching for metastable states.

\begin{figure*}
    \includegraphics[width=\linewidth]{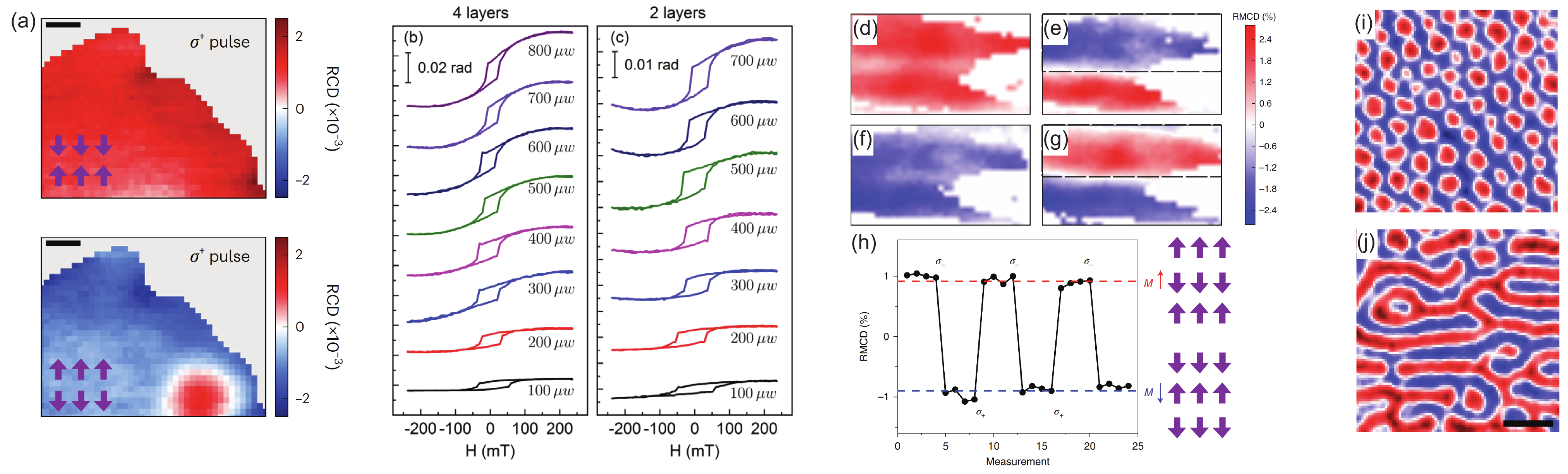}
    \caption{\label{fig:25}Optical switching of magnetism. (a) Reflective magneto circular dichroism (RMCD) mapping of opposite AFM domains of 8-layer MnBi$_2$Te$_4$ upon ultrafast left-handed circularly polarized 1.2 eV pulse excitation. The insets show the schematic of the A-type AFM structure. (b),(c) Magneto optical Kerr rotation (MOKE) as a function of magnetic field induced and detected by 3.1 eV ultrafast pulses with various fluences in 4-layer and 2-layer Fe$_3$GeTe$_2$. (d)-(g) RMCD mapping at zero field for trilayer CrI$_3$ in spin-up or spin-down states before [(d), (f)] and after [(e), (g)] circularly polarized pulsed light exposure at 2.03 eV in areas enclosed by the dashed box. (h) Repeated switching measurements with the schematic of the A-type AFM structure showing on the right. (i),(j) Wide-field Kerr imaging of spin textures in CrGeTe$_3$ after 100 and 10 pulsed laser irradiation, respectively. The scale bar is 5 $\mu$m. Panel (a) is adapted from \citet{RN459}. Panels (b) and (c) are adapted from \citet{RN460}. Panels (d)-(h) are adapted from \citet{RN461}. Panels (i) and (j) are adapted from \citet{RN462}.}
\end{figure*}

\subsection{Optical switching of magnetic phases}

Beyond static tuning knobs such as external magnetic fields \cite{RN39,RN177}, electric fields \cite{RN463,RN118}, strain \cite{RN464,RN465}, pressure \cite{RN466,RN120}, and electrostatic doping \cite{RN117}, optical pulses offer a dynamic route for manipulating vdW magnetism. This capability, known as all optical switching (AOS), holds promise for non-volatile magneto-optical memory devices with ultrafast data storage and processing functionalities. Light-enhanced ferromagnetism was observed in the few-layer ferromagnetic metal Fe$_3$GeTe$_2$ using linearly polarized pulses \cite{RN460}. MOKE probes an increase in net magnetization and $T_{\rm{C}}$, accompanied by a quenching of coercivity, with increasing light intensity [Figs.~\ref{fig:25}(b),(c)]. The enhancement of $T_{\rm{C}}$ and exchange arises from light-excited holes that downshift the Fermi level, increasing the density of states that strengthens the Stoner effect. Simultaneously, the uniaxial magnetic anisotropy decreases, reducing the coercivity. In this case, an ultrafast laser is required to generate enough carriers per pulse, and time-resolved measurements show that the ferromagnetism develops within $\sim$100 ps and persists beyond $\sim$1 ns.

Intuitively, it should be possible to reverse magnetic orders in ferromagnets using circularly polarized pulses, as they naturally break time reversal symmetry, allowing linear coupling to magnetic orders. However, such reversal was only achieved in a heterostructure composed of a thick film of ferromagnetic CrI$_3$ interfaced with WSe$_2$. The experiments first showed the wide tunability of the valley polarization ($\pm40\%$) and the valley Zeeman splitting (effective exchange field over 20 T) in WSe$_2$ by varying the optical excitation power over an order of magnitude \cite{RN467}. Such tunability stems from flipping of the magnetization of the adjacent layer CrI$_3$. This was later corroborated by Kerr imaging of CrI$_3$: optical excitation leads to intrinsically helicity-dependent partial and reproducible switching between up and down spin states only in regions where CrI$_3$ interfaces with WSe$_2$, suggesting the essential role of interfacial spin-polarized charge transfer \cite{RN468}. Surprisingly, linearly polarized light can also be used to toggle between the spin state, due to the imbalance in lifetime and population of spin-polarized electrons in the two valleys in the heterostructure. Later, a breakthrough was achieved with a reversible deterministic AOS of ferromagnetism in bare atomically thin CrI$_3$ layers, triggered by circularly polarized pulses [Figs.~\ref{fig:25}(d)-(h)]. The strong dependence on photon energy and polarization further indicates that this process is related to angular momentum transfer from the photoexcited carrier spins to the local moments of the valence Cr $d$ orbitals, rather than to mechanisms such as the inverse Faraday effect, where circularly polarized light induces a magnetic field that flips the magnetization, or the thermal process of the magnetic circular dichroism, where different domains exhibit helicity-dependent absorption \cite{RN461}. 

Surprisingly, selective AOS has also been realized in a fully compensated A-type antiferromagnet. In even-layer MnBi$_2$Te$_4$ with zero net magnetization, circularly polarized light induces circular dichroism exclusively in reflection but not in transmission \cite{RN459}. Such circular dichroism originates from the quantum geometric properties of the material, namely axion magnetoelectric coupling. Experiments further showed that unlike continuous laser irradiation, which deterministically induces antiferromagnetic order upon cooling down from the paramagnetic phase, ultrafast pulses enable direct and reversible switching of magnetic domains without the need for thermal cycling [Fig.~\ref{fig:25}(a)].

Because of the existence of magnetic domains, optical pulses can also create nontrivial spin textures with particular topology or chirality. Wide-field Kerr microscopy demonstrates the formation of magnetic bubbles and stripes upon exposure to different numbers of pulses in the vdW ferromagnet CrGeTe$_3$, which stems from coalescence or isolation of magnon droplets after heating-induced demagnetization [Figs.~\ref{fig:25}(i),(j)] \cite{RN462}. Moreover, magnetic bubbles can be switched to stripe domains, while the reversed transition requires the application of external fields. Spin dynamics simulations also predict the emergence of skyrmions and antiskyrmions in this material and merons and antimerons in another ferromagnet CrCl$_3$, pointing to a path for stable, reproducible, and reversible topological magnetic switching \cite{RN469}. 

These results highlight the diverse mechanisms and functionalities of optical control in vdW magnets, offering a versatile platform for ultrafast magneto-optical applications. By enabling deterministic, reversible engineering of magnetic and topological states, AOS in vdW magnets advances prospects for next-generation spintronic technologies.

\subsection{Floquet engineering}

Floquet engineering has emerged as a promising approach for controlling microscopic interactions in solids, resulting in changes to macroscopic properties or even entirely new phases of matter. Several excellent reviews have discussed the principles of Floquet engineering and applications to a variety of material systems \cite{RN418,RN419,RN420,RN470,RN471}. Here, we focus exclusively on vdW magnets whose high crystalline qualities and tunability make them a promising playground for Floquet engineering.

\subsubsection{Theory}

\begin{figure*}
    \centering
    \includegraphics[width=\linewidth]{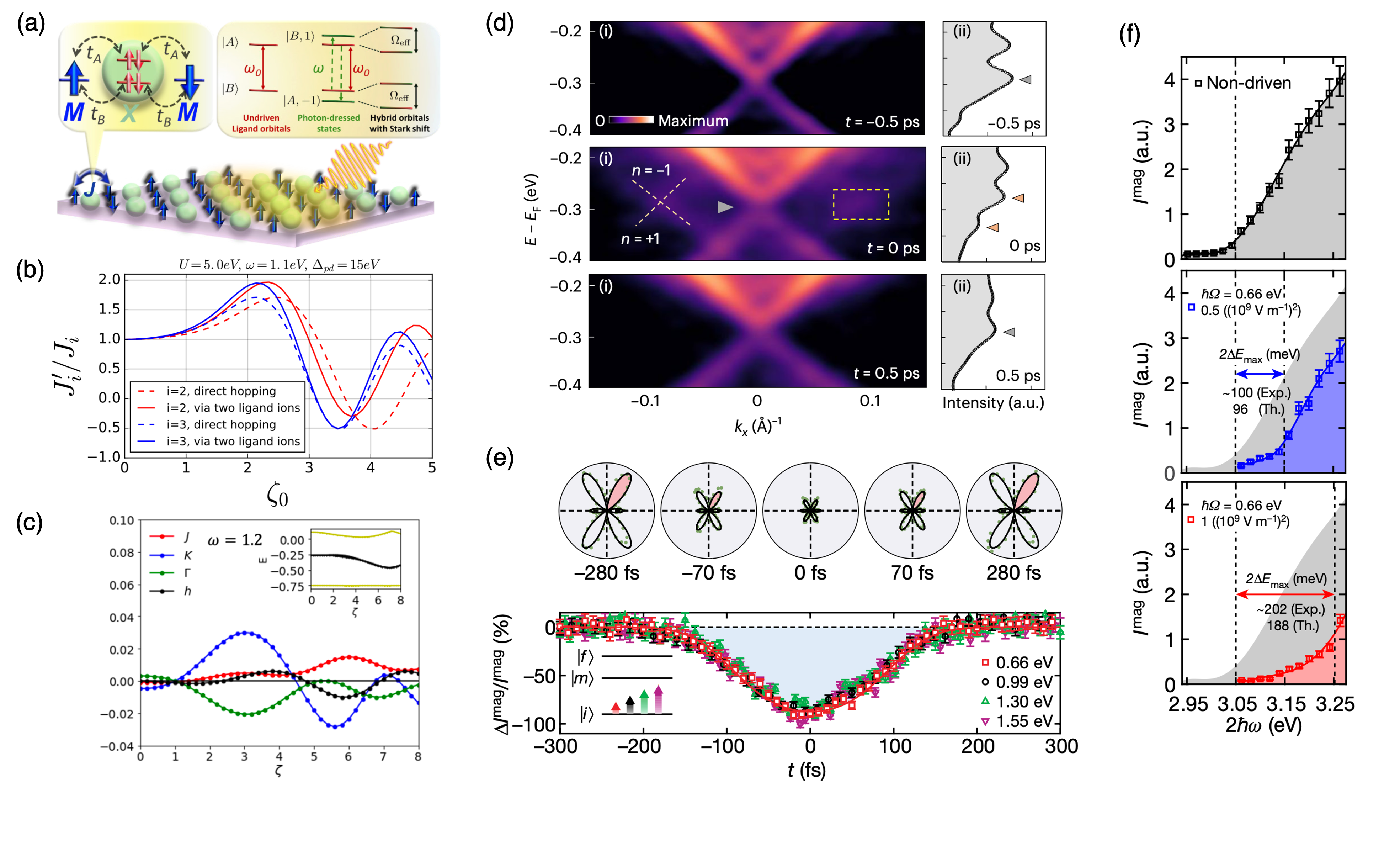}
    \caption{\label{fig:26}Floquet engineering. (a) Schematic diagram of the Floquet engineering of spin exchange interactions in 2D transition metal trichalcogenides mediated by ligand orbitals. Left: Virtual hopping of electrons between the magnetic ion  ($M$) and ligand ion ($X$). Right: Under a periodic drive, the ligand orbitals are replaced by hybridized photon-dressed orbitals, which shifts the energy levels of virtual excitations and thereby modifies the exchange interactions. (b) Changes in magnetic coupling strength for second and third nearest neighbor as a function of drive parameter $\zeta_0=eEa/\omega$ for a periodically driven Fermi-Hubbard model on a honeycomb lattice. (c) Spin exchange couplings $J$, $K$, and $\Gamma$ and emergent magnetic field $h$ as a function of drive strength $\zeta$ for a multi-orbital strongly spin orbit coupled Hubbard model. (d) (i) Tr-ARPES spectra of MnBi$_2$Te$_4$ for various pump-probe delay times $t$ at 8~K. Crystal momentum $k_x$ is along the $M$-$\Gamma$-$M$ direction. Sidebands are indicated by the dashed lines and the induced Dirac gap is indicated by the gray triangle for $t=0$~ps. (ii) Energy distribution curves along the Dirac point ($k=0$) for the spectra shown in (i). The triangles indicate the position of the peaks closest to the Dirac point. (e) Top: Time-resolved rotational anisotropy SHG patterns of MnPS$_3$ at 10~K using a subgap drive with a photon energy of 0.66~eV and electric field strength of $10^9$~V/m. The black curves are fits to a Floquet model. Bottom: Change in SHG intensity $\Delta I/I$ as a function of time for different pump photon energies. (f) Top: Non-driven SHG spectrum at 10~K. Middle and Bottom: SHG spectra for different pump intensities compared to the non-driven spectrum (gray shaded region). Panel (a) is from \citet{RN472}. Panel (b) is from \citet{RN473}. Panel (c) is from \citet{RN474}. Panel (d) is from \citet{RN475}. Panels (e),(f) are from \citet{RN75}.}
\end{figure*}

There have been many theoretical proposals demonstrating the Floquet engineering of magnetic exchange interactions. One theory work \cite{RN472} investigated 2D transition metal trichalcogenides, whose spin interactions are mediated by nonmagnetic ligand ions. They showed that the superexchange interactions can be modified by using a periodic Floquet drive to manipulate the orbital degrees of freedom of the ligand ions. Figure~\ref{fig:26}(a) depicts the process of the ligand orbitals being replaced by hybridized photon-dressed orbitals during a periodic drive, which shifts the energies of the virtual excitations involved in the exchange process. Another theoretical work \cite{RN473} considers realistic models of 2D magnetic materials and provides a framework for Floquet engineering of the exchange interaction via changes in the hopping parameter due to photon-assisted tunneling and virtual excitations between different Floquet sectors. The drive-induced changes to the spin exchange interactions between nearest-neighbors and further neighbors differ as a result of different bond angles and the number of ligand ions involved in each exchange process. Since different magnetic ground states are predicted to be realized under different ratios of nearest- to further-neighbor exchange, Floquet engineering was proposed to be a route to dynamically tune between them [Fig.~\ref{fig:26}(b)]. In addition to modifying the exchange, it was shown that pumping on resonance with orbital transitions between crystal field split levels of magnetic ions can also dynamically alter the magnetic anisotropy in the transition metal trichalcogenide NiPS$_3$ (see the Coherent Magnon Generation section for further details), further broadening the utility of Floquet engineering for designing magnetic Hamiltonians.  

Other theoretical works have proposed that the Floquet engineering of vdW Kitaev quantum magnets can be used to realize a quantum spin liquid state \cite{RN474,RN476}. These studies consider multi-orbital Mott insulators with strong spin-orbit coupling, in particular ruthenates and iridates. The application of circularly polarized light is shown to modify the magnitudes and signs of the three relevant exchange interactions $J$ (Heisenberg exchange), $K$ (Kitaev exchange) and $\Gamma$ (symemtric off-diagonal exchange) in these systems, as shown in Fig.~\ref{fig:26}(c). In certain cases, they can be tuned so that $K$ is the dominant spin interaction, thereby providing conditions favorable to the stabilization of the Kitaev quantum spin liquid phase.

In topological magnetic systems, Floquet engineering has been proposed to induce exotic topological phenomena. In one theoretical study \cite{RN477}, driving a 2D vdW honeycomb ferromagnet with circularly polarized light is shown to induce an optically tunable synthetic scalar spin-chirality interaction that can stabilize topological magnons. They find that the driven system can be described by a magnon Haldane model with a topological gap and chiral magnon edge states. Applying this theory to monolayer CrI$_3$ predicts a gap of $\sim2$~meV in the magnon spectrum for an experimentally realistic electric field strength of $10^9$~V/m and a photon energy of 1~eV for the Floquet drive that can induce non-zero Chern numbers and chiral magnon edge states.

Another theoretical work investigated the magnetic topological insulator MnBi$_2$Te$_4$ using Floquet engineering \cite{RN478}. MnBi$_2$Te$_4$ is an antiferromagnet consisting of septuple layers (SL) coupled by weak vdW interactions. Odd-SL MnBi$_2$Te$_4$ films host a quantum anomalous Hall (QAH) state, while even-SL MnBi$_2$Te$_4$ films exhibit an axion insulator state with a zero Hall plateau. This work demonstrates using a combination of first principles and Floquet theory that circularly polarized light can induce topological phase transitions in MnBi$_2$Te$_4$. For odd-SL films, the right circularly polarized light can drive the QAH state into a normal insulator and then to another QAH state with a reversed sign of the Chern number; on the other hand, the left circularly polarized light enhances the band gap by an amount that scales with the amplitude of the driving light. For even-SL films, circularly polarized light can tune the axion insulator state to a QAH state.

\subsubsection{Experiment}

On the experimental side, Floquet-Bloch engineering of the band structure of MnBi$_2$Te$_4$ has been demonstrated using time- and angle-resolved photoemission spectroscopy (tr-ARPES) \cite{RN475}. The experiment used circularly polarized mid-IR 155~meV pump pulses that lie below the bulk band gap of 200~meV. Snapshots of the tr-ARPES spectra as a function of pump-probe delay reveal photon-dressed replicas of the original Dirac cone as well as a gap opening at the Dirac point [Fig.~\ref{fig:26}(d)]. Above $T_N$, opposite helicities of the circularly polarized pump yield the same gap magnitude, while below $T_N$ they induce substantially different Dirac mass gaps. This can be explained by a non-uniform Dirac mass on the surface of MnBi$_2$Te$_4$.

However, several challenges have precluded the realization of Floquet engineering of magnetic exchange interactions. To produce sizeable Floquet effects, the electric field of the driving laser pulse must be sufficiently large. However, strong driving is usually accompanied by runaway heating and decoherence, which is detrimental to the observation of Floquet-induced changes. An emerging strategy is to drive magnetic insulators at frequencies far below their charge gap to minimize linear and nonlinear absorption. An experimental study on the vdW antiferromagnet MnPS$_3$ \cite{RN75} showed that heating could indeed be minimized with subgap driving even at peak fields up to 10$^9$ V/m. In this strong coherent driving regime, they showed using time-resolved SHG that the bandgap is widened by up to 10$\%$ as shown in Figs.~\ref{fig:26}(e),(f). By choosing an SHG photon energy that resonates with the band gap of MnPS$_3$, they further demonstrated that the SHG efficiency can be coherently tuned by a factor of 10 based on the modulated band gap. Floquet theory calculations using a single-ion picture reproduces the experimental results and confirms that the amount of band gap widening is proportional to the electric field of the Floquet drive.

\section{\label{sec:V}Spintronics}

\subsection{Tunable magnetic properties by diverse multi-field approaches}

\begin{figure*}
    \centering
    \includegraphics[width=\linewidth]{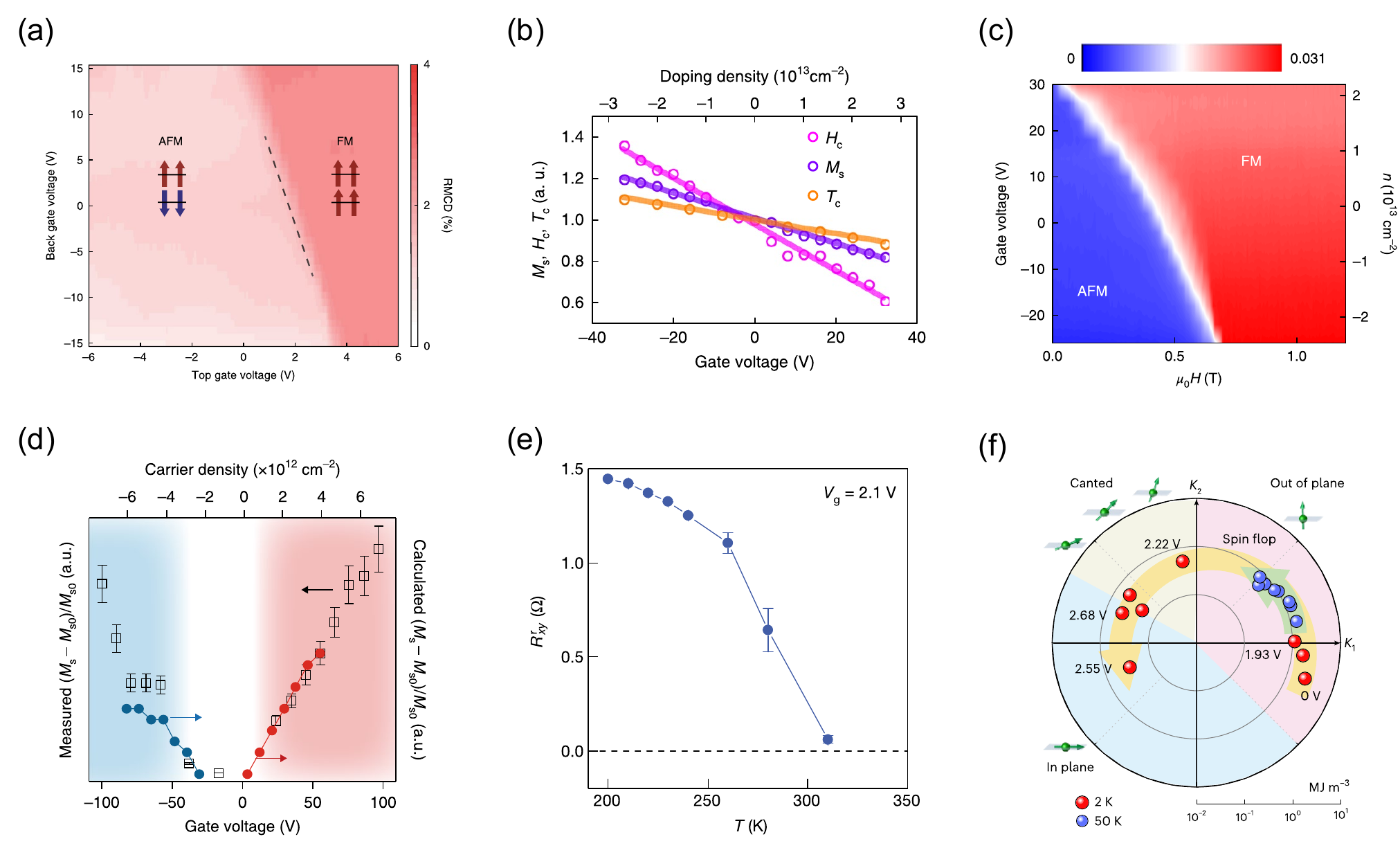}
    \caption{\label{fig:27}Gating-tuned magnetism in several classic vdW magnets. (a) Top and bottom gate voltage mapping of the RMCD in bilayer CrI$_3$ under a fixed magnetic field, where an antiferromagnetic (AFM) to ferromagnetic (FM) transition can be realized via gating. (b) Linear response of \( M_s \), \( H_c \), and \( T_c \) versus gate voltage (or equivalently doping density) in monolayer CrI$_3$. (c) Gate voltage and magnetic field mapping of the magnetic signal in bilayer CrI$_3$, where the AFM to FM transition is likely to be realized under zero magnetic field by applying a large gate voltage. (d) Enhancement of magnetization in few-layer Cr$_2$Ge$_2$Te$_6$ by both positive and negative gate voltages. (e) \( R_{xy} \) versus \( T \) under an ionic gating voltage of \(2.1\,\mathrm{V}\) for a four-layer Fe$_3$GeTe$_2$, whose Curie temperature is significantly boosted to \(310\,\mathrm{K}\). (f) Continuous magnetic anisotropy tuning from out-of-plane to in-plane easy axes by ionic gating for Fe$_5$GeTe$_2$ nanoflakes. Figs.~(a--f) are from ~\citet{RN118}; Fig.~2c in ~\citet{RN117}; Fig.~3b in ~\citet{RN117}; Fig.~5d in ~\citet{RN103}; Fig.~4b in ~\citet{RN40}; and Fig.~4a in ~\citet{RN479}, respectively.}
\end{figure*}

Magnetic vdW materials naturally inherit the high tunability of 2D vdW materials- the most generic advantages of the huge 2D layered materials family. In 2017, the MOKE technique discovered intrinsic ferromagnetism in few-layer CrI$_3$ and CrGeTe$_3$. Soon after, in 2018, three independent groups performed the gating experiment to modulate the magnetic properties of few-layer CrI$_3$: Huang $et$ $al.$ employed the electrostatic gate control technique and magneto-optical Kerr effect (MOKE) microscopy to modulate and detect the magnetization in bilayer CrI$_3$ \cite{RN118}. They discovered the gate-controlled transition between antiferromagnetism and ferromagnetism at finite and zero magnetic fields and found the time-reversal pair with spin-layer locking and consequent linear dependence of MOKE signals on gate voltage with opposite slopes [Fig. \ref{fig:27}(a)]. Jiang $et$ $al.$ combined both high top-gate and back-gate \cite{RN117} and successfully controlled the magnetism with hole/electron doping in monolayer CrI$_3$ [Fig. \ref{fig:27}(b)]. Additionally, an antiferromagnetic to ferromagnetic transition was realized for a bilayer CrI$_3$ by a high electron doping above $\sim$2.5$\times$10$^{13}$ cm$^{-2}$ [Fig. \ref{fig:27}(c)]. Wang $et$ $al.$ applied gating on a few-layer CrGeTe$_3$ \cite{RN103}. They demonstrated a bipolar tunable magnetization by both hole and electron doping due to the moment rebalance in the spin-polarized band structure [Fig. \ref{fig:27}(d)]. These three works unambiguously demonstrate the gating tuneability of 2D materials and open the door to novel optoelectronic and magnetic devices where an electric field radically and controllably changes the material from having no/small net magnetization to a large one \cite{RN119}.

Compared to the ferromagnetic insulator CrI$_3$, vdW ferromagnetic metal, e.g., Fe$_3$GeTe$_2$ (FGT), is much more difficult to exfoliate into nanoflakes due to the strong metal bonds within the layers and probably a greater interlayer coupling than CrI$_3$. In 2020, Deng \textit{et al.} developed a new Al$_2$O$_3$-assisted exfoliation technique \cite{RN40}, allowing obtaining the few-layer FGT even down to the monolayer. Moreover, the ionic gating technique was applied to a four-layer FGT. It significantly enhanced its Curie temperature to $\sim$310 K above room temperature [Fig. \ref{fig:27}(e)], even higher than bulk FGT's Curie temperature of $\sim$205 K. Another interesting gating experiment was reported for a pristine room-temperature ferromagnetic metal Fe$_5$GeTe$_2$ in 2023. Tang \textit{et al.} employed the same ionic gating approach and discovered that magnetic anisotropy can be continuously modulated from a perpendicular to an in-plane anisotropy [Fig. \ref{fig:27}(f)] via a spin-flop pathway \cite{RN479}. These electrostatic or ionic gating experiments \cite{RN118,RN117,RN103,RN119,RN40,RN479,RN480} generally manipulate the carrier density, shift the Fermi level, and consequently tune the magnetic interactions in the systems, providing a unique mediating tool for layered magnets.

Besides the generic gating tuning with 2D few-layer devices, the inherent high-tuneability merit also allows one to modulate the physical properties, particularly the magnetic characteristics, using multiple external physical fields. For example, light illumination can increase saturated magnetization and reduce the coercivity of FGT, and even raise its Curie temperature from 200 to 300 K due to the optical doping effect and the consequent change in electronic structure \cite{RN460}. High pressure could also modulate the magnetization and corresponding anomalous Hall conductivity in bulk FGT crystal \cite{RN146}. In addition to the high-pressure technique, Wang \textit{et al.} has developed an effective strain setup by adopting the bending geometry to apply the extensible strain to FGT \cite{RN481}. Using their techniques, they could considerably modulate the Curie temperature and coercive field through this applied strain field. These diverse scientific cases demonstrate huge opportunities for regulating magnetic properties in the bulk, few-layer, and monolayer vdW magnets by different methods.

\subsection{Giant spin filter effect and its gate-tunable tunneling}

\begin{figure*}
    \centering
    \includegraphics[width=\linewidth]{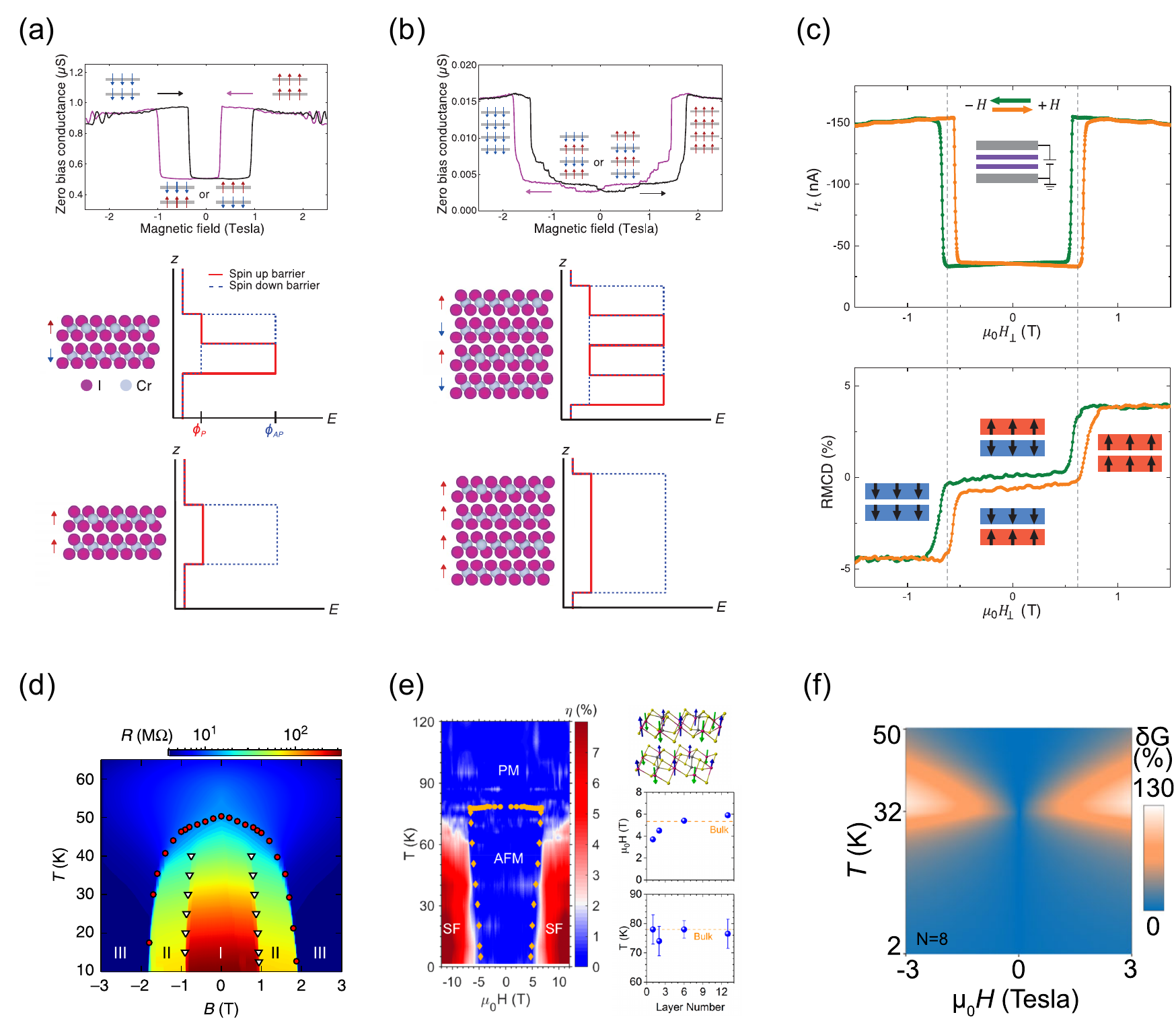}
    \caption{\label{fig:28}Spin filter effect in several classic vdW magnets. (a) Zero-bias conductance versus magnetic field for bilayer CrI$_3$, where a giant spin filter effect is observed depending on the antiparallel and parallel spin configurations in CrI$_3$, leading to different tunneling barriers. (b) Same as (a) but for a four-layer CrI$_3$. (c) Consistent magnetic field dependence of the tunneling current and the RMCD for a bilayer CrI$_3$. (d--f) Tunneling magnetoconductance for CrI$_3$ (d), MnPS$_3$ (e), and CrBr$_3$ (f), reflecting spin-flip, spin-flop, and critical fluctuations, respectively. Figs.~6.2(a--f) are reproduced from Fig.~2A--C in ~\citet{RN51}; Fig.~2D--F in ~\citet{RN51}; Fig.~2A--B in ~\citet{RN50}; Fig.~4a in ~\citet{RN116}; Fig.~TOC in ~\citet{RN482}; and Fig.~2a in ~\citet{RN483}, respectively.}
\end{figure*}

The spin filter effect is usually studied using a metal/ferromagnetic-insulator/metal sandwiched structure. In such a magnetic tunnelling junction, the conduction carrier with specific spin polarization can be selected depending on the spin direction of the ferromagnetic insulator spacer. As a representative vdW ferromagnetic insulator, CrI$_3$ is suitable for such a spin filter device in several aspects, as detailed below [Fig. \ref{fig:28}(a-c)].

For instance, two groups \cite{RN51} \cite{RN50} studied this case independently at the very beginning of this research field. When the tunneling carrier passes through the junction from the top metallic graphene to the bottom metallic graphene, the carrier with different spin polarizations sees a low/high barrier if the carrier spin aligns parallel/antiparallel to the magnetization of the ferromagnetic insulator CrI$_3$. For bilayer CrI$_3$, the neighboring two layers have spins aligned antiparallel in the A-type antiferromagnetic configuration and thus, in principle, host the lowest tunneling current \cite{RN51,RN50}.

When the spin configuration of the ferromagnetic insulator can be tuned by external fields/perturbations, the spin filter effect will also be changed correspondingly. For example, in the bilayer CrI$_3$ case with an out-of-plane magnetic field, the spin configurations change from parallel to antiparallel and then to parallel again while sweeping the magnetic field, and so does the filtered spin-dependent tunneling current \cite{RN51,RN50}. This spin filter effect modulated by the external magnetic field also applies to the general even/odd-layered CrI$_3$ \cite{RN51,RN50}. Please note that the spin filter process is extremely efficient in CrI$_3$, impressively reaching a giant record ratio of 19,000 \% \cite{RN50}. More interestingly, solid gating could even tune the even-layer CrI$_3$ from antiferromagnetic to ferromagnetic states and thus tune the spin filter effect accordingly. From another perspective, such a spin filter effect can be an effective electrical tool to detect the spin information of nanoscale magnetic insulators beyond the optical probes such as nano-MOKE, etc.

The ideal spin-filter effect above can be generalized to the tunneling spectrum of magnetic insulators, offering a unique transport probe to study diverse magnetic insulators. One group did a series of works along this direction in a systematic manner [Fig. \ref{fig:28}(d-f)] starting from 2018 as the above two papers, including the very beginning work of spin-flip transitions in few-layer CrI$_3$ \cite{RN116}, i.e., the anisotropic layered antiferromagnet. They discovered the spin-flop transition from antiferromagnetic ground states by a magnetic field in the vdW magnet MnPS$_3$ \cite{RN482}. Such phenomena were universally demonstrated from bulk to monolayer, indicating the persistence of magnetism down to the monolayer in this intralayer antiferromagnet of the isotropic Heisenberg Hamiltonian \cite{RN482}. They also observed phase boundaries induced by critical fluctuations in the ferromagnet CrBr$_3$, reproduced by the spin-dependent Fowler-Nordheim model for tunneling, and is a direct manifestation of spin splitting of the CrBr$_3$ conduction band \cite{RN483}. Due to all these efforts, together with many works unmentioned, it is clear now that tunneling spectroscopy establishes an effective electrical way to investigate the nanoscale vdW magnetic insulators, adding more flexibilities and opportunities beyond the specific optical approaches.

\subsection{Magnon Transport}

\begin{figure*}
    \centering
    \includegraphics[width=\linewidth]{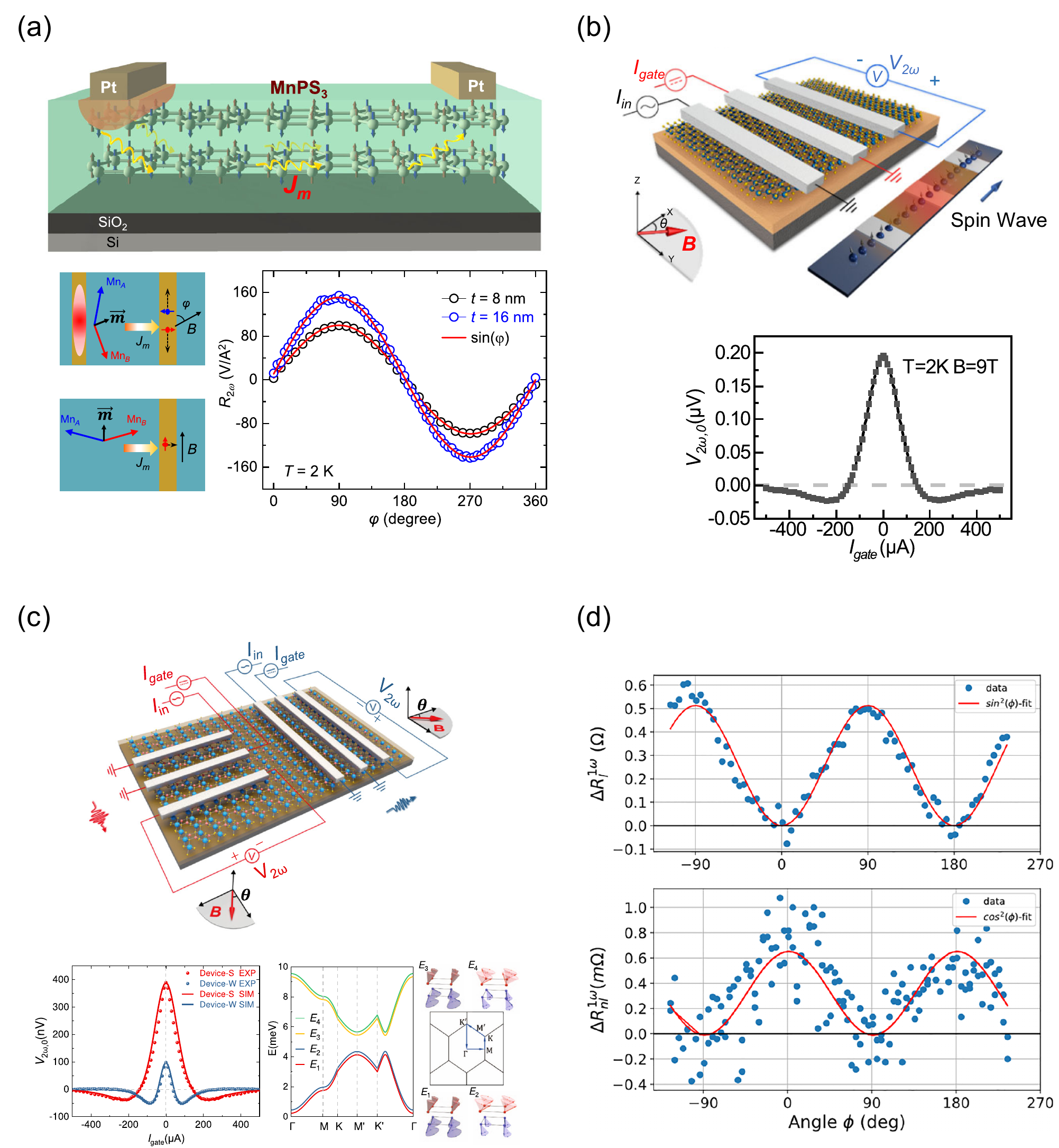}
    \caption{\label{fig:29}Magnon transport in magnetic vdW insulators. (a) Illustration of the measurement schematic for MnPS$_3$ magnon transport and the nonlocal SHG measurement result. (b) Schematic of gating control of MnPS$_3$ magnon transport. The $V_{2\omega}$ signal is first sharply reduced to a negative value and then restored to zero under a gating current. (c) Measurement schematic for anisotropic magnon transport in CrPS$_4$, which is intimately related to its anisotropic magnon dispersion and Seebeck effect. (d) The local and nonlocal first harmonic signal for CrPS$_4$ magnon transport under a large magnetic field. Figs.~6.3(a--d) are reproduced from ~\citet{RN489}; Fig.~1c and Fig.~3a from~\citet{RN490}; Fig.~1c and Fig.~2a--b from~\citet{RN491}; and Fig.~3 from~\citet{RN492}, respectively.}
\end{figure*}

In addition to vertical tunneling spectroscopy, one can also investigate magnetism by in-plane electrical transport. However, for vdW magnetic insulators, in-plane electrical transport is not ideal as a result of badly behaved electronic conduction determined by their large bandgap and intrinsic insulator characteristics. In the literature, there are only a few attempts in this direction. One group performed in-plane transport measurements in multilayer CrPS$_4$ and found spin-flip and spin-flop transitions in its magnetotransport and gate-controlled electrostatic modulation \cite{RN484}. Anyway, in a general sense, in-plane electrical transport is not a well-suitable method for investigating magnetic insulators compared to tunneling spectroscopy.

Fortunately, the magnetic insulator is ideal for another kind of spintronic study, magnon transport \cite{RN485,RN486,RN487,RN488}. In a classic protocol, the magnon transport consists of magnon generation, diffusion, and detection. At the generation end, the magnon density imbalance is produced by a current flowing through a platinum electrode via the spin Hall effect and/or thermal activation. Then, the accumulated magnon will diffuse far away toward the detection end. The transverse voltage arising from the inverse spin Hall effect will be measured at the detection end. In 2019, Xing \textit{et al.} reported the first magnon transport experiment on a vdW magnetic insulator \cite{RN489}. They used A.C. current excitations and nonlocal measurement geometry as described above and observed long-distance magnon transport in MnPS$_3$. As the thickness of MnPS$_3$ nanoflakes decreases from 40 to 8 nm, the magnon diffusion length monotonically reduces from 4.7 to 1~$\mu$m. Meanwhile, the magnon diffusion length increases at lower temperatures [Fig. \ref{fig:29}(a)] \cite{RN489}. Interestingly, only the SHG signal is observed, while the first harmonic signal is absent. Since the first harmonic response comes from the spin Hall effect, while the second benefits from the thermal activation, the magnon signal detected in MnPS$_3$ is dominantly from the thermal activation by the current-driven Joule heating effect at the end of the generation. Similar phenomena were reproduced by another work on MnPS$_3$ \cite{RN490}, and also in another vdW antiferromagnetic insulator CrPS$_4$ \cite{RN491}. However, very recently, in a 100~nm-thick CrPS$_4$, de Wal $et$ $al.$ focused on the first harmonic signal and observed an abrupt increase of the nonlocal spin signal over distances exceeding 1~$\mu$m under an in-plane magnetic field above 6~T \cite{RN492} [Fig. \ref{fig:29}(d)]. Since there are some differences in the sample status, especially in the thickness range, it is not yet easy to unify a consistent picture. Obtaining a uniform and reasonable understanding of this issue requires further exploration and clarification.

On the other hand, Chen \textit{et al.} did two main systematic works on both MnPS$_3$ \cite{RN490} and CrPS$_4$ \cite{RN491}, and reported very instructive results. First, they applied a D.C. current to a second Pt electrode near the end of the generation to find that the detected SHG signal can be reduced even to a negative signed value and then recover further in MnPS$_3$ [Fig. \ref{fig:29}(b)]. The D.C. current flowing in the second-generation electrode works as a gate, in analogy to the gate technology of the field-effect transistor. They discussed the change in thermal gradient caused by the two-generation currents, which combined with the Seebeck coefficients to account for this observation \cite{RN490}. Second, in another work on CrPS$_4$, they discovered giant anisotropy in its magnon transport and explained this behavior from the anisotropies in magnon dispersions \cite{RN491}. This magnon dispersion perspective provides useful information that one may control the magnon transport by tuning/selecting different magnon bands [Fig. \ref{fig:29}(c)]. In this way, the magnon transport research field can be significantly expanded in analogy to the abundant essential concepts of electronic bands such as flat band, band dispersion/velocity/mass, and topological band crossings, to name only a few.

There have been many interesting theoretical predictions on magnon transport recently. For example, several groups did a series of works focusing on this topic, including magnon accumulation in chirally coupled magnets \cite{RN93}, magnon trap by chiral spin pumping \cite{RN494}, spin-wave Doppler shift by magnon drag \cite{RN495}, chiral-damping-enhanced magnon transmission \cite{RN496}, and giant magnon transport enhancement via the superconductor Meissner screening \cite{RN497}. In contrast, the experimental reports are rather few, especially on the magnetic vdW materials, which may require a large effort in the future.

\subsection{Large spin valve effect and the modulation of tunnelling magnetoresistance}

\begin{figure*}
    \centering
    \includegraphics[width=\linewidth]{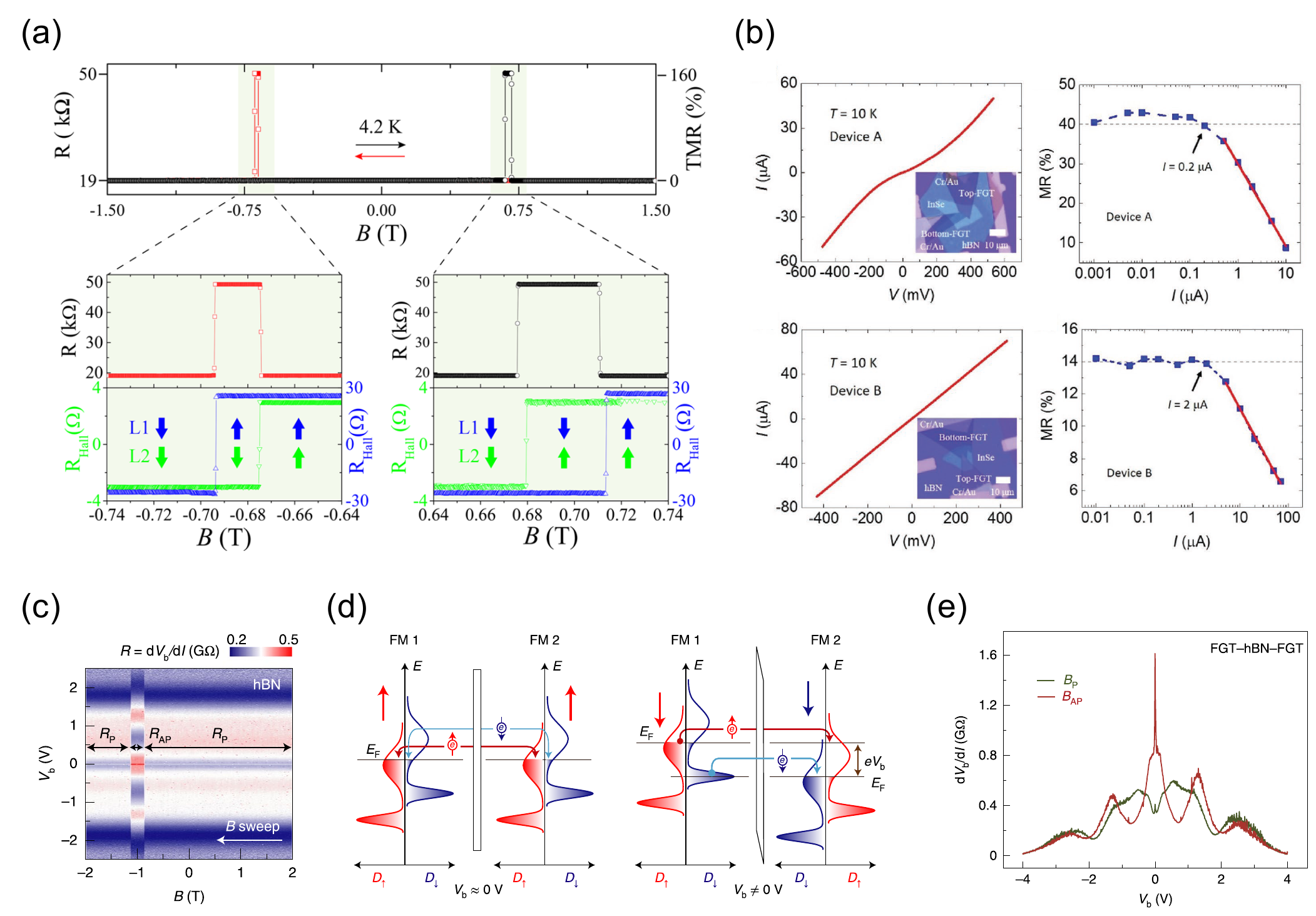}
    \caption{\label{fig:30}Large spin valve effect and the modulation of tunneling magnetoresistance using magnetic vdW metals. (a) Large TMR in the FGT/hBN/FGT heterostructure. The parallel (antiparallel) spin configuration gives a low (high) tunneling resistance plateau. (b) Two types of FGT/InSe/FGT heterostructures with nonlinear (Device A, pinhole-free) and linear (Device B, pinhole) I--V curves. In general, the TMR ratio increases while lowering the tunneling current. However, Device A hosts a higher TMR ratio with a slightly monotonic change at low tunneling current, whereas Device B shows a saturation for low tunneling current. (c) Bias and magnetic field mapping of TMR for the FGT/hBN/FGT heterostructure, where both positive and negative TMR ratios are observed. (d) Spin-polarized bands and Fermi level alignment without and with a bias. With a bias, negative TMR is possible due to Fermi level shift and high spin density of states. (e) TMR varies significantly depending on the bias voltage, so the resistance for BP and BAP oscillates and shows positive/negative TMR upon bias. Figs.~6.4(a--e) are reproduced from ~\citet{RN157}; Fig.~2a,d and Fig.~3c,d from ~\citet{RN159}; Fig.~2a from ~\citet{RN156}; Fig.~1b,c from~\citet{RN156}; and Fig.~3a from~\citet{RN156}, respectively.
}
\end{figure*}

Unlike the spin-filter effect, the spin valve effect employs a sandwiched structure of magnetic-metal/non-magnetic-insulator-spacer/magnetic-metal. In a conventional scheme, the magnetic-metal layer is adopted from a ferromagnetic metal, and the magnetic tunneling junction can be denoted as free-layer-ferromagnet/insulator/pinning-layer-ferromagnet. The magnetization of the free-layer ferromagnet can be changed by external stimuli such as a magnetic field, while the pinning-layer ferromagnet remains unchanged in its ferromagnetic magnetization direction. When current flows through the free-layer ferromagnet, the carrier's spin will be polarized parallel to the magnetization direction of the free-layer ferromagnet. This spin-polarized current further tunnels through the insulating spacer, which physically separates the free-layer and pinning-layer ferromagnet, and finally passes through the pinning-layer ferromagnet. Depending on whether the spin polarization of the tunneling current, i.e. the free layer's magnetization direction is parallel/antiparallel to the pinning-layer's magnetization orientation, the tunneling carrier would see a low/high barrier considering the spin-dependent scattering. Such a process would lead to the low/high tunneling resistance representing the magnetic information of ``0'' and ``1'' states in a magnetic memory, which is the so-called tunnelling magnetoresistance (TMR) effect. The TMR ratio is defined as
\[
\text{TMR ratio} = \frac{R(\text{high}) - R(\text{low})}{R(\text{low})}.
\]

Although the above physical picture is concise and clear, there has been a long way in history to understand it in all details and, most importantly, to improve the spin-valve effect in a practical device. For decades, researchers have investigated each part of the sandwiched structures, including the free and pinning layer ferromagnet, the insulating spacer, and different growth/annealing conditions, etc. With the advent of 2D magnetic vdW materials, one can now easily achieve a large spin-valve effect in vdW heterostructures, which also exhibits great advantages to advance the TMR topic, including high gate tuneability, high modulation potential by insulating material and thickness, and easy fabrication, etc.

In 2018, Wang \textit{et al.} reported a spin valve device using a vdW magnet, that is the FGT/hBN/FGT heterostructure, and obtained a record-high TMR ratio of $\sim$160\% at 2 K [Fig.~\ref{fig:30}(a)]~\cite{RN157}. At the same time, another group also conducted very systematic investigations on the vdW TMR topic. In the beginning, they fabricated the FGT/MoS$_2$/FGT vdW heterostructure~\cite{RN158}, and obtained a TMR ratio of 3.1\% at 10 K. This TMR ratio is around 8 times larger than that of the reported spin valves based on MoS$_2$ sandwiched by conventional ferromagnetic electrodes and monotonically decreases upon increasing temperature following Bloch's law. As the bias current decreases exponentially, the MR increases linearly to a maximum of 4.1\%. In 2021, they changed the spacer to InSe and observed two distinct behaviors: tunneling and metallic [Fig.~\ref{fig:30}(b)]. Each behavior was attributed to a pinhole-free tunnel barrier at the FGT/InSe interface, and metallic behavior with pinholes in the InSe spacer layer, respectively~\cite{RN159}. Tunneling devices host a large TMR ratio of 41\% under an applied bias current of 0.1~$\mu$A at 10 K, three times greater than metallic devices. Furthermore, the tunneling device exhibits a lower operating bias current and a more sensitive bias current dependence than the metallic device. In 2023, the insulating spacer was replaced by GaSe, and the TMR ratio was boosted to 192\% at 10 K. The TMR ratio was significantly modulated by varying the thickness of the spacer layer and also the bias voltage~\cite{RN160}. So far, it has been very popular to use a vdW magnet and a vdW heterostructure for spin valve devices and large TMR.

In addition to the larger TMR ratio, interesting manipulation and new physics beyond the conventional spintronics were also enabled by the magnetic vdW materials. Min \textit{et al.} discovered the gate-tunable TMR ratio and, most importantly, the emergence of a negative TMR effect in FGT/hBN/FGT and FGT/MoS$_2$/FGT devices [Fig.~\ref{fig:30}(c)]. Through a physical picture and DFT + DMFT calculations, they ascribed this effect to the bias-modulated energy-band alignments with localized minority spin states. As shown in [Fig.~\ref{fig:30}(d)], with a bias voltage $V_b$ applied, spin polarized electrons in the energy window ($E_F - eV_b$, $E_F$) of FM 1 are injected into the empty states of FM 2 in ($E_F$, $E_F + eV_b$). In particular, when a sharp density of states (DOS) peak of the filled states with minority spins in FM 1 aligns with the relatively large DOS hump of the empty states with majority spins in FM 2, the TMR ratio is expected to reverse its polarity to negative~\cite{RN156}. They demonstrated electrically tunable, highly transparent spin injection and detection across the vdW interfaces. More interestingly, the net spin polarization of the injected carriers can be modulated and reversed in polarity via electrical bias [Fig.~\ref{fig:30}(e)], beneficial from the high-energy localized spin states in the metallic ferromagnet FGT~\cite{RN156}. A similar effect was also found in Fe$_3$GaTe$_2$/GaSe/Fe$_3$GaTe$_2$ vdW spin valve device by Zhu \textit{et al.}~\cite{RN160}. Such new phenomena and physics are extremely difficult to achieve in conventional spintronics, if not impossible, but are easily and more universally produced in vdW magnet-based heterostructures. With the topic continuing, one can expect more research opportunities in this direction. Interestingly, if the insulating spacer is removed, one may also observe plateau-like magnetoresistance in the FGT/FGT homojunction, probably due to weak interlayer coupling and a larger interface gap~\cite{RN161,RN162}.

\subsection{Gigantic spin-orbit torque, field-free switching, and all-vdW memory}

\begin{figure*}
    \centering
    \includegraphics[width=\linewidth]{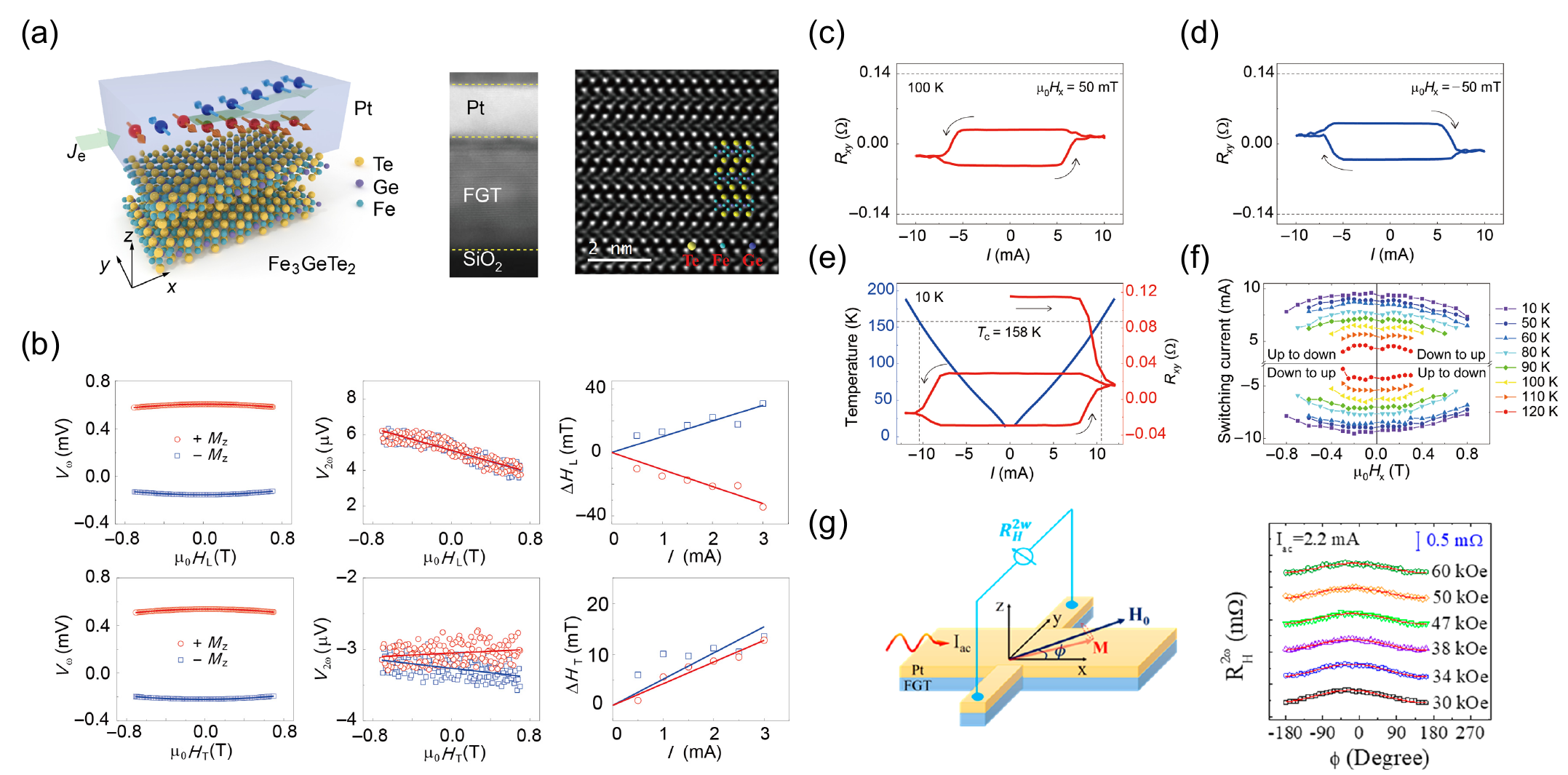}
    \caption{\label{fig:31}Spin-orbit torque in FGT/heavy-metal-Pt composite devices. (a) Schematic and TEM image of the FGT/Pt system. (b) First and second harmonic Hall voltages versus two orthogonal in-plane magnetic fields, from which spin-orbit torque effective fields are extracted. (c--f) Current-driven switching by spin-orbit torque under positive and negative in-plane magnetic fields. The Joule heating situation is analyzed, and the switching current is obtained at various magnetic fields and temperatures. (g) Another second harmonic Hall voltage measurement scheme and its dependence on the angle between the in-plane magnetic field and current. Figs.~6.5(a--g) are reproduced from ~\citet{RN163}; Fig.~3 in ~\citet{RN163}; Fig.~4A in ~\citet{RN163}; Fig.~4B in ~\citet{RN163}; Fig.~4C in ~\citet{RN163}; Fig.~4D in ~\citet{RN163}; and Fig.~4A--B in ~\citet{RN164}, respectively.}
\end{figure*}

In addition to the TMR effect, recently developed spin-orbit torque is the most central topic of spintronics, as it hosts massive advantages and possibilities and, most importantly, is closest to practical next-generation memory applications and in-memory computing \cite{RN196,RN197}. Current can generate spin-orbit torque in traditional ferromagnet/heavy-metal hybrid systems, mainly coming from the spin-orbit coupling. The two mainstream physical mechanisms are the spin Hall effect and the interfacial Edelstein effect. Specifically, charge current can generate spin current by injecting it into the ferromagnet or accumulating spin imbalance at the interface. Both can produce a torque onto the magnetization of the ferromagnet layer, eventually leading to the magnetization switching.

The spin-orbit torque related to vdW magnets was first and extensively investigated using a magnetic vdW material FGT since it is the first and only vdW ferromagnetic metal for the previous several years and hosts the most favorable properties for spintronics including perpendicular hard ferromagnetism, high Curie temperature up to $\sim200$ K, etc. \cite{RN143,RN139}. In 2019, Wang \textit{et al.}~\cite{RN163} and Alghamdi (Alghamdi~\cite{RN164} adopted the above classic SOT architecture and made the FGT/Pt heterostructure [Fig.~\ref{fig:31}]. They investigated the SOT of the device by field-dependent and angle-dependent second-harmonic measurement, respectively, and also demonstrated the SOT-induced switching by current under an in-plane magnetic field. The effective SOT field per current is around $\sim$5 Oe/(mA/$\mu$m$^2$), 10 times greater than conventional devices such as Pt/Fe, Ta/Fe, etc.

\begin{figure*}
    \centering
    \includegraphics[width=\linewidth]{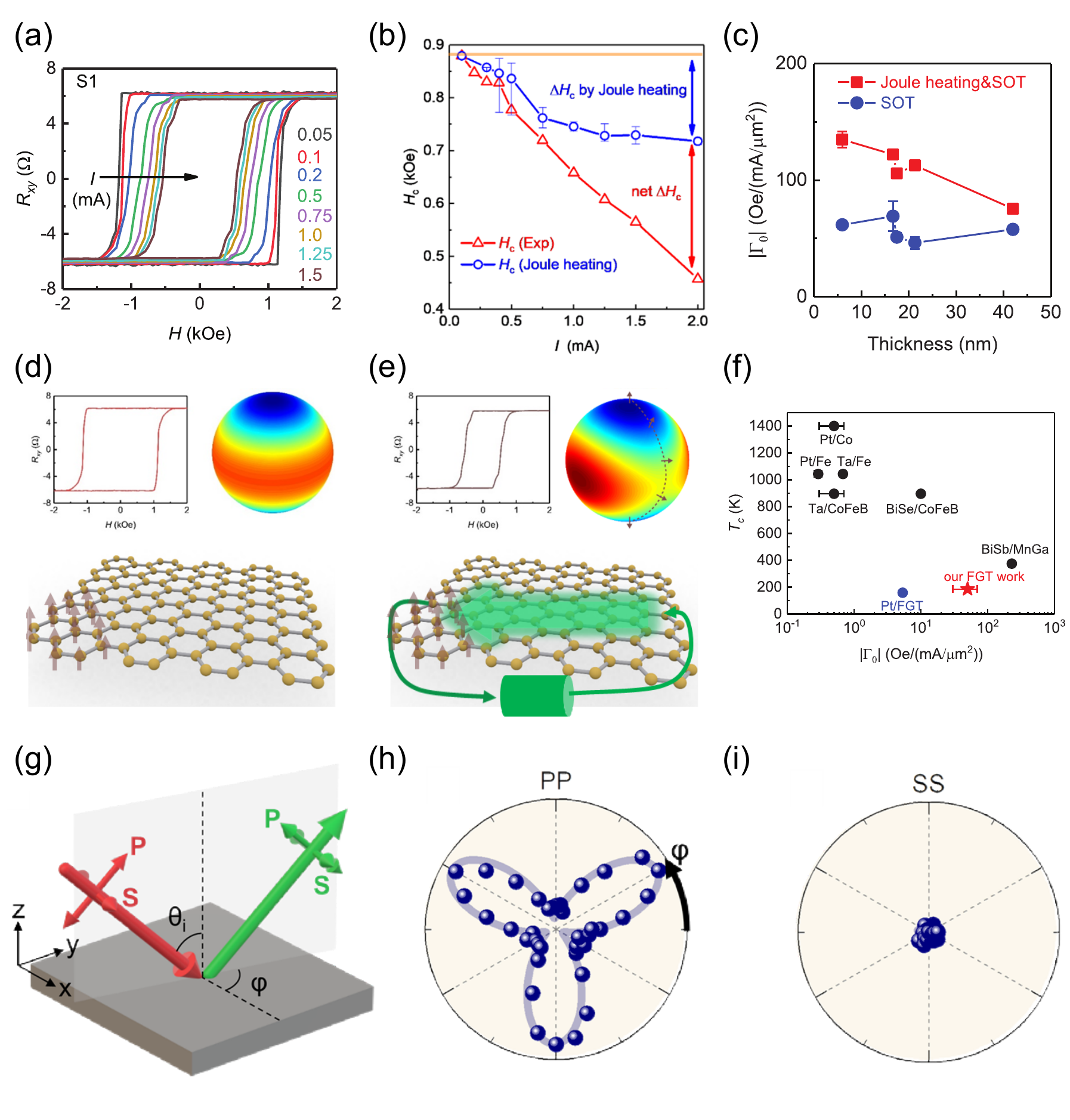}
    \caption{\label{fig:32}Gigantic intrinsic spin-orbit torque and inversion symmetry breaking in FGT. (a) Significant coercivity reduction upon increasing current. (b) Large coercivity reduction after subtracting the contribution from Joule heating. (c) Spin-orbit torque magnitude extracted from current-driven coercivity reduction before (red) and after (blue) subtraction of the Joule heating contribution. (d--e) Physical mechanism without (d) and with (e) current: current-driven spin-orbit torque can directly reduce the magnetic anisotropy and the switching barrier. (f) The intrinsic spin-orbit torque magnitude is nearly 100 times greater than that of heavy metals such as Pt and Ta. (g--i) Schematic and results of SHG measurement. Fe defects related to inversion symmetry breaking occur in FGT, predominantly along the out-of-plane direction. Figs.~6.6(a--i) are reproduced from Fig.~2a in ~\citet{RN52}; Fig.~S4f in ~\citet{RN52}; Fig.~3e in ~\citet{RN52}; Fig.~3b in ~\citet{RN52}; Fig.~3b in ~\citet{RN52}; Fig.~3f in ~\citet{RN52}; Fig.~1a in ~\citet{RN170}; Fig.~1b in ~\citet{RN170}; and Fig.~1c in ~\citet{RN170}, respectively.
}
\end{figure*}

Note that FGT is a topological nodal-line material with large Berry curvature \cite{RN165}, which leads to the large anomalous Hall Effect \cite{RN165} and anomalous Nernst effect \cite{RN166}. Related to this, another new kind of SOT, i.e., gigantic intrinsic SOT in a single FGT \cite{RN45} itself without any heavy-metal layer, is also reported in several studies. In 2019, Johansen \textit{et al.}, discussed the possible SOT in the FGT monolayer considering special geometrical symmetries \cite{RN167}. They found that under these symmetries, most of the SOT coefficients are canceled out, and only one parameter $\Gamma_0$ is left alone. The spin-orbit torque term is so simple that it can be incorporated into its free energy expression, indicating that SOT can directly change the magnetic anisotropy. At the same time, Zhang \textit{et al.} performed a comprehensive study that combined theory and experiment and published their results in 2020 \cite{RN52}. In this work, the authors developed the theory further but differently: They started from the Kubo formula and derived the SOT magnitude expression, with which they calculated the SOT strength $\Gamma_0$ to be about 30 Oe/(mA/$\mu$m$^2$) from the Berry curvature of FGT's topological bands \cite{RN52}. This $\Gamma_0$ tells how much magnetization was accumulated per applied current, reflecting the deeply underlying atomic magnetoelectric effect in FGT \cite{RN49}. In the experiment, Zhang \textit{et al.} found that the coercivity of the hysteresis loops is greatly suppressed upon increasing current. For example, a small current of 1.5 mA can reduce the coercivity by more than 50\% [Fig.~\ref{fig:32}(a)], much larger than the previously reported value of 5\% for the 3-dimensional ferromagnet \cite{RN52}. One should be concerned with Joule heating while dealing with a current-dependent experiment. They employed three independent methods to assess Joule heating and concluded that Joule heating cannot fully explain our experiment result. For example, after subtracting the Joule heating's contribution, the remaining net coercivity reduction is still large, roughly half the total coercivity reduction [Fig.~\ref{fig:32}(b)]. They also performed a detailed assessment of the Joule heating by three different approaches -- Method I: fabricating NbSe$_2$/FGT heterostructure and then use of NbSe$_2$ as a nanofabricated thermometer; Method II: use of the longitudinal resistance $R_{xx}$ of FGT itself as an internal thermometer; Method III: using COMSOL simulations to estimate Joule heating. All three methods give the same result with temperature indifference within 10 K and support the remaining large coercivity reduction after subtracting the Joule heating contribution \cite{RN52}.

In their work, the authors attribute this net reduction in coercivity to the intrinsic spin-orbit torque of the FGT. Based on the SOT scenario, they explained the experimental observations: FGT is a hard ferromagnet without current. If the magnetization changes from the up to the down directions, the spin needs to overcome a uniform perpendicular magnetic anisotropy energy barrier ($M_s \times K_z/2$). When current is applied, magnetic anisotropy can be suppressed along a particular direction by SOT (via the term $\Gamma_0 \times J$), leading to a heavily reduced coercivity [Fig.~\ref{fig:32}(d-e)]. In other words, current-driven SOT can induce the hard-to-soft transition in FGT \cite{RN52}. Such in-plane anisotropy in the free-energy landscape was also supported by the in-plane AMR effect in their work. They also obtained the SOT magnitude from the coercivity reduction experimental data [Fig.~\ref{fig:32}(c)], close to the calculated value. It is about 100 times larger than conventional heavy metals Platinum and Tantalum [Fig.~\ref{fig:32}(f)]. Note that after subtracting the Joule heating's contribution, the net SOT field remains similar within the FGT’s thickness range from 6 to 40 nm [Fig.~\ref{fig:32}(c)] \cite{RN52}.

Several unique points are emphasized for FGT's intrinsic SOT \cite{RN52}. First, SOT emerges within a single FGT itself without a heavy-metal layer. Secondly, such an SOT can be incorporated into its free energy and change magnetic anisotropy directly. Third, although the energy barrier is reduced, the SOT term formula does not involve $m_z$, so up/down magnetization states are still degenerate in energy under current, meaning that such intrinsic self-SOT cannot make deterministic switching by pure current without a magnetic field in principle. Finally, the gigantic SOT value is due to the large Berry curvature of its topological bands, providing a guide to exploring new large SOT systems with band topology.

There is another symmetry issue on FGT's SOT: Generally, SOT requires inversion symmetry breaking because it can cause asymmetric carrier/spin scattering, and thus net imbalanced spin at two opposite directions, i.e., spin polarization. Monolayer FGT is indeed noncentrosymmetric. However, multilayer or bulk FGT was previously believed to be centrosymmetric since an FGT unit cell consists of two neighboring layers that are mutually inversion partners, causing a contradiction between symmetry and SOT. The previous work \cite{RN52} regards bulk FGT's SOT as a hidden SOT in analogy to the similar hidden Rashba effect in centrosymmetric systems \cite{RN498}: the intralayer interaction is about 1 eV, which is about 1000 times larger than the interlayer interaction, and predominates the spin-orbit torque so that the SOT in neighboring layers does not cancel out. Very recently, another contribution of inversion symmetry breaking was added to the intrinsic SOT of FGT \cite{RN170}. Using the SHG technique \cite{RN170}, they found that inversion symmetry breaking does exist in FGT due to defect, dominantly along the out-of-plane direction [Fig.~\ref{fig:32}(g-i)]. This SHG signal is independent of temperature, but sensitive to Fe deficiency. Higher Fe deficiency tends to stabilize the inversion-symmetry-breaking status and reduce the space group from centrosymmetric $P6_3/mmc$ to noncentrosymmetric $P3m1$, consistent with a recent XRD report of noncentrosymmetry in FGT \cite{RN499}. Soon after, Wu \textit{et al.} also demonstrated similar phenomena in another material Fe$_5$GeTe$_2$ \cite{RN500}. The significance of this discovery affects much more beyond FGT or SOT itself as it can be generalized to understand the contradictions between pristine centrosymmetric systems and inversion-symmetry-breaking-required phenomena in similar systems with layer inversion. Moreover, broken inversion symmetry also implies accompanying more emergent phenomena and numerous rich physical properties, including nonreciprocal electronic transport, the nonlinear Hall effect, the nonlinear optical response, and topological vdW polar metal, to name only a few. However, one needs to be careful when starting those researches in FGT since the discovered symmetry breaking is not strong, so the induced effect may not be prominent enough for observation (see more details in Ref.~\cite{RN170}).

\begin{figure*}
    \centering
    \includegraphics[width=\linewidth]{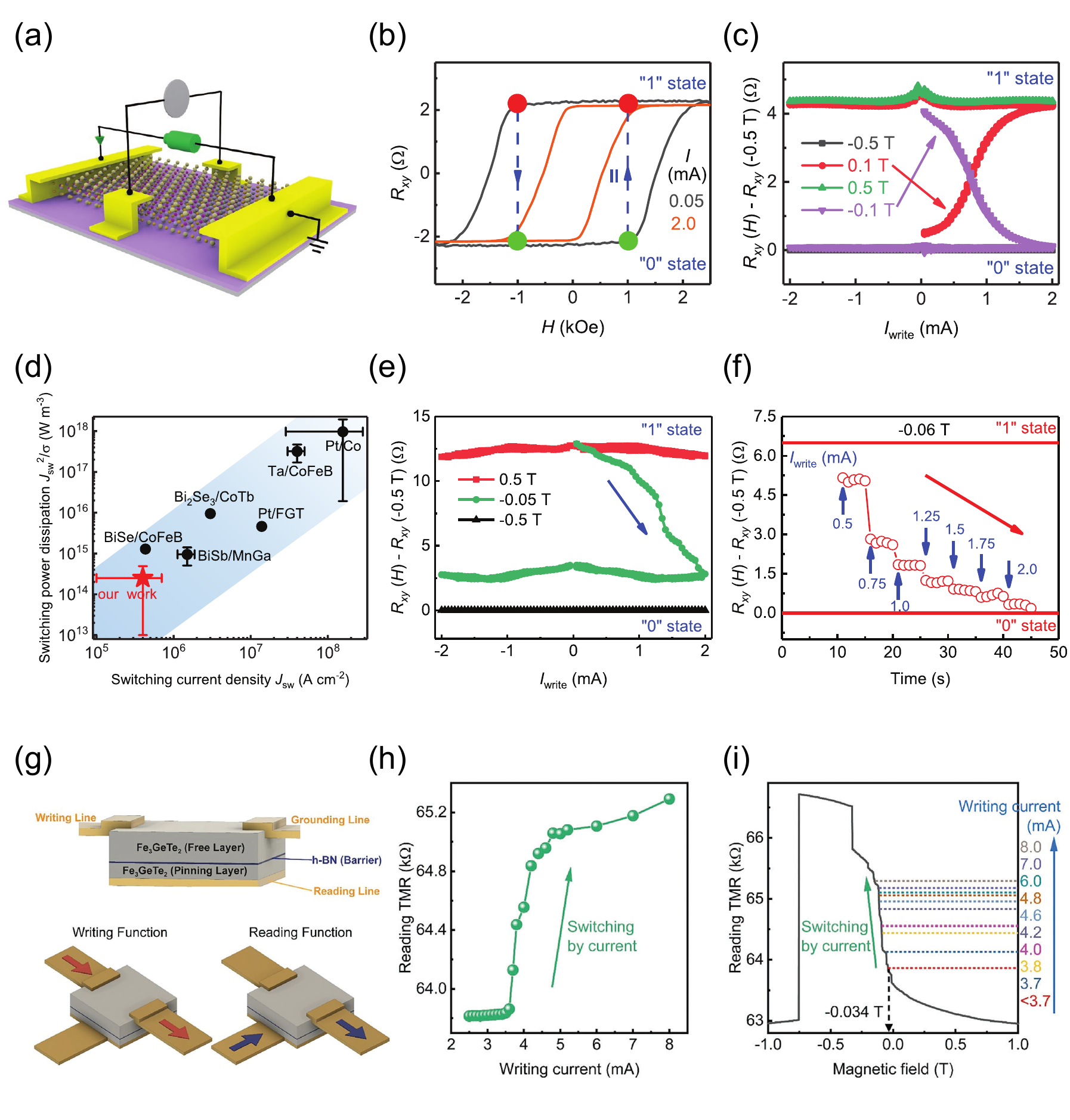}
    \caption{\label{fig:33}Spintronic applications based on the gigantic intrinsic spin-orbit torque of FGT. (a) Schematic of the first device demonstration, where information is read out via a simple Hall measurement. (b) Schematic of current-driven switching. (c) Experimental realization of current-driven magnetization switching. (d) The required switching current density and power dissipation are about 400 and 4000 times smaller, respectively, than those in Pt-based systems. (e) Controlling the writing current enables the realization of an intermediate state. (f) Multi-level states controlled by writing current as illustrated in (e). (g) Schematic of the second device demonstration: an all-vdW 3-terminal SOT-MRAM composed of FGT/hBN/FGT heterostructure. Writing current flowing in the free-layer FGT modifies its magnetization via SOT. Reading current through the tunneling junction detects information via the TMR effect. This decoupling of writing and reading paths enhances design and optimization flexibility, as well as device endurance. (h--i) The writing and reading principles are successfully demonstrated. Figs.~6.7(a--i) are reproduced from Fig.~2a in ~\citet{RN168}; Fig.~2c in ~\citet{RN168}; Fig.~2d in ~\citet{RN168}; Fig.~3d in ~\citet{RN168}; Fig.~4c in ~\citet{RN168}; Fig.~5c in ~\citet{RN168}; Fig.~1b--c in ~\citet{RN169}; Fig.~4b in ~\citet{RN169}; and Fig.~4c in ~\citet{RN169}, respectively.
}
\end{figure*}

Based on gigantic intrinsic SOT in FGT, magnetic memory devices \cite{RN168} were made, and current magnetization switching is highly efficient and nonvolatile [Fig.~\ref{fig:33}(a--f)]. Switching current density and power consumption are about 400 and 4000 times lower compared to platinum-related devices [Fig.~\ref{fig:33}(d)]. Moreover, multilevel states were controlled by current, of roughly 8 states corresponding to 3 bits, which can enhance information capacity and reduce computing cost in a single device [Fig.~\ref{fig:33}(e--f)]. Taking one more step forward, they further develop the memory for the vdW three-terminal SOT-MRAM composed of an FGT/hBN/FGT heterostructure [Fig.~\ref{fig:33}(g--i)] \cite{RN169}. Note that such a bulky self-SOT was also further supported by the SHG measurement from Martin \textit{et al.} independently \cite{RN501}. In addition, Hejazi \textit{et al.} did imaging on current-driven switching in FGT \cite{RN54}, and three more groups replaced the FGT with the same-structured Fe$_3$GaTe$_2$, pushing the intrinsic SOT and the corresponding switching application above room temperature \cite{RN173,RN174,RN175}. As a passing remark, very recently self-SOT was also discovered in the vdW polar metal material Fe$_{2.5}$Co$_{2.5}$GeTe$_2$, benefiting from a different source of inherent polarization \cite{RN502,RN503,RN504}.

Despite the interesting physics of the unconventional gigantic intrinsic SOT in FGT, its understanding does not seem straightforward. Recently, researchers have become more interested in using it as a ferromagnet combined with other intriguing vdW materials, e.g., topological materials, to achieve better performance in hybrid heterostructures. In 2020, two groups built the FGT/Bi$_2$Te$_3$ heterostructure. They raised the Curie temperature to 400~K by interfacial exchange coupling \cite{RN505} and performed efficient SOT switching by current under an in-plane magnetic field \cite{RN505}. Furthermore, an emerging interfacial antiferromagnetism has been artificially created in the FGT/MnPS$_3$ heterostructure, producing an intermediate Hall plateau, and is well manipulated under high pressure \cite{RN506}. More importantly, vdW semimetal WTe$_2$ with low-crystal symmetries was originally used in the 3D ferromagnetic metals spintronics community and has recently expanded to vdW magnet spintronics. Taking advantage of the out-of-plane spin polarization induced by the unconventional spin Hall effect in WTe$_2$, magnetic field-free switching was achieved in the WTe$_2$/FGT device with current applied along the $a$-axis (no switching along the $b$-axis) of WTe$_2$ \cite{RN507}. Recently, Keum \textit{et al.} combined vdW magnet and oxide spintronics and achieved novel magnetic-field-free switching in FGT/SrTiO$_3$ heterostructure~\cite{RN508}, replacing Ge with Ga makes the Curie temperature increase to \(350\text{--}380\,\mathrm{K}\)~\cite{RN155}, well above the room temperature, which drives one's attention to Fe$_3$GaTe$_2$, a room temperature ferromagnet sharing the same physical characteristics as Fe$_3$GeTe$_2$ in structures, symmetries, band topologies, etc. Soon after, magnetic field-free switching was also reported in Fe$_3$GaTe$_2$/WTe$_2$~\cite{RN509} and Fe$_3$GaTe$_2$/wedged-Pt~\cite{RN510} devices. Note that those Fe$_3$GeTe$_2$ and Fe$_3$GaTe$_2$-related heterostructures generally have lower switching current density than conventional devices. We suspect that the gigantic intrinsic SOT in the FGT itself may also facilitate reducing the switching current density during the switching process of these heterostructures.

A final matter in vdW magnet spintronics is whether a real SOT-MRAM can be made experimentally. SOT-MRAM hosts massive inherent advantages, including the non-volatility of flash memory, high speed of SRAM, high density of DRAM, low cost, low power consumption, non-destructive read-out, permanent retention, radiation hardness, etc. Regarding this part, Cui \textit{et al.} made a three-terminal SOT-MRAM unit using an all-vdW heterostructure of top-FGT/h-BN/bottom-FGT [Fig.~\ref{fig:33}(g--i)]~\cite{RN169}. The current-driven gigantic intrinsic SOT can write the magnetic information in the top FGT, which can then be read out by the industry-practical TMR effect of the whole tunnel junction, with the bottom FGT as the magnetic pinning layer. The writing and reading processes and paths are physically separated or decoupled in this three-terminal SOT-MRAM, naturally enhancing the device's design and optimization flexibility and endurance [Fig.~\ref{fig:33}(g--i)]. Furthermore, Pan \textit{et al.} integrates all the reported best metrics in vdW magnet spintronics, realizing the magnetic-field-free SOT-MRAM of WTe$_2$/Fe$_3$GaTe$_2$/hBN/Fe$_3$GaTe$_2$ working at room temperature~\cite{RN511}. These works demonstrate that the three-terminal SOT-MRAM concept in conventional spintronics can be readily expanded in vdW magnet spintronics and push forward the currently flourishing 2D magnetic materials toward the subsequent march of industrial application.

An important question that still remains is: What are the advantages of vdW magnet spintronics over conventional non-vdW magnet spintronics? It is still challenging to provide a definitive answer, because while vdW magnets offer numerous opportunities for spintronics, concrete beneficial examples are fast emerging. Here are some of our perspectives:

\begin{enumerate}
    \item Previously, there was uncertainty about how easily spintronics concepts and techniques could transition to vdW magnets. However, recent research shows that with advances in 2D materials technology, spintronics principles can be effectively and efficiently applied to vdW magnets, expanding the materials available for spintronics research.

    \item vdW magnets introduce new spintronics ideas such as intrinsic spin-orbit torque (SOT) without heavy metals, topological Berry curvature, and magnon transport in nanoscale vdW antiferromagnets. Being inherently 2D materials, vdW magnets possess the following characteristics typical of 2D materials: atomic thickness, high tuneability, controllability, compatibility, and flexibility in design and tailoring. This facilitates efficient exploration and development of spintronic functionalities, such as rapid iteration and optimization of spintronic functions, like achieving high TMR ratios through diverse heterostructures built in hours using 2D-materials transfer techniques.

    \item Here, we name a few advantages of vdW magnet devices. Concrete advantages highlighted in recent references include achieving TMR ratios of up to \(\sim160\%\) or more with vdW magnet-based heterostructures tunable by gating---a feat not easily achieved in conventional spintronics. Switching current densities for SOT are generally lower (\(1\text{--}10\,\mathrm{MA/cm^2}\)) than many conventional systems.

    \item However, industrial applications of vdW magnets remain uncertain, pending the discovery of an ideal 2D magnet-much-above-room-temperature ferromagnetism, low current densities, high speed, large TMR ratios, and most importantly, ease of handling and robustness.

    \item Recently, vdW magnet spintronics has been extended to employing spin-orbit torque to tune exotic quantum magnets, e.g., current-driven collective control of helical spin texture in antiferromagnets~\cite{RN512}.
\end{enumerate}

At present, it is important to note the broad opportunities that vdW magnets offer for advancing spintronics, generating new ideas, concepts, and physics that accelerate the field's development.

\section{\label{sec:VI}Outlook and Summary}

This section shares our perspective on the rapidly evolving field of vdW magnets with a brief summary. It is remarkable how much progress has been made since the first report on vdW monolayer magnets in 2016. Over the past decade, the field has transitioned into a new phase that requires more focused efforts on specific challenges. Here, we summarize some past successes and highlight outstanding problems, hoping to inspire further discussion and future advancements.

\subsection{Materials}
Section \ref{sec:II} provided an overview of various vdW magnets reported so far, which exhibit a wide range of magnetic and electronic ground states \cite{RN59}. The field has matured significantly, encompassing a diverse set of materials with diverse magnetic and electronic ground states as discussed in Section \ref{sec:II}. Additionally, vdW systems allow the realization of all four fundamental two-dimensional lattice types: honeycomb, triangular, Kagome, and square.

Although the diversity of vdW magnets is already impressive, a key challenge remains: achieving higher transition temperatures, particularly those exceeding room temperature. A promising approach involves the Fe$_n$GaTe$_2$ family ($n = 3, 4, 5$) \cite{RN150}, where increasing the number of Fe atoms leads to higher transition temperatures. However, two fundamental questions remain:

Can pristine monolayer vdW ferromagnets achieve transition temperatures above room temperature?
The answer depends on the chemical bonding strength between the magnetic ions and the surrounding chalcogens or halogens. These bonds tend to be weaker than those in magnetic oxides, leading to lower transition temperatures. Optimizing these bonds could be the key to overcoming this limit.
    
What is the practical upper limit for transition temperatures in vdW magnets?
Materials must sustain transition temperatures well above 500 K for industrial applications to ensure operational stability. Whether magnetic chalcogenides or halides can achieve such robustness remains an open question.

Another critical issue is material stability, as many vdW magnets are highly air-sensitive. For example, CrI$_3$ degrades within minutes of exposure. However, some materials, such as TMPS$_3$, CrSBr, and CrPS$_4$, exhibit exceptional air stability. Improving transition temperatures and stability is essential for the advancement of fundamental research and practical applications of vdW magnets.

\subsection{Spin Hamiltonian}

The initial motivation behind the study of vdW magnets was to determine whether stable monolayer magnetism could exist in real materials. However, a broader objective was to experimentally realize the three fundamental spin Hamiltonian models: Ising, XY, and Heisenberg~\cite{RN4}. Today, multiple vdW systems serve as the realization of these models, aligning well with theoretical predictions.

Further exploration of vdW magnets in the context of the Ising and Heisenberg models may yield diminishing returns, as these models are well understood. However, the XY model presents an exciting avenue for future research. Although attempts have been made to realize the XY model in Josephson junctions \cite{RN513} and liquid helium \cite{RN514}, these studies have largely focused on static studies. The dynamics of vortex-antivortex pairs, a defining feature of the XY model, remains largely unexplored.

Identifying an ideal vdW material that closely approximates the XY model remains a challenge. Many proposed candidates have been found to align more closely with the XXZ model due to residual out-of-plane fluctuations \cite{RN208}. One possible approach is applying an external magnetic field, which could selectively turn a system into an ideal XY regime. This strategy has been successfully applied in quantum critical point studies, such as the transverse Ising model, and could prove similarly useful for vdW magnets.

\subsection{Optical Studies}

Optical techniques, including SHG, Raman spectroscopy, MOKE, THz spectroscopy, and optical absorption, have provided valuable insight into vdW magnets. However, several unresolved questions remain.

First, the ultrasharp magneto-exciton peak in NiPS$_3$ is still not fully understood. Despite theoretical efforts, the peak's extraordinary sharpness, resembling a coherent excitation, remains a mystery. Second, the coupling between phonons, nonlinear optical properties, and the magnetic states in vdW magnets is largely unexplored. Third, quasiparticle interactions, such as magnon-exciton and magnon-phonon interactions, are emerging topics that require further study, particularly in the vdW magnet-superconductor interfaces.

Magneto-optic measurements are still in their early stages, with much work needed to fully characterize linear dichroism (optical rotation), linear birefringence (ellipticity), circular dichroism, and circular birefringence.

\subsection{Light-induced out-of-equilibrium dynamics}

Driving vdW magnets out of equilibrium using pulsed electromagnetic fields is an emerging research direction. This approach has potential applications in next-generation electronic, photonic, and spintronic devices. Although optical switching of magnetic order via photocarrier excitation has been demonstrated, discovering new nonequilibrium phases remains an open challenge. 

One promising method is the selective excitation of magnons and phonons to large amplitude. This could drive a system into an adjacent local minimum in the free-energy landscape or produce quasi-static rectified responses via nonlinear coupling to other modes, which modify the Hamiltonian \cite{RN421}. Lattice and magnetic anharmonicities have recently been observed in vdW magnets, setting the stage for exploring new out-of-equilibrium quantum phases of matter through phononic or magnonic control \cite{RN357,RN359}. 

Another emerging technique is Floquet engineering \cite{RN419}. Pushing beyond recent demonstrations of Floquet band engineering, the next key challenge is to demonstrate Floquet engineering of many-body phases. In the context of vdW magnetism, this includes coherent tuning of spin exchange interactions and magnetic anisotropies in an effort to drive materials into exotic magnetic ground states such as quantum spin liquids \cite{RN472,RN476,RN474,RN446}. Realizing this vision requires the development of driving protocols that minimize heating and decoherence. A promising approach is to drive vdW insulators at frequencies below the band gap while carefully selecting Keldysh parameters to minimize multiphoton absorption and quantum tunneling effects \cite{RN515,RN516}.

The exciting frontier involves extending the concepts of cavity quantum electrodynamics to vdW magnets, bridging quantum optics and quantum materials \cite{RN517,RN518}. 2D cavities integrated with vdW materials provide a versatile platform for polaritonics \cite{RN519,RN306}, enabling strong light-matter coupling and the potential realization of magnon-polaritons or Bose-condensed exciton polaritons in vdW magnets. Leveraging the tunability of moir\'e heterostructures of vdW magnets can potentially unveil additional cavity-induced quantum collective phenomena that involve the delicate interplay of electronic correlations, topology, and light \cite{RN520}.

\subsection{Spintronics}

vdW magnets offer unique opportunities to advance spintronics concepts and applications. Here are a few potentially interesting directions: Developing new spintronic concepts, The most significant feature of vdW magnets to spintronics is their 2D layered nature. They belong to a representative class of 2D materials, inheriting all the advantages such as exfoliable atomically thin structures, high compatibility and design flexibility, efficient assembling with a clean interface, etc. These merits render the efficient testification of novel ideas quicker than conventional 3D spintronics. From this perspective, one can foresee the burgeoning conceptional developments in different topics of traditional spintronics such as TMR \cite{RN156,RN160}, magnon transport, and spin-orbit torque. Indeed, several intriguingly new discoveries are pushing in this direction, such as bias-tunable positive/negative TMR, anisotropic and gate-tunable magnon transport \cite{RN490,RN491}, gigantic intrinsic spin-orbit torque \cite{RN52,RN167,RN170,RN501}, unconventional out-of-plane spin-orbit torque \cite{RN507,RN508}, current-driven tuning of exotic spin texture in intriguing quantum magnets \cite{RN512}, to name only a few.

Enhancing energy efficiency limit and response speed, Another prominent merit of vdW magnet spintronics is their high energy efficiency. Although one did not initially expect this point, many experiments have demonstrated the low current density and power dissipation in spintronic devices built on vdW magnets. For example, the spin-orbit torque switching current density can be easily achieved on the order of $10^6$--$10^7$ A/cm$^2$, which is generally hard to achieve in conventional spintronics with orders of magnitude lower values. One can expect even higher energy efficiency by carefully fabricating high-quality 2D magnetic heterostructure or homostructure with a superior interface. Moreover, integrating vdW antiferromagnets can facilitate robust, high-speed, and high-density in-memory computing due to their stability to the external magnetic field, terahertz spin dynamics, and absence of stray field \cite{RN512,RN521,RN522,RN523,RN524}.

Engineering and optimizing diverse technology routes, As mentioned earlier in this section, vdW magnets endow the efficient handling and fabrication of various spintronic devices, benefiting from advanced 2D material transfer techniques \cite{RN525,RN526,RN161,RN27}. This can also expedite the engineering and optimization process of spintronics device technologies. Several examples have already been documented in the literature, such as the twisted FGT homostructure with plateau resistance \cite{RN161}, large TMR ratio and spin filter effect \cite{RN157,RN50}, and most importantly, all vdW three-terminal SOT-MRAM architecture \cite{RN169} and unconventional magnetic-field-free switching \cite{RN507,RN508}.

Quantum spintronic physics and devices, 2D vdW nonmagnetic materials also bear emergent quantum characteristics with enormous cases in the previous literature, as do the 2D vdW magnets. When vdW magnet spintronics meets the fundamentally important concepts in modern condensed matter physics, such as symmetry, topology, correlation, twisted moir\'e pattern, magnetoelectric coupling, magnetic/superconducting proximity, and also nonequilibrium many-body interactions, tremendous opportunities lay ahead by designing and fabricating the layered spintronics-incorporated devices.

Predicting the trajectory of vdW magnets is challenging, but promising avenues include multiferroicity, magnetoelectric (ME) effects, and heterostructures \cite{RN528,RN529,RN506,RN49}. With their inherent thinness, the advent of 2D vdW multiferroic materials offers a transformative solution to some of the existing technical problems, opening new horizons for application and research \cite{RN31}.

\subsection{Others}

Interestingly, some vdW materials exhibit ME effects that far exceed those of their oxide-based counterparts. For example, Fe$_3$GeTe$_2$ has demonstrated a current-induced spin-orbit torque (SOT) with a substantial ME coefficient ($\Gamma_0 = 50$ Oe/mA/$\mu$m$^2$), among the highest reported in real materials \cite{RN12}. This characteristic highlights significant potential for spintronics applications. Remarkably, the ME effect in Fe$_3$GeTe$_2$ is accompanied by a combination of three simple symmetries: $3z$, $m_y$, and $m_z$. This simplicity suggests that other naturally occurring materials may exhibit similarly intriguing behaviors.

The recent concept of sliding ferroelectricity is therefore interesting \cite{RN530}. When two vdW materials are stacked in a manner that breaks inversion symmetry, ferroelectricity naturally arises. This concept, both straightforward and generalizable, can be extended to vdW magnets that are not inherently ferroelectric. Beyond increasing the pool of vdW multiferroic systems, this approach also introduces opportunities for entirely new physical phenomena. For example, when two thin layers of vdW magnets are stacked and twisted, flat-band physics may emerge, offering an exciting platform for exploring quantum phenomena. These twisted systems could enable studies of quantum geometry and other frontier topics in condensed matter physics, marking a transformative step for the field \cite{RN531}.

Another exciting area for future exploration that has not been much discussed in this review for page limits is heterostructures using vdW magnets. We think that this topic deserves a separate review. With many 2D materials reported so far, including graphene, many TMDC (transition metal dichalcogenide), and several exciting ground states like superconductivity, magnetism, and even ferroelectricity, the field is wide open for numerous opportunities. In a sense, one can call this area `playing quantum cards'. Given how many discoveries have been made even with simple vdW magnets, one can only imagine how many new and thought-provoking discoveries await us.

We end this review on a very high note. Our review has amply shown that since the field of vdW magnet started in the mid-2010s, many exciting discoveries have been made one after another. Some of the highlights discussed in this review from the past few years have justified how useful and exciting these vdW magnets have become. What we outlined in the Outlook section is the tip of the iceberg, with numerous exciting developments awaiting us years ahead. Every new idea and/or sample will create another fruitful subfield and offer an exciting journey.

\begin{acknowledgments}
We are thankful to all those who have helped us come up with an understanding of the fast-moving field of vdW magnets, which is presented in this review, through close collaborations or generous and friendly discussions. We express our special thanks to Beom Hyun Kim, Young-Woo Son, Jong Seok Lee, and Hyun-Woo Lee for their ongoing discussions. We also appreciate the people: Vinh Nguyen, Jihoon Keum, Siwon Oh, Sungjin Park, Suhan Son, Youjin Lee, and Rahul Kumar, who helped us at various stages of editing the manuscript. The work at Seoul National University was supported by the Samsung Science \& Technology Foundation (Grant No. SSTF-BA2101-05) and the Leading Researcher Program of the National Research Foundation of Korea (Grant No. 2020R1A3B2079375). The work of Sogang University was supported by a National Research Foundation project (RS-2024-00450714) funded by the Ministry of Science and ICT and the G-LAMP project (RS-2024-00441954) funded by the Ministry of Education. J.H.K. acknowledges the Samsung Science and Technology Foundation (Grant No. SSTF-BA2102-04) and the National Research Foundation of Korea (Grant No. NRF-2021R1A2C3004989) and the Nano \& Material Technology Development Program through the National Research Foundation of Korea funded by the Ministry of Science and ICT(RS-2023-00281839). The work at Caltech was supported by the Institute for Quantum Information and Matter (IQIM), an National Science Foundation Physics Frontiers Center (PHY-2317110), the Brown Investigator Award, a program of the Brown Science Foundation, and the Gordon and Betty Moore Foundation through Grant GBMF11564 to Caltech to support the work of D.H. One of the authors (C.B.) acknowledges support from a Caltech Presidential Postdoctoral Fellowship and the Clark B. Millikan Prize Postdoctoral Fellowship. One of the authors (J.G.P.) acknowledges the hospitality of the Indian Institute of Science, particularly D. D. Sarma, where some of the manuscript was written during visits supported by the Infosys Foundation.

\end{acknowledgments}

\providecommand{\noopsort}[1]{}\providecommand{\singleletter}[1]{#1}%

\end{document}